\documentclass[traditabstract]{aa} 
\usepackage{graphicx,graphics,amsmath}
\usepackage{color}
\usepackage{amssymb}
\newcommand{\squeezeup}{\vspace{-3.0mm}}
\begin{document}

\squeezeup
   \title{SFXTs versus classical SgXBs: Does the difference lie in the companion wind?}
   \author{P. Pradhan
          \inst{1,2}
        \and E. Bozzo
          \inst{3}
        \and  B. Paul
           \inst{4}        
          }
   \institute{Pennsylvania State University, State College, Pennsylvania, 16801, US ; \email{pup69@psu.edu}
   \and 
   St. Joseph's College, Darjeeling-734104, West Bengal, India
         \and
        Department of Astronomy, University of Geneva, Chemin d’Ecogia 16, Versoix, 1290, Switzerland
        \and 
        Raman Research Institute, Sadashivnagar, Bangalore-560080, India.
             }
   \date{Submitted: -; Accepted -}
\squeezeup
\abstract{We present a comparative study of stellar winds in classical supergiant high mass X-ray binaries (SgXBs) 
and supergiant fast X-ray transients (SFXTs) based on the analysis of publicly available out-of-eclipse observations performed 
with \emph{Suzaku} and \emph{XMM-Newton}. Our data set includes 55 observations of classical SgXBs and 21 observations of SFXTs. 
We found that classical SgXBs are characterized by a systematically higher absorption and luminosity compared to the SFXTs, confirming the results 
of previous works in the literature. Additionally, we show that the equivalent width of the fluorescence K$_\alpha$ iron line in 
the classical SgXBs is significantly larger than that of the SFXTs (outside X-ray eclipses). Based on our current understanding of the physics of accretion in 
these systems, we conclude that the most likely explanation of these differences is ascribed to the presence of mechanisms inhibiting accretion most of the 
time in SFXTs, thereby leading to a much less efficient photoionization of the stellar wind compared to classical SgXBs. 
We do not find evidence for the previously reported anticorrelation between the equivalent width of the fluorescence iron line and the luminosity  
of SgXBs.} 
\squeezeup
\maketitle
\squeezeup
\section{Introduction}
Supergiant X-ray binaries (SgXBs) are usually divided into classical systems and supergiant fast X-ray transients (SFXTs). 
The SFXTs share many properties in common with classical systems (e.g., similar supergiant companions and orbital period distribution) and in all these sources   
high energy emission is mostly due to the accretion of stellar wind from the massive companion onto the compact object. 
Compared to classical systems, SFXTs show a much more pronounced variability, comprising sporadic short 
X-ray outbursts and fainter flares with fast rise times (tens of minutes) and typical durations of a few hours. 
Outside these events, the SFXTs have average X-ray luminosities that are 2-3 orders of magnitude lower than the classical 
systems with similar orbital periods \citep[see, e.g.,][for a recent review]{walter2015}. 
The presence of a neutron star (NS) as a compact object has been established by the detection 
of X-ray pulsations in classical SgXBs and in a few intermediate objects  
between SFXTs and classical SgXBs. There are no confirmed detections of pulsations for any of the known SFXTs\footnote{\scriptsize{Different tentative spin period detections of  
the SFXT IGR\,J17544-2619 have been reported but never confirmed \citep{drave2012_17544,drave2014_17544, romano2015_17544}. Similar cases are those of  
the SFXT IGR\,J18483-0311 \citep{sguera07,ducci13} and IGR J18410-0535 \citep{bamba01,bozzo2011}.}}. 
Cyclotron lines, probing the strength of the NS magnetic field, have been detected in many classical systems, but  only in one SFXT has some evidence been reported 
for a cyclotron feature at $\sim$17~keV  \citep{bhalerao2015}. This was not confirmed by more recent observations \citep{bozzo2016}. \\
The few models proposed to explain the extreme X-ray variability of the SFXTs are still a matter of debate.  
These include extremely clumpy stellar winds \citep{intzand2005_sfxts}, magnetic or centrifugal gates \citep[][]{grebenev2007,bozzo2008}, or the settling 
of a long-lasting quasi-spherical accretion regime \citep{shakura2012}. In the latter two cases, it was shown that reasonably limited clumpy winds 
are needed to achieve the dynamic range of the SFXTs if their activities are sporadically boosted by the effect of the NS rotating magnetosphere or its interaction 
with the magnetized wind from the supergiant companion or both. 

\noindent
Both the magnetic or centrifugal gates and settling accretion regime are likely to inhibit accretion in SFXTs, explaining their subluminosity compared to 
classical SgXBs. Our current limited understanding of the SFXT phenomenology makes any comparative study between these sources and the classical SgXBs particularly interesting. 
In this paper, we exploit archival \emph{Suzaku} \citep{M07} and \emph{XMM-Newton} \citep{jansen2001} 
observations to carry out a comparative analysis of the stellar wind properties in these two classes of systems. 
We focus on the measurement of the average absorbing column density associated with the stellar wind and 
the properties of the fluorescence iron line (centroid energy and equivalent width). 
\squeezeup
\section{Observations and data reduction}
We only included SgXBs that are believed to be primarily wind-fed systems  in our data set. Sources for which strong evidence was reported 
in the literature for the presence of an accretion disk were not included. A summary of all observations used for the present work is provided in Table~\ref{obslog}. 
Recent reviews by \citet{walter2015} and \citet{unified2017} provide an overview of the most relevant properties of each system. 
In contrast with other studies investigating the spectral variability on timescales comparable with the clumpy wind dynamics 
\citep[see, e.g.,][]{bozzo2011,bozzo2016,bozzo2017}, in the present case we are interested in evaluating the wind properties on a larger scale. As clumps in the wind 
are known to give rise to variability over a hundred to thousand seconds \citep[see, e.g.,][and references therein]{walter2007}, their effect can be neglected 
when using integration times as long as several tens of kiloseconds. To quantitatively investigate the properties of the stellar winds in classical SgXBs and SFXTs, 
we thus extracted their average spectra from all publicly available \emph{Suzaku} and \emph{XMM-Newton} observations. We focused on deriving, from the fits to these spectra, 
a measurement of the absorption column density in excess of the Galactic value and fluorescence iron line properties. The first parameter 
provides an estimate of the average stellar wind density from which the compact object is accreting. The centroid energy and 
equivalent width (EW) of the fluorescence iron emission line are also key probes of the stellar wind properties. 
This feature originates from the fluorescence of the X-rays from the compact object onto the surrounding stellar wind and it is known that 
larger EWs correspond to denser winds (outside X-ray eclipses; see, e.g., \citealt{torrejan2010}). 

\noindent
We processed \emph{Suzaku} data from one of the X-Ray Imaging Spectrometer (XIS) units - XIS0 \citep[0.2-12~keV;][]{K07}, using filtered cleaned event 
files obtained from the application of predetermined screening criteria\footnote{\scriptsize{http://heasarc.gsfc.nasa.gov/docs/suzaku/analysis/abc/}}. 
For sources that showed jitters in the detector image, the event files were corrected via the \texttt{aeattcorr} and \texttt{xiscoord} tools to update the attitude information. For those sources affected by pileup, we discarded 
photons collected within the portion of the point spread function (PSF) where the estimated pileup fraction was 
greater than 4 \%. This was carried out with the FTOOLS task \texttt{pileest}. The XIS0 spectra were extracted by
choosing circular regions of $2^{'}$, $3^{'}$, or $4^{'}$ radius centered around the best-known source position, depending on whether the observation was made in 
1/8, 1/4, or 0 window mode, respectively. Background spectra were extracted by selecting regions of
the same size, as mentioned above, in a portion of the CCD that was not significantly contaminated by the source X-ray emission. 
Response files were created using the CALDB version `20150312'. 

\noindent
\emph{XMM-Newton} observation data files (ODFs) were processed using the standard 
Science Analysis System (SAS 14.0) and following the procedures given in the online analysis threads\footnote{\scriptsize{http://www.cosmos.esa.int/web/xmm-newton/sas-threads}}. 
We primarily used data from the PN (0.5-12~keV) whenever available, as this instrument provides a better statistics compared to the Metal Oxide Semi-conductor (MOS) 
cameras (0.5-10~keV). The latter were used in all those cases in which the PN data were not collected or not usable. We did not make use of the  Reflection Grating Spectrometers 
(RGS) data because of the limited band pass 
of this instrument and the need for the results to be comparable with those obtained from the \emph{Suzaku} data. All European Photon Imaging Camera (EPIC) spectra were corrected for pileup, 
whenever required. The correction was carried out using an annular extraction region whose inner 
radius was determined via the SAS tool {\sc epaplot}. Background spectra were extracted from a 
region located on the same CCD as that used for the target source, thereby avoiding any contamination from its emission. 
The difference in extraction areas between source and background was accounted for with the SAS {\sc
backscale} task. All spectra were rebinned in order to have at least 25 counts per energy bin and, at the same 
time, to prevent oversampling of the energy resolution by more than a factor of three. 
Individual \emph{Suzaku} and \emph{XMM-Newton} spectra were fit with a power-law model corrected for line-of-sight Galactic and local absorption with 
\texttt{phabs} and 
additional Gaussian components to take into account the presence of iron emission lines. The addition of a partial covering (\texttt{pcfabs}) or a thermal 
blackbody component was required in a few cases to account for the soft excesses in the X-ray spectra. 
Since the data we used for the present work are not affected by low statistics issues, the detection of 
soft excess does not alter the measured values of the absorption column densities. We are thus confident about the representative nature of the obtained average values of the absorption column 
densities.
Spectral fits were performed in all cases with \texttt{XSPEC} v12.9.0. 
For the eclipsing SgXBs, such as 4U 1700-37, Vela X-1, OAO 1657-415, XTE J1855-026, EXO 1722-363, 
IGR J16195-4945, IGR J16479-4514, and 4U 1538-522, we did not use data collected during the X-ray eclipses. 
All spectra of the various sources are shown in Appendix~\ref{app:spectra}, together with the 
best-fit models and the residuals from the fits. We note that for the absorption 
models \texttt{phabs} and \texttt{pcfabs}, we used the default element abundances and cross sections in \texttt{XSPEC} \citep[]{anders1989,verner1996} as the S/N of the data at 
the lower energies does not allow us to discriminate between different possibilities.
\begin{table}
\scriptsize
\caption{\scriptsize{Log of all used observations. The effective exposure 
time for each observation is also indicated.}}
\squeezeup
\resizebox{\columnwidth}{!}{ %
\begin{tabular}{@{}llll@{}} 
\hline
Source & \multicolumn{2}{c}{OBSID} & Effective Exp. (ks)\\
& \multicolumn{1}{c}{\emph{XMM-Newton}} & \multicolumn{1}{c}{\emph{Suzaku}}\\
\hline
\multicolumn{3}{c}{Classical SgXBs} \\
\hline
IGR J00370+6122 & 0501450101 & --- & 16.2 \\
\noalign{\smallskip}
4U 0114+65 & --- & 406017010 & 106.6 \\
\noalign{\smallskip} 
Vela X-1 & 0406430201, 0111030101 & 403045010 & 118.9, 53.3, 104.7 \\
\noalign{\smallskip}
GX 301-2 & 0555200301, 0555200401 & 403044020 & 58.1, 46.0, 61.8 \\
& & 403044010 & 11.4\\
\noalign{\smallskip} 
4U 1538-522 & 0152780201 & 407068010 & 38.1, 25.1\\
\noalign{\smallskip} 
IGR J16207-5129 & 0402920201 & 402065020 & 30.6, 32.7\\
\noalign{\smallskip} 
IGR J16318-4848 & 0154750401, 0201000201, 0201000301, & 401094010 & 23.3, 17.8, 21.0, 97.3\\
& 0201000401, 0742270201 & --- & 15.1,64.4 \\
\noalign{\smallskip} 
IGR J16320-4751 & 0128531101, 0556140101, 0556140201, & --- & 17.4, 7.9, 6.9  \\
& 0556140301, 0556140401, 0556140501, & & 6.1, 9.8, 1.9  \\ 
& 0556140601, 0556140701, 0556140801, & & 10.8, 7.8, 8.4  \\ 
& 0556141001, 0201700301 & & 6.5, 44.5  \\
\noalign{\smallskip}
IGR J16393-4643 & 0206380201, 0604520201 & 404056010 & 8.5, 6.3, 50.5   \\
\noalign{\smallskip} 
IGR J16418-4532 & 0405180501, 0679810201, & --- & 22.7, 11.1  \\ 
\noalign{\smallskip} 
IGR J16493-4348 & --- & 401054010 & 21.1\\
\noalign{\smallskip} 
OAO 1657-415 & --- & 406011010 & 84.7 \\
\noalign{\smallskip} 
4U 1700-37 & 0083280101, 0083280201, 0083280301 & 401058010 & 24.9, 31.2, 19.5, 40.7  \\
\noalign{\smallskip} 
EXO 1722-363 & 0405640301, 0405640401, 0405640701, & --- & 4.2, 5.6, 19.2    \\
& 0405640801, 0405640901 & 0206380401 & 12.3, 12.0, 0.6   \\
\noalign{\smallskip} 
SAX J1802.7-2017 & 0206380601, 0745060401, 0745060501,  & --- & 9.5, 37.3, 14.3   \\ 
& 0745060601, 0745060801 & & 15.2, 13.7  \\
\noalign{\smallskip} 
XTE J1855-206 & --- & 409022010 & 42.2 \\
\noalign{\smallskip} 
4U 1909+07 & --- & 405073010 & 29.3 \\
\noalign{\smallskip} 
IGR J19140+0951 & 0761690301 & ---& 34.1\\
\noalign{\smallskip} 
\hline
\multicolumn{3}{c}{SFXTs} \\
\hline
IGR J11215-5952 & 0405181901   & --- & 15.2 \\
\noalign{\smallskip} 
IGR J16195-4945 & --- & 401056010 & 39.2\\
\noalign{\smallskip} 
IGR J16328-4726 & 0728560201, 0728560301, 0654190201   & --- & 29.6, 11.2, 14.9   \\
\noalign{\smallskip}
IGR J16479-4514 & --- & 406078010 & 51.8 \\
\noalign{\smallskip}
IGR J17354-3255 & 0701230101, 0701230701  & --- & 22.2, 18.3 \\
\noalign{\smallskip} 
IGR J17391-3021 & 0554720101, 0561580101 & --- & 34.0, 24.3 \\
\noalign{\smallskip} 
IGR J17544-2619 & 0744600101, 0679810401  & 402061010 & 117.6, 10.5, 103.8   \\ 
& 0154750601 & & 2.5 \\
\noalign{\smallskip}
SAX J1818.6-170 & 0693900101   & --- & 21.5\\
\noalign{\smallskip}
IGR J18410-0535 & 0604820301 & 505090010 & 37.1, 49.6 \\
\noalign{\smallskip} 
IGR J18450-0435 & 0728370801, 0306170401   & --- & 14.9, 15.2  \\
\noalign{\smallskip}
IGR J18462-0223 & 0651680301 & & 12.6   \\
\noalign{\smallskip}
IGR J18483-0311 & 0694070101   & --- & 36.9 \\
\noalign{\smallskip}
\hline
\hline
\multicolumn{3}{l}{}
\end{tabular}
}
\squeezeup
\label{obslog}
\end{table} 
\squeezeup
\begin{figure*}
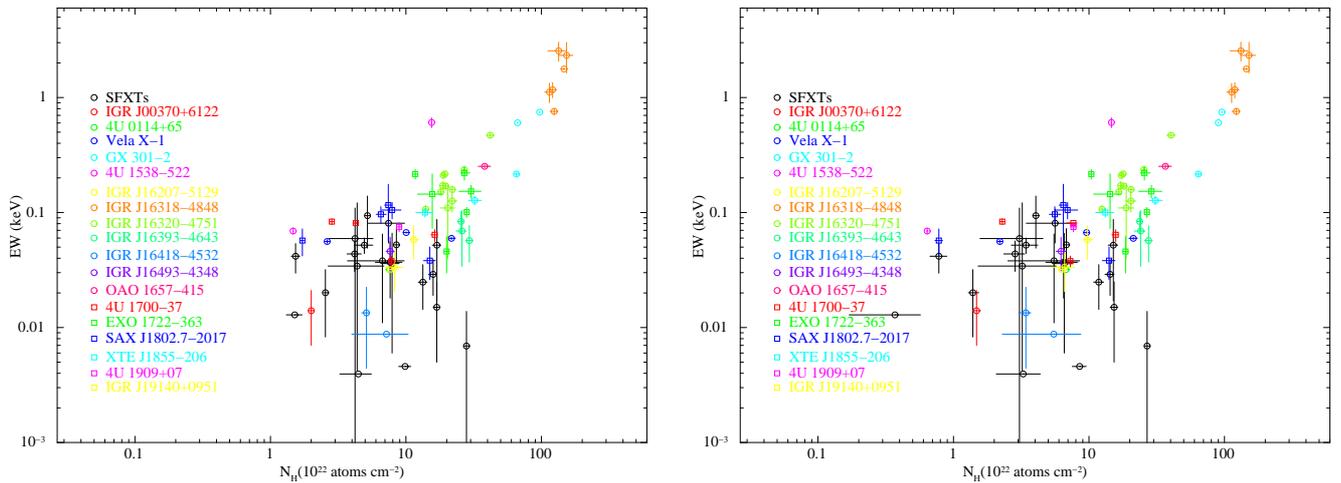

\centering
\includegraphics[width=6.3cm,angle=-90]{nh_eqw_xmm_suzaku-nolmxbs_new_ref1.ps}
\includegraphics[width=6.3cm,angle=-90]{eqw_nh_xmm_suzaku-nolmxbs_new_subtractednh.ps}
\caption{\scriptsize{\it Left}: Equivalent width vs. total column density of the Fe K$\alpha$ line measured from all sources considered in this paper. 
The SFXTs are indicated in black, while a variety of colours have been used for classical SgXBs. {\it Right}: Same as the left plot, but in this case the 
contribution of the Galactic absorption to the total $N_{\rm H}$ has been removed for all sources.}
\label{iron}
\end{figure*}
\section{Results}
We plot the main findings of our analysis in Fig.~\ref{iron}, showing the measured values of the iron K$\alpha$ equivalent width (EW) as a 
function of ${N_{\rm H}}$. For those sources in which a partial covering component was required, the value of the absorption 
column density reported in left panel of Fig.~\ref{iron} includes all contributions 
\footnote{\scriptsize{Total ${N_{\rm H}}$ = ${N_{\rm H1}}$+${N_{\rm H2}}$*${C_{V}}$, where ${N_{\rm H1}}$ is the 
hydrogen column density along our line of sight to the source, ${N_{\rm H2}}$ accounts for local absorption, and ${C_{V}}$ is the covering fraction. 
The parameter ${C_{V}}$ represents the fraction 
of the radiation from the NS that escapes from the variable and strong local absorptions. Large variations of this parameters have been recorded in different sources and 
ascribed to the presence of a largely variable and unstable accretion environment around the compact object \citep[see, e.g.,][]{malacaria2016}. 
See also Table~A.1 and A.2.}}.

We also show in the right panel of Fig.~\ref{iron} that the results do not change significantly if we remove for 
each source the expected Galactic contribution from the total absorption column density. \footnote{\scriptsize{We note that this test is carried out because if the measured value of Galactic absorption is several times 
larger than the online value, as is the case now, it is speculated to be due to the presence of a complex multicomponent absorber local to the source.}}
The value of the Galactic 
absorption for all sources were estimated using the HEASARC online tool\footnote{\scriptsize{https://heasarc.gsfc.nasa.gov/cgi-bin/Tools/w3nh/w3nh.pl}}.

\noindent
We also show a plot of the iron K$\alpha$ line EW versus the X-ray luminosity in Fig. 2. The uncertainty on this last parameter  
is dominated for all sources by their poorly known distances. 
We do not find any indication of the anticorrelation between the two parameters represented in this plot, which is at odds with the findings reported  
by \citet{torrejan2010} and \citet{garcia2015}. The red points on the top left side of the plot, which  give the appearance of an anticorrelation, 
are from one source, IGR\,J16318-4848, in which the compact object is believed to be obscured for most of the time by a dense cocoon of 
material \citep{chaty12}. 
\squeezeup

\section{Discussion}

The results reported in the two panels of Fig.~\ref{iron} show that there is a clear direct correlation between the absorption column density and the iron line 
EW in classical SgXBs and SFXTs. This is expected because the iron lines in these systems is produced by fluorescence of the X-rays from the compact object onto the 
surrounding stellar wind \citep{george1991}, which is also the material giving rise to the measured local absorption column density \citep{inoue}. A larger $N_{\rm H}$ indicates a denser 
environment and thus also a larger amount of material that is involved in the fluorescence emission. This result has been known since the 
previous studies presented by \citet{torrejan2010} and \citet{garcia2015}. Compared to these works, we extended the sample of measurements by including \emph{Suzaku} 
data, which provide consistent results in a broader range of X-ray luminosity. 

\begin{figure}
\includegraphics[width=5.8cm,angle=-90]{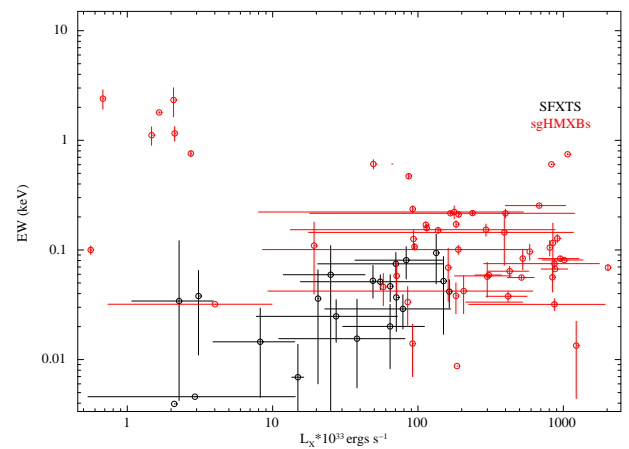}
\caption{\scriptsize{Plot of the equivalent width of the Fe K$\alpha$ line vs. the X-ray luminosity for all sources analyzed in this paper. We indicated classical SgXBs in red and SFXTs in black.}}
\label{iron-lumin}
\end{figure}

The plots in Fig.~\ref{iron} also confirm the interesting feature mentioned by \citet[][see their Fig.~10]{garcia2015} that all SFXTs are systematically less 
absorbed than most of the classical SgXBs, but we show here in addition that the SFXTs are characterized, on average, by K$\alpha$ iron lines 
with significantly lower EWs \footnote{\scriptsize{The bulk of the SFXT observations we analyzed are in the quiescent or, at most, in the intermediate state, 
in which the luminosity is lower than 10$^{35}$~erg~s$^{-1}$.}}. These two results together indicate that the accretion environment around the 
compact objects in the SFXTs is systematically less dense than that in classical systems. As mentioned by \citet{garcia2016}, this difference can be explained 
either by assuming that the stellar winds in the SFXTs are less powerful than those in classical SgXBs, or that the interaction between the compact object and 
the stellar wind in these two classes of sources is not the same. 

\noindent
A search for systematic differences in the winds of the supergiant companions 
in SFXTs and in classical SgXBs has been attempted by several authors in the literature, but there is no strong evidence in 
favor of this hypothesis \citep{unified2017}. 
A complication that has so far prevented a detailed study of the stellar winds in these systems is the fact that they are highly absorbed and located 
at much larger distances with respect to the close-by supergiants for which UV and optical observations provided a great wealth of information on 
the structure and composition of their winds \citep[see, e.g.,][for a recent review]{Sundqvist11}. 
As there does not seem to be an evident dichotomy between spectral classes of supergiants in SFXTs and classical systems, 
we concentrate in the following paragraphs on the idea that the different interaction between the compact object with the wind of the companion drives 
the discrepancy between the average absorption column density and iron line EW in these systems.     

\noindent
Detailed studies of classical SgXBs have demonstrated that the irradiation of  high energy emission 
from the compact object can significantly affect the velocity of the surrounding wind, as the latter is radiatively driven and the photoionization 
by the compact object reduces the main acceleration force of the wind \citep{ho87,Manousakis2012,kri15,kri16}. 
In the case of Vela\,X-1, the prototype of classical SgXBs, a drop of the wind velocity as large as a 
factor of $\sim$3 has been inferred from the measured velocity shifts of the emission lines from highly ionized ions close to the NS \citep[using 
observations performed with X-ray gratings spectrometers;][]{watanabe2006}. 

\noindent
In the simplistic case of a smooth and symmetrical stellar wind, it is expected that a reduction in the velocity from $v$ to $v'$=$f$$v$, with $f$$<$1, 
leads to an increase in the local density by a factor of $1$/$f$ \citep[assuming that the mass loss rate from the supergiant does not change; see, e.g.,][]{sako2003}. 
These variations can be even larger in the case of structured winds, in which dense clumps transport the bulk of the wind material and could be already endowed with much lower 
velocities compared to the surrounding inter-clump medium \citep{oskinova12}. As the 
photoionization of the stellar wind steeply increases with the X-ray luminosity and it is particularly effective above $\gtrsim$10$^{35}$-10$^{36}$~erg~s$^{-1}$ \citep[see, e.g.,]
[and references therein]{ducci2010}, we suggest that the most likely explanation for the lower density medium around the compact objects in the SFXT is due 
to the lack of an efficient photoionization of the stellar wind compared to classical systems. This is in line with the widely agreed scenario that accretion in the SFXTs 
is inhibited for most of the time either by centrifugal or magnetic barrier or by an inefficient settling accretion regime. Their X-ray emission is thus only sporadically 
achieving the required intensity to substantially slow down the stellar wind and increase the density around the compact object. We note that this would contribute to reducing 
the X-ray luminosity of the SFXTs even further because the cross section of the compact object for the capture of the stellar wind is inversely proportional to the square 
of the wind velocity and a smaller cross section implies a reduced mass accretion rate \citep{frank2002}. In classical SgXBs, the mechanisms inhibiting 
accretion are unlikely to be at work for a substantial amount of time, and thus we expect these systems to have larger X-ray luminosities and slower 
winds close to the compact object. 

\noindent
An important assumption in the considerations above is that the bulk of the fluorescence emission leading to the measurable iron K$\alpha$ lines in both the SFXTs and SgXBs 
is provided by material around the compact object rather than from the rest of the stellar wind surrounding the binary. 
This assumption is supported by the rapid variability of the iron line EW measured in several of these sources, as commented in the correspondingly published works 
\citep[see the cases of, e.g., OAO\,1657-415, IGR\,J17544-2619, and IGR\,J18410-0535;][]{pradhan2014,rampy2009,bozzo2011} 

\noindent
Our plot of the iron line EW versus the X-ray luminosity for both classical SgXBs and SFXT (Fig.~\ref{iron-lumin}) does not confirm the anticorrelation previously 
reported by \citet{torrejan2010} and \citet{garcia2015}. We argue that this is most likely due to the larger number of observations and broader range of the 
X-ray luminosity exploited in the present work thanks to the addition of all available \emph{Suzaku} data. \\

\squeezeup
\scriptsize
\begin{acknowledgements}
The data for this work has been obtained through the High Energy Astrophysics Science Archive (HEASARC) Online Service provided by NASA/GSFC. 
We would also like to thank the anonymous referee for his/her invaluable comments and suggestions.
\end{acknowledgements}

\bibliographystyle{aa}
\bibliography{comparative_arxiv}


\newpage 
\appendix
\section{Spectral fits results}
\label{app:spectra}

\begin{figure*}
\caption{\emph{Suzaku} spectra of all considered classical SgXBs. The source name and the observation ID is indicated in each figure.}
\includegraphics[height=4.2cm,width=4.3cm,angle=-90]{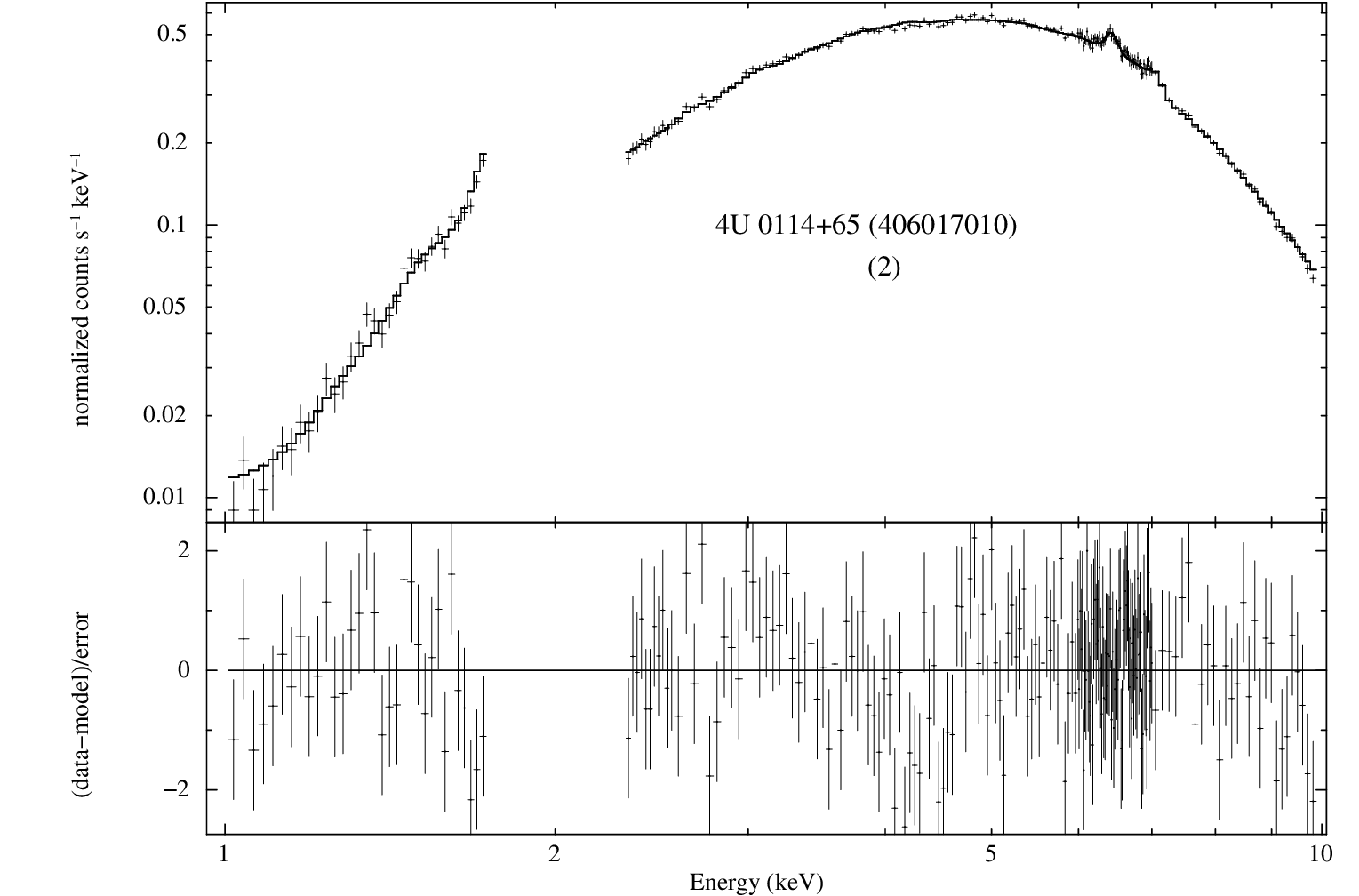}
\includegraphics[height=4.2cm,width=4.3cm,angle=-90]{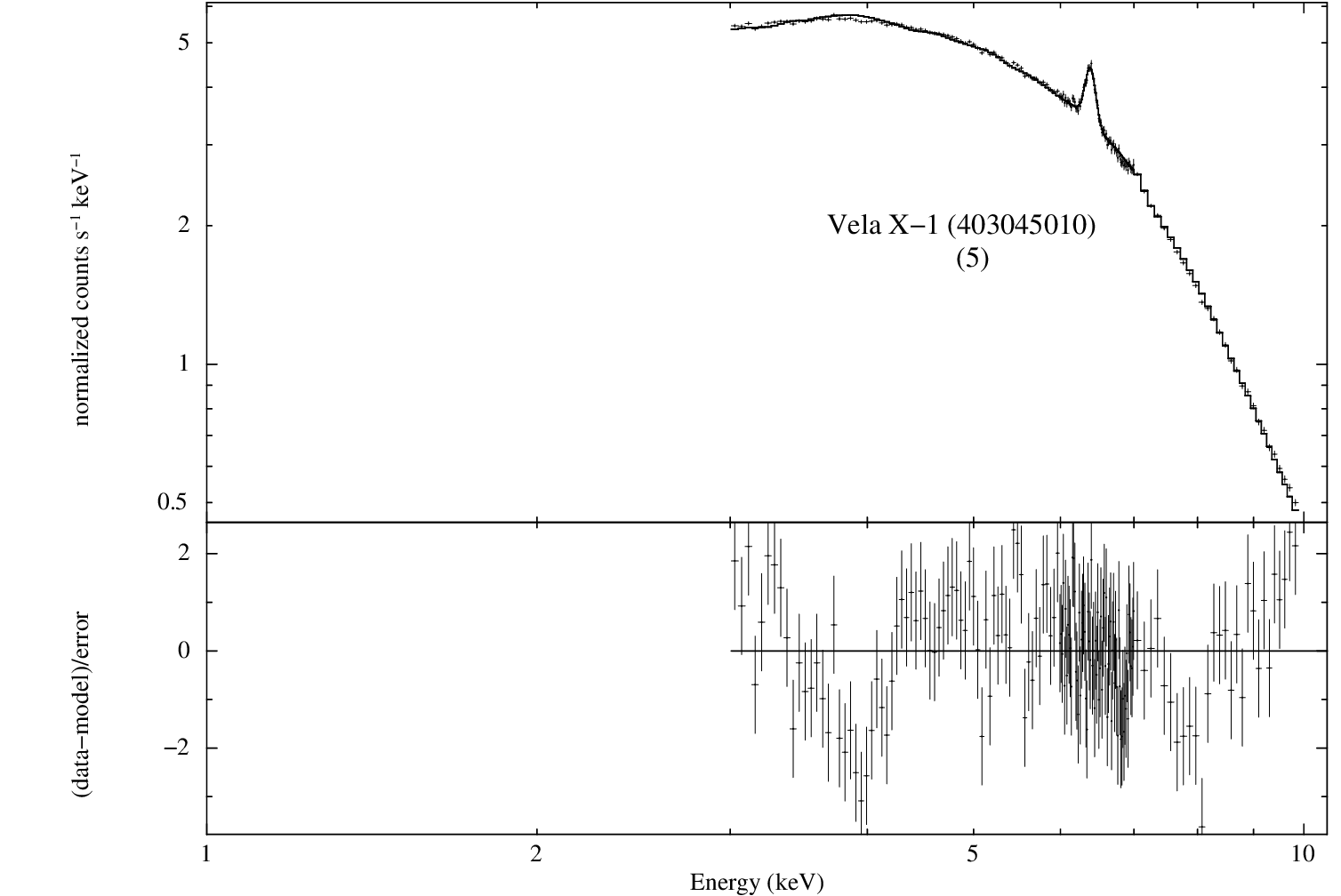}
\includegraphics[height=4.2cm,width=4.3cm,angle=-90]{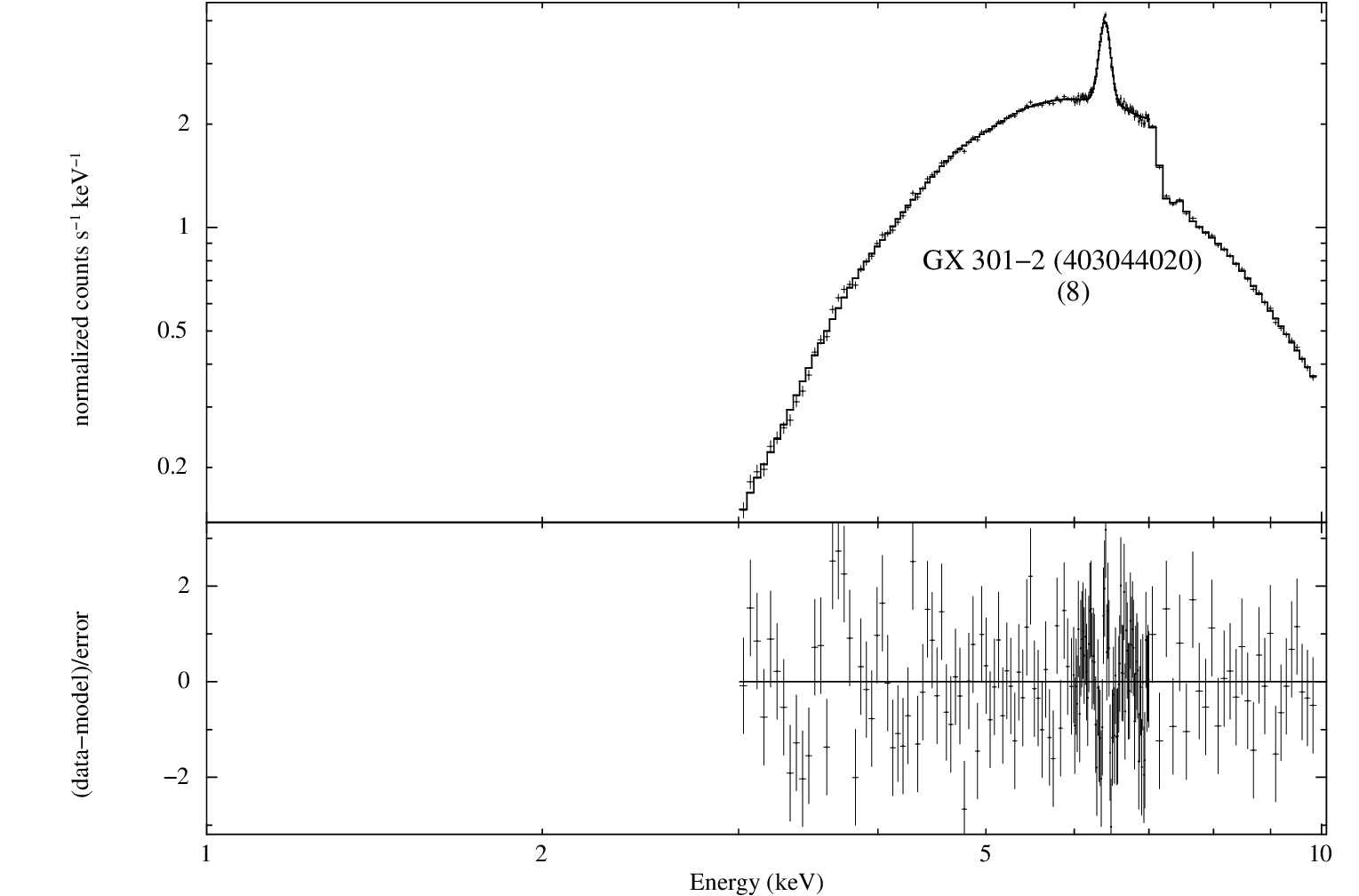}
\includegraphics[height=4.2cm,width=4.3cm,angle=-90]{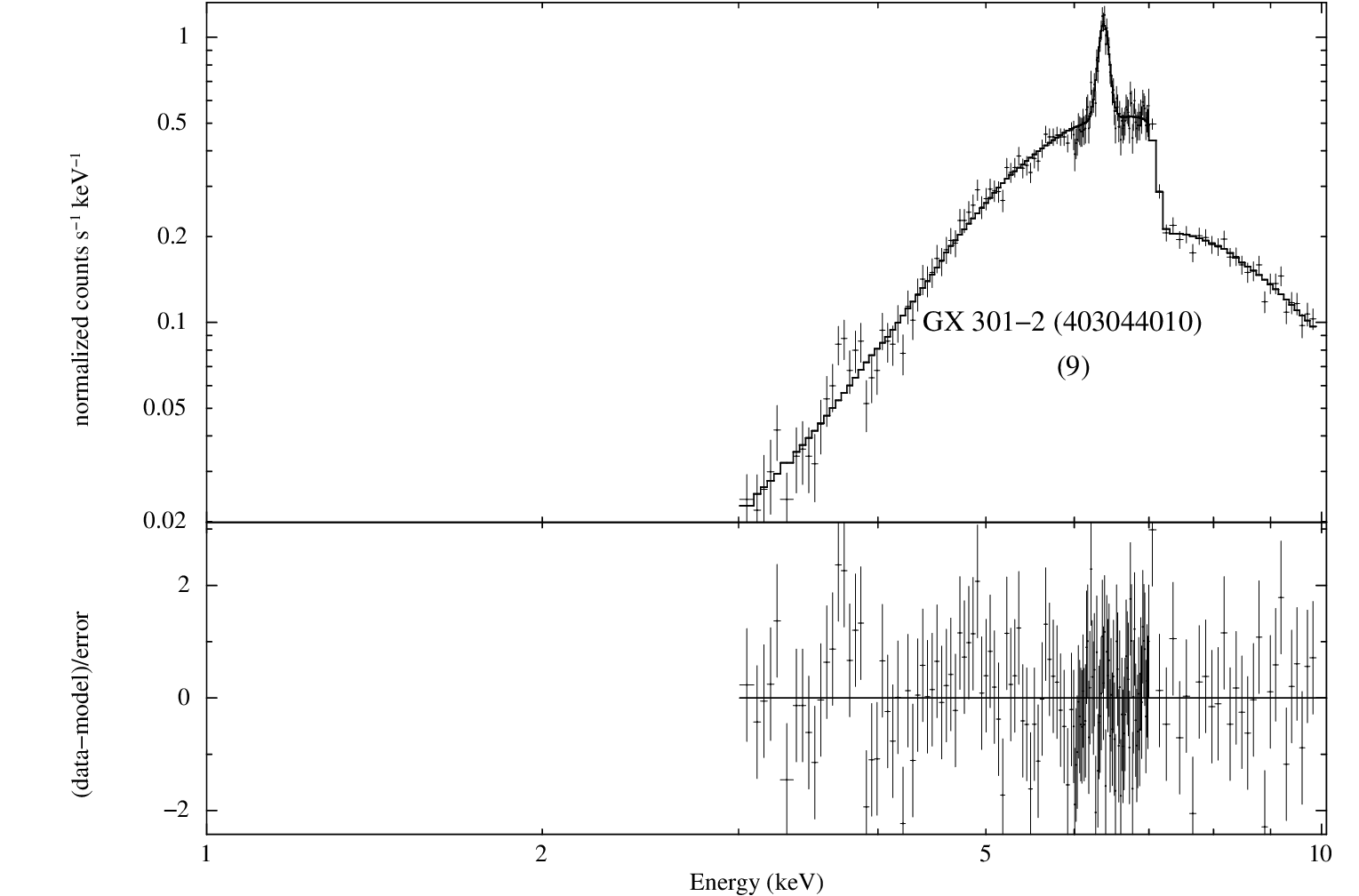}
\includegraphics[height=4.2cm,width=4.3cm,angle=-90]{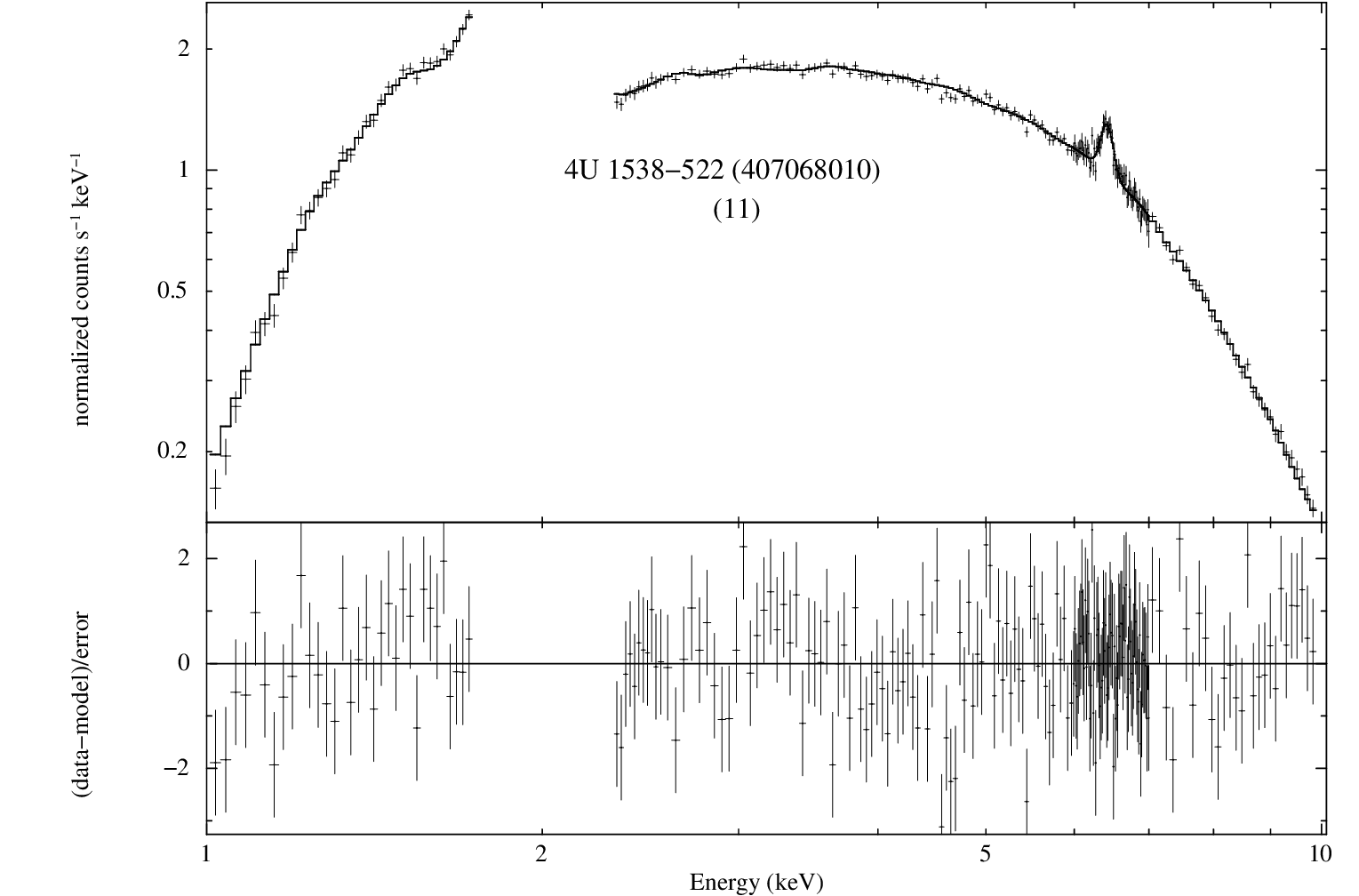}
\includegraphics[height=4.2cm,width=4.3cm,angle=-90]{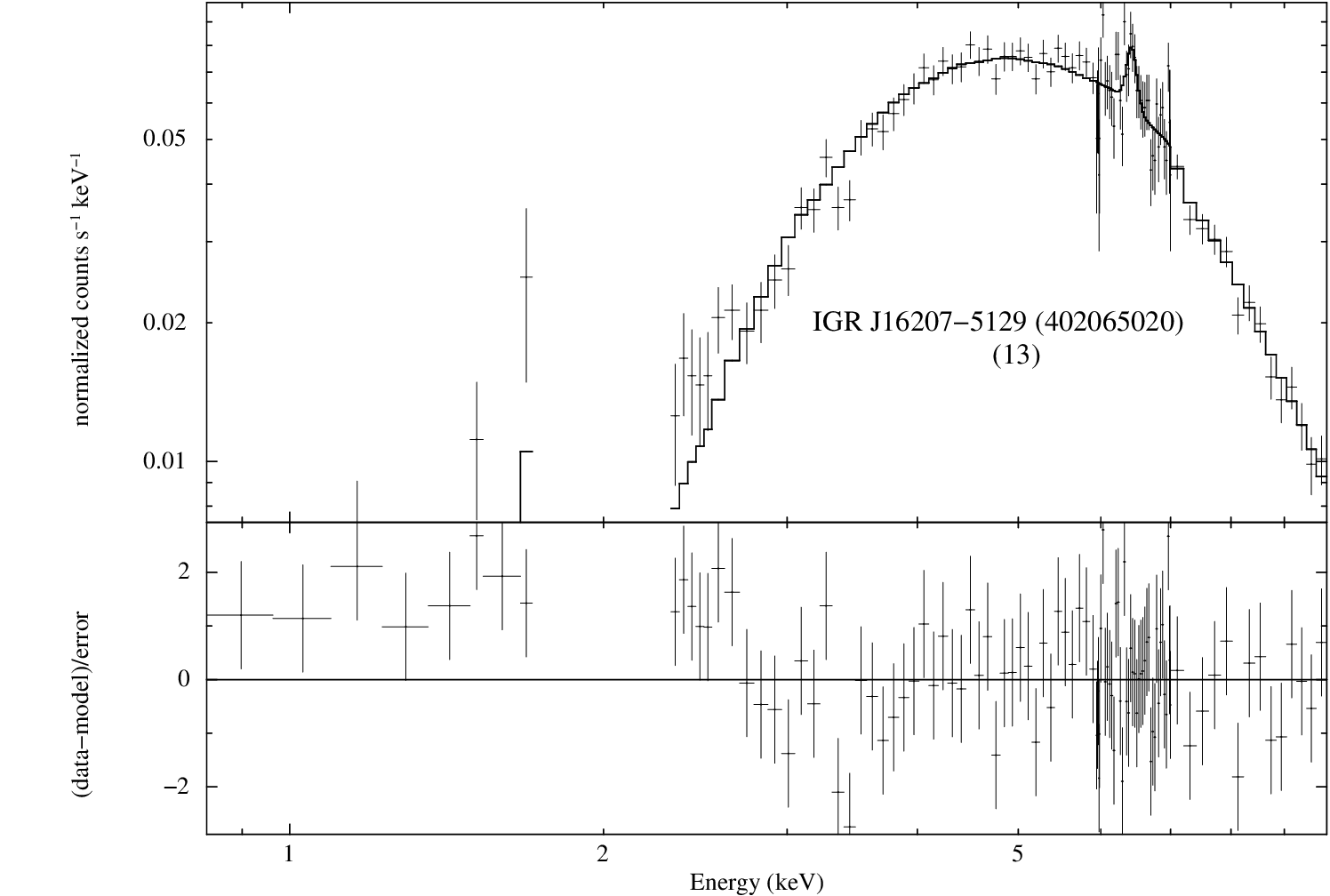}
\includegraphics[height=4.2cm,width=4.3cm,angle=-90]{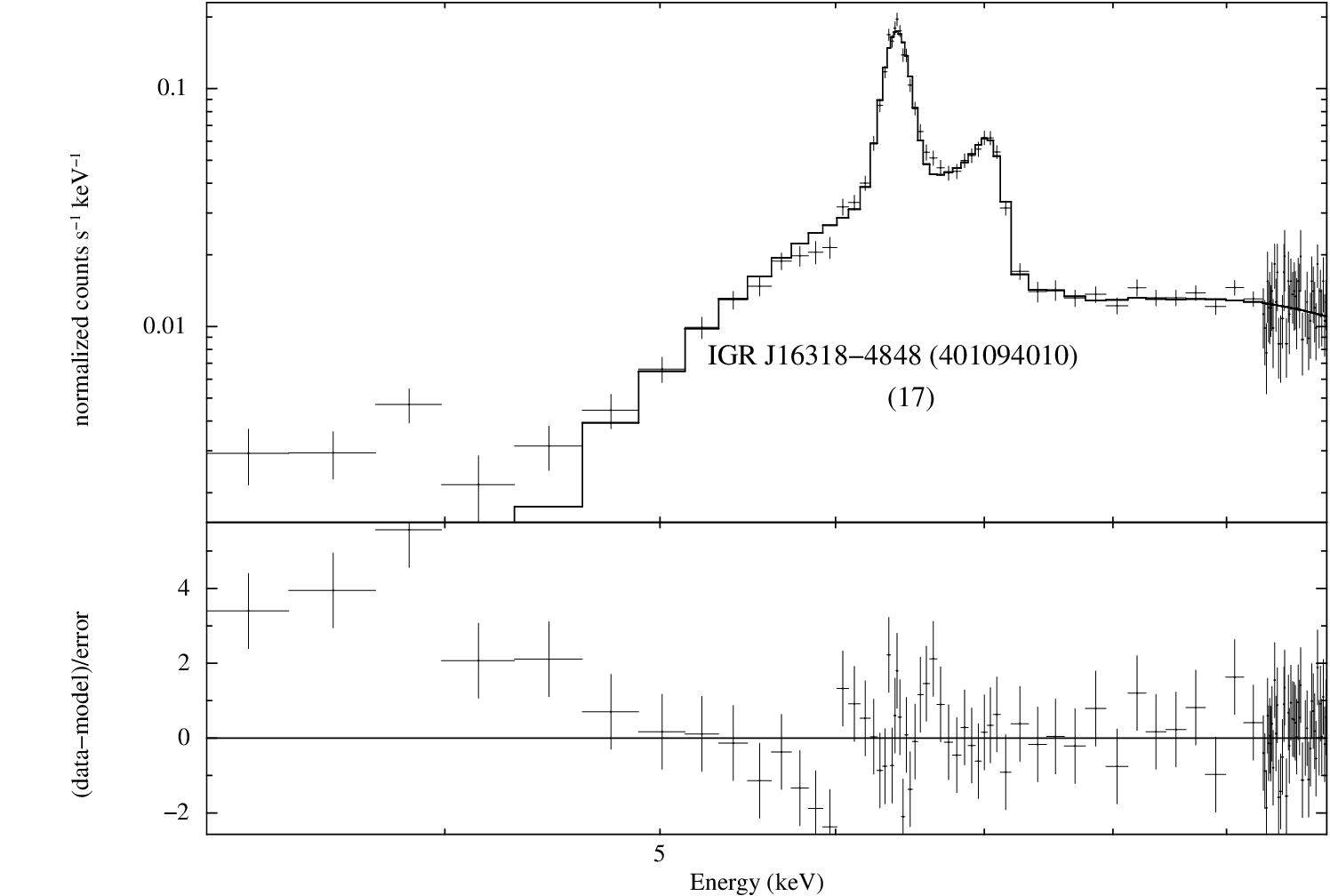}
\includegraphics[height=4.2cm,width=4.3cm,angle=-90]{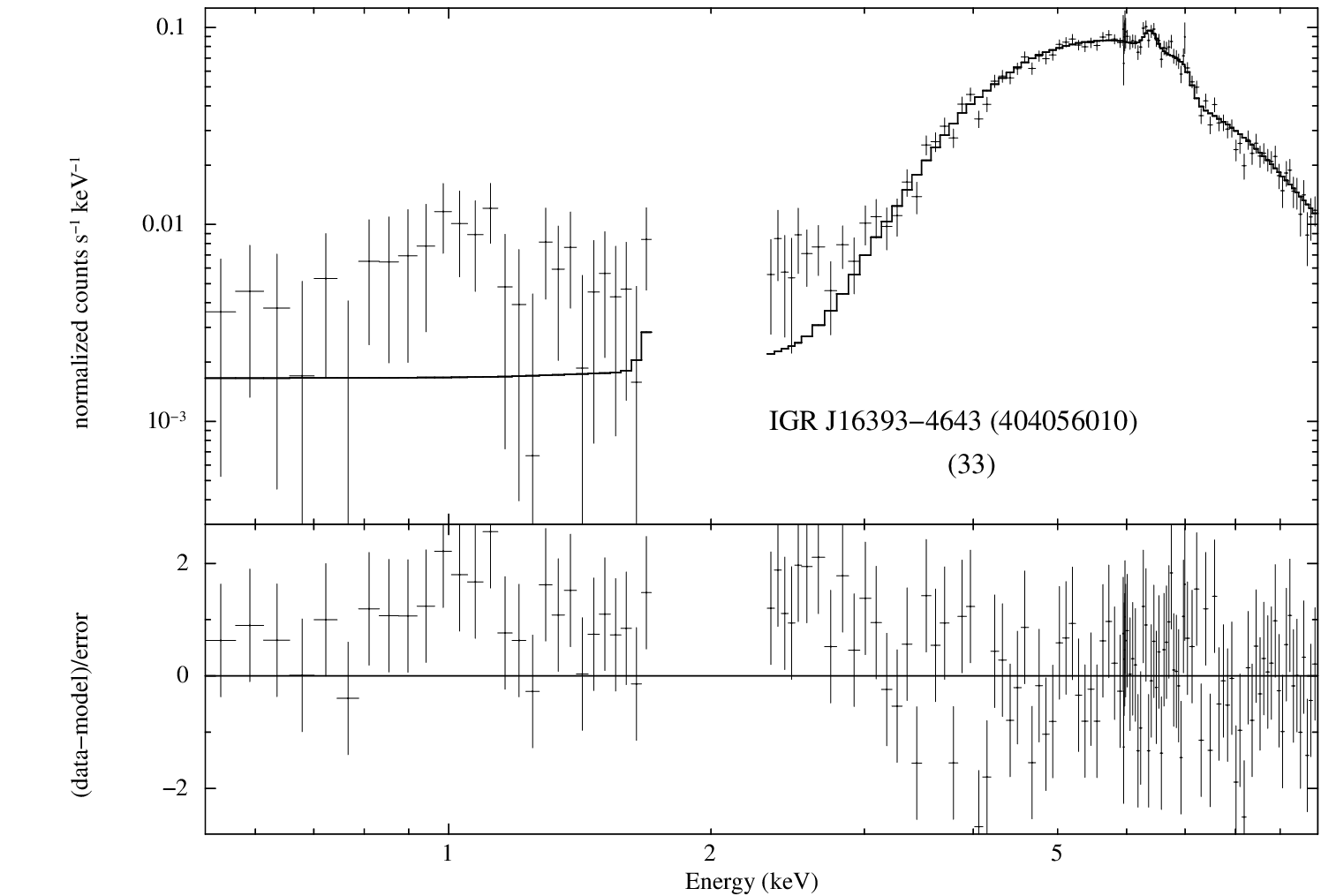}
\includegraphics[height=4.2cm,width=4.3cm,angle=-90]{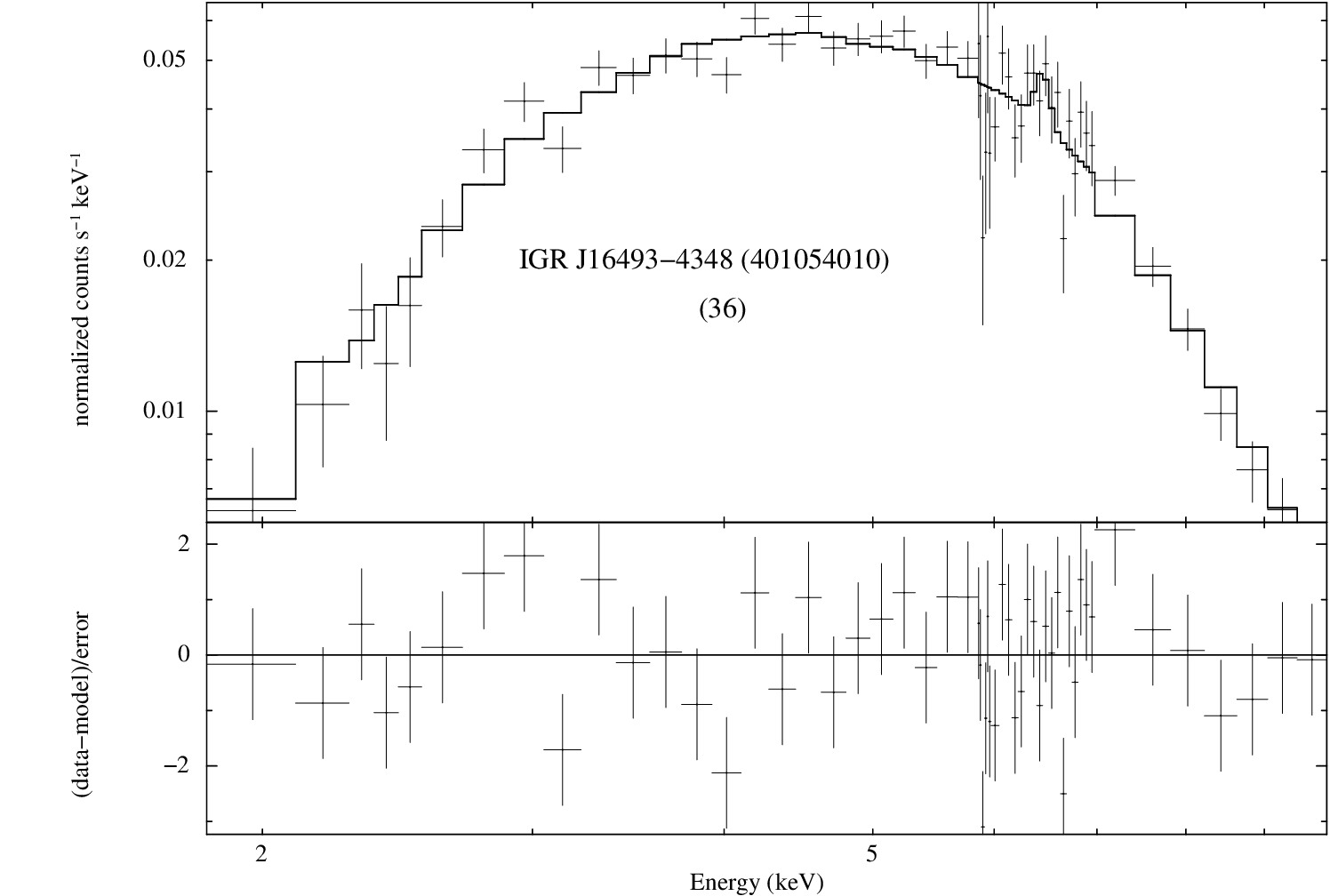}
\includegraphics[height=4.2cm,width=4.3cm,angle=-90]{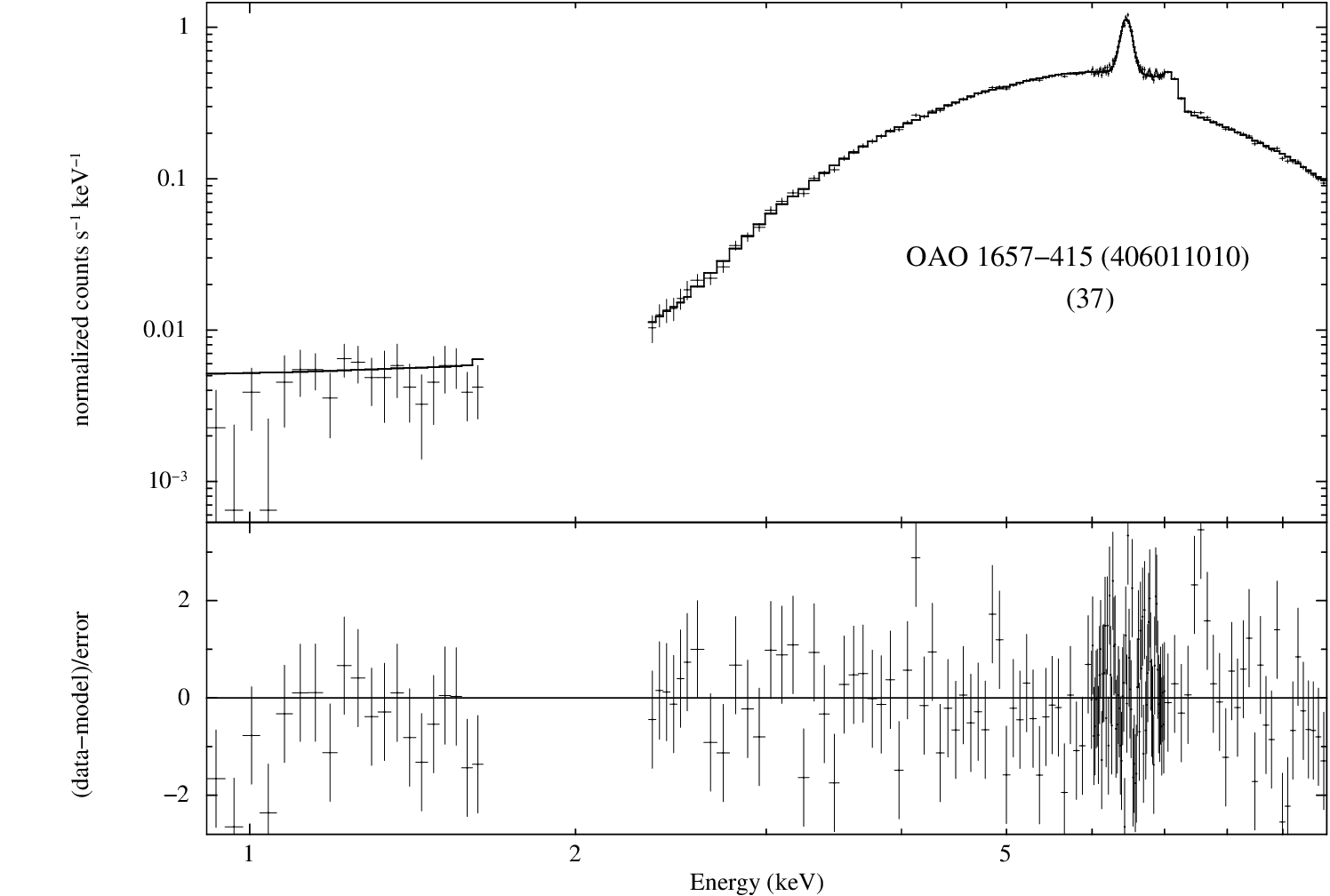}
\includegraphics[height=4.2cm,width=4.3cm,angle=-90]{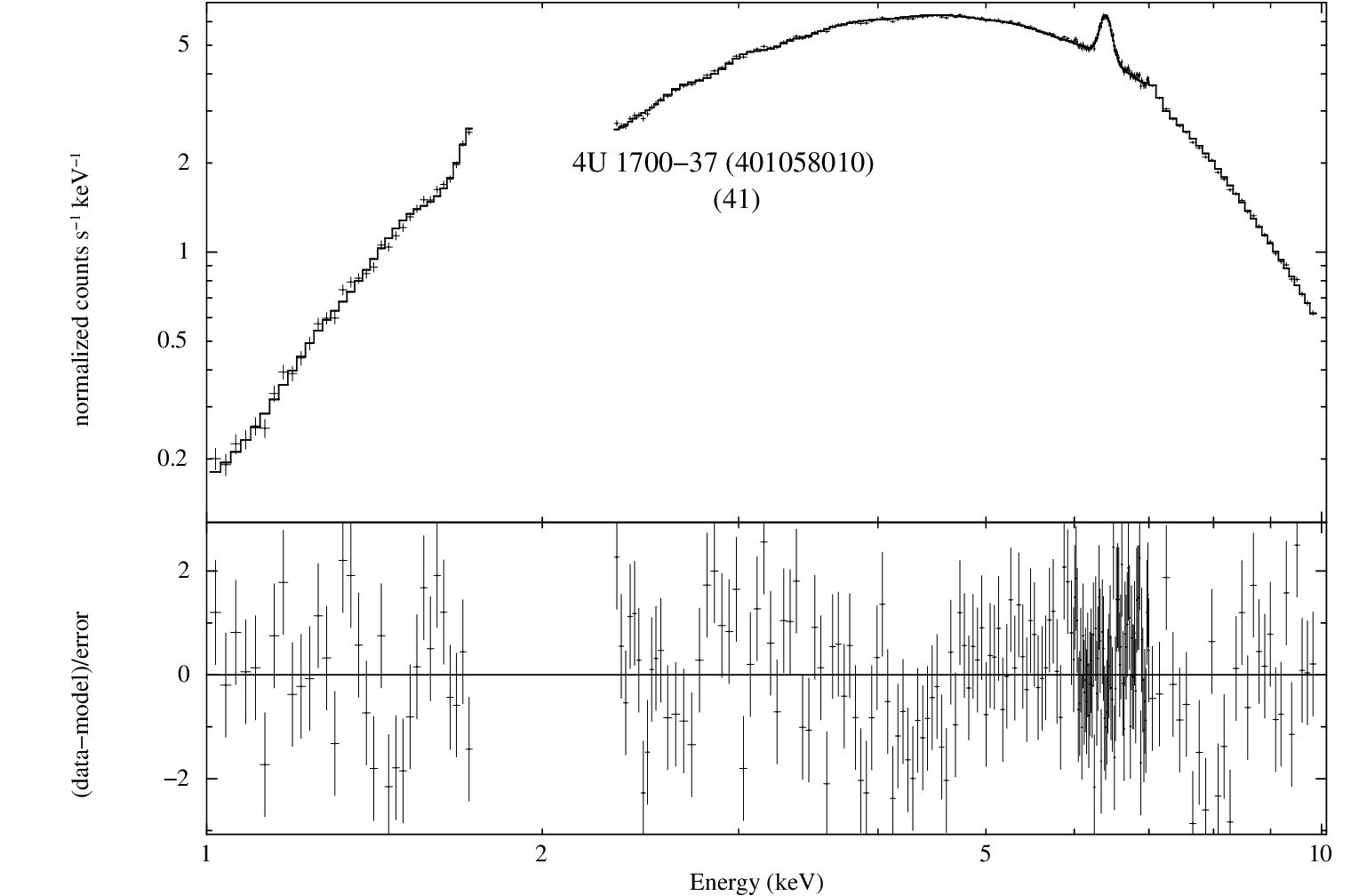}
\includegraphics[height=4.2cm,width=4.3cm,angle=-90]{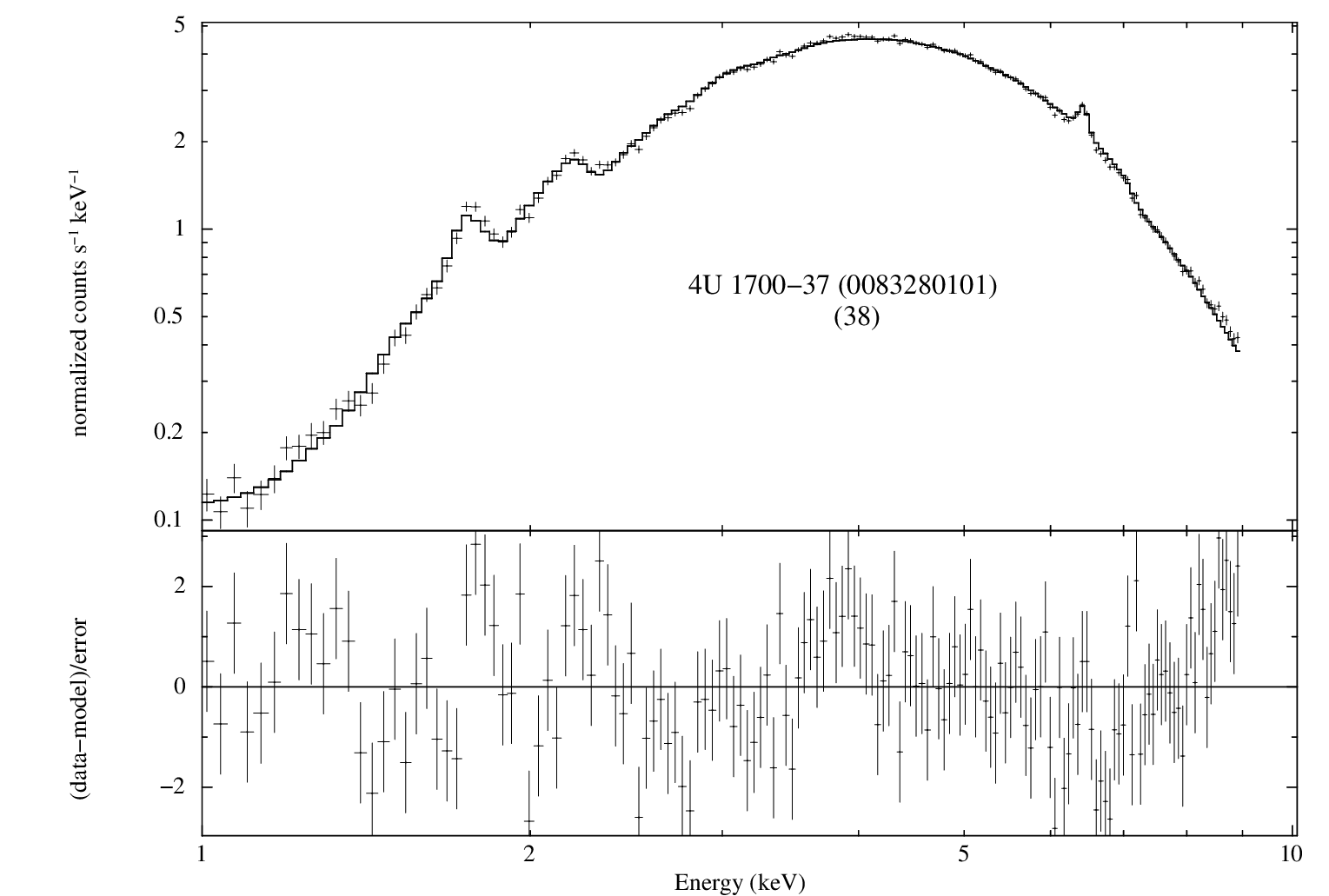}
\includegraphics[height=4.2cm,width=4.3cm,angle=-90]{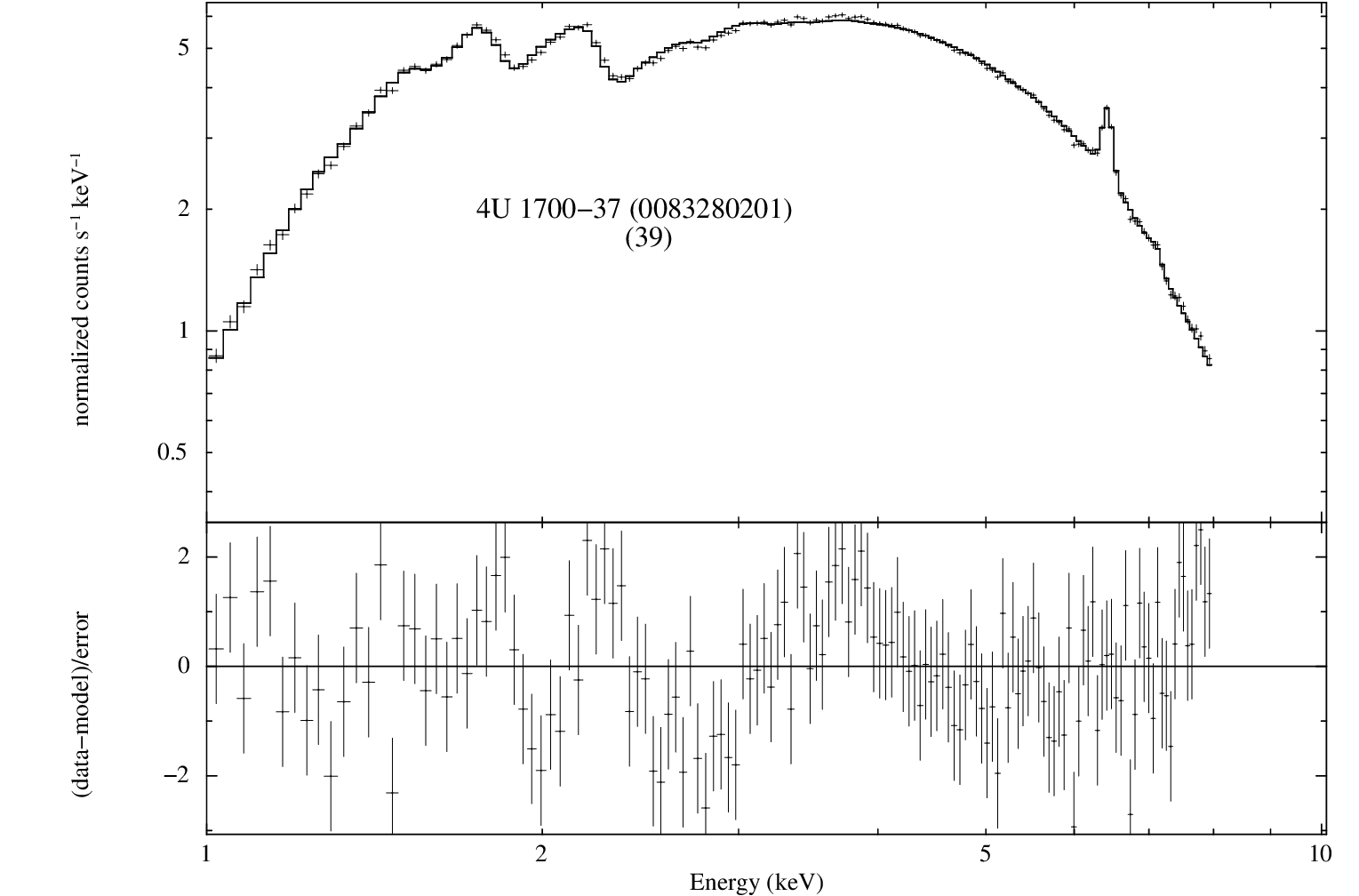}
\includegraphics[height=4.2cm,width=4.3cm,angle=-90]{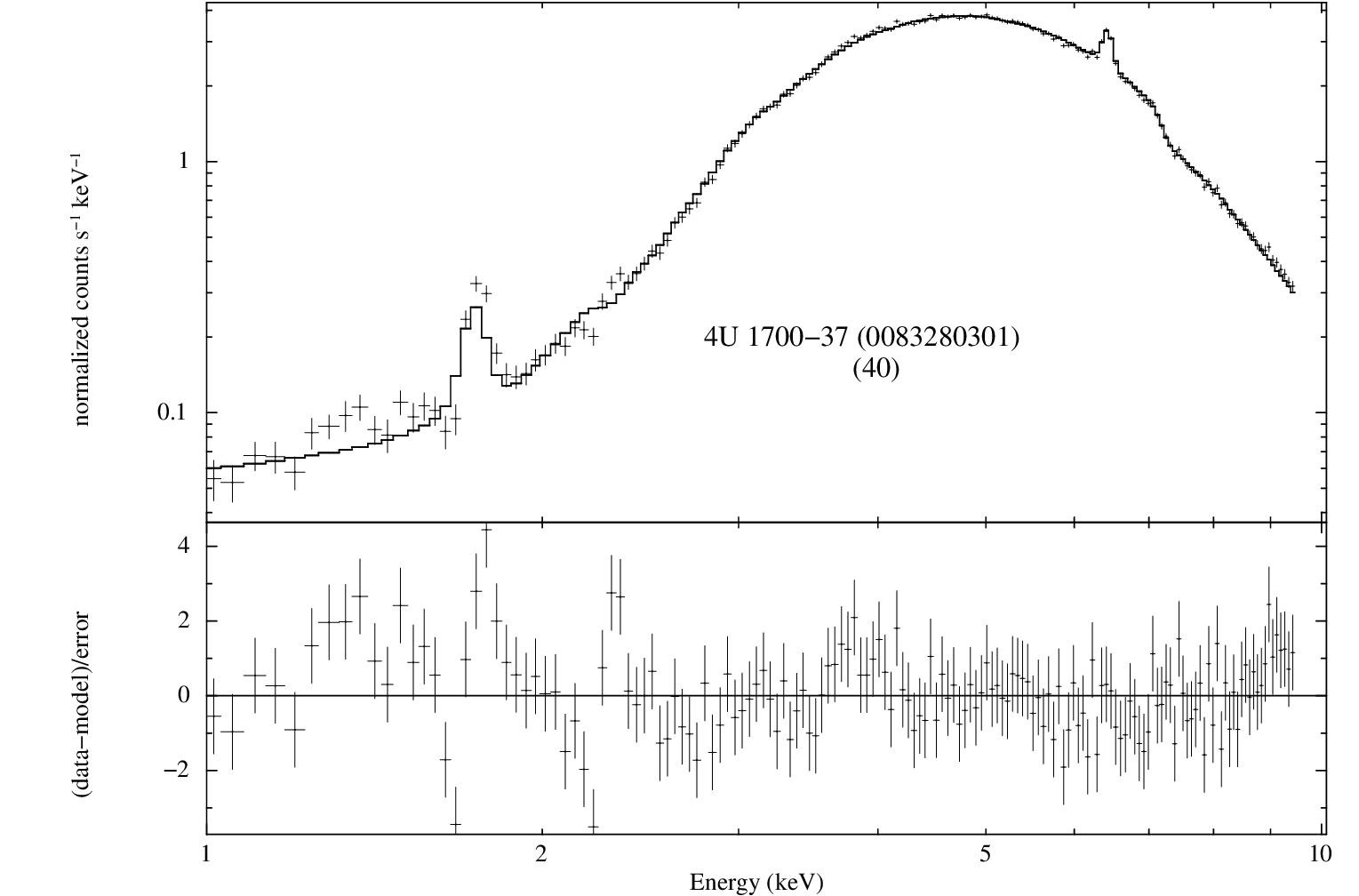}
\includegraphics[height=4.2cm,width=4.3cm,angle=-90]{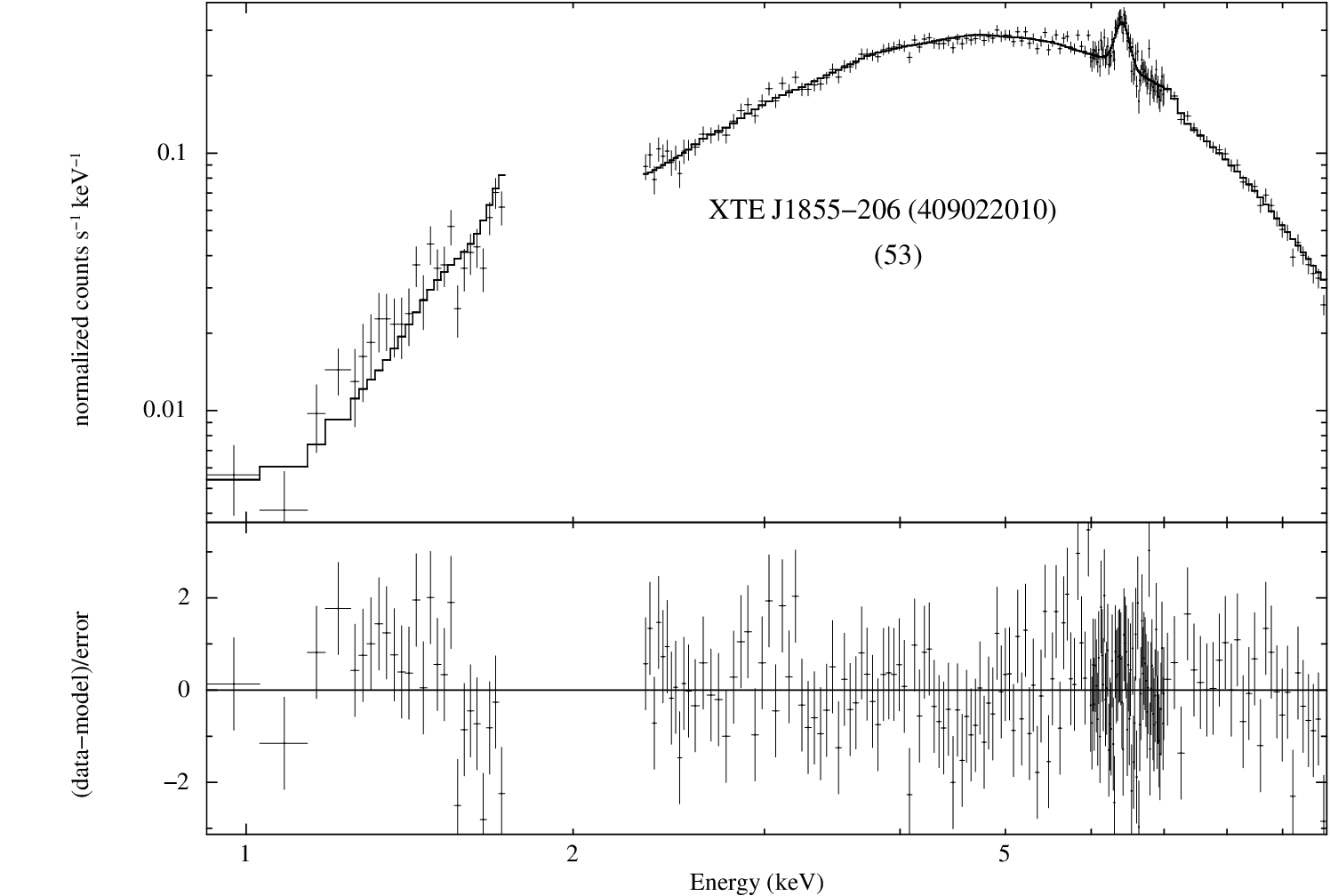}
\includegraphics[height=4.2cm,width=4.3cm,angle=-90]{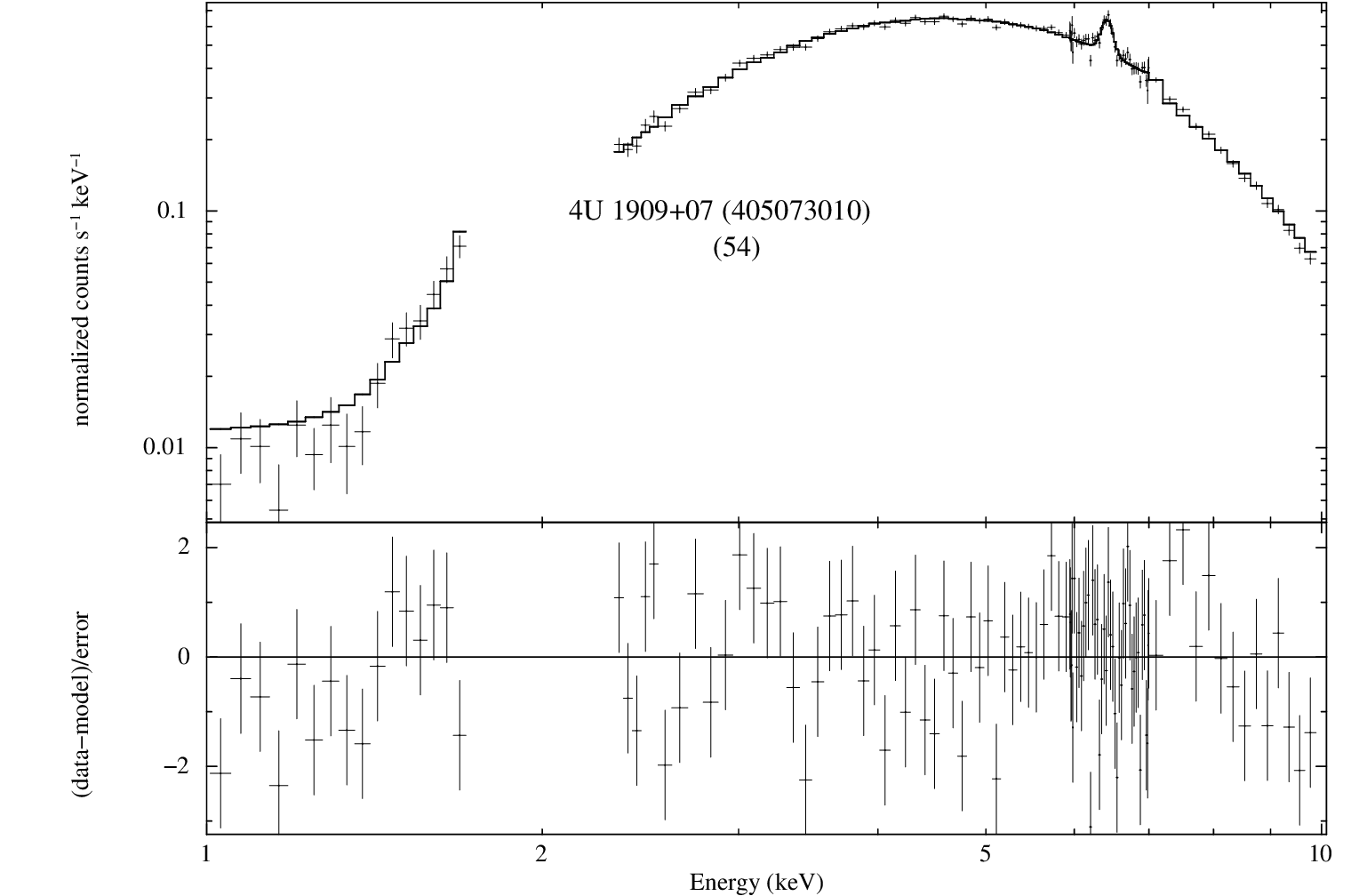}
\label{spectra1}
\end{figure*}

\begin{figure*}
\caption{\emph{Suzaku} spectra of all considered SFXTs. The source name and the observation ID is indicated in each figure.}
\includegraphics[height=4.2cm,width=4.3cm,angle=-90]{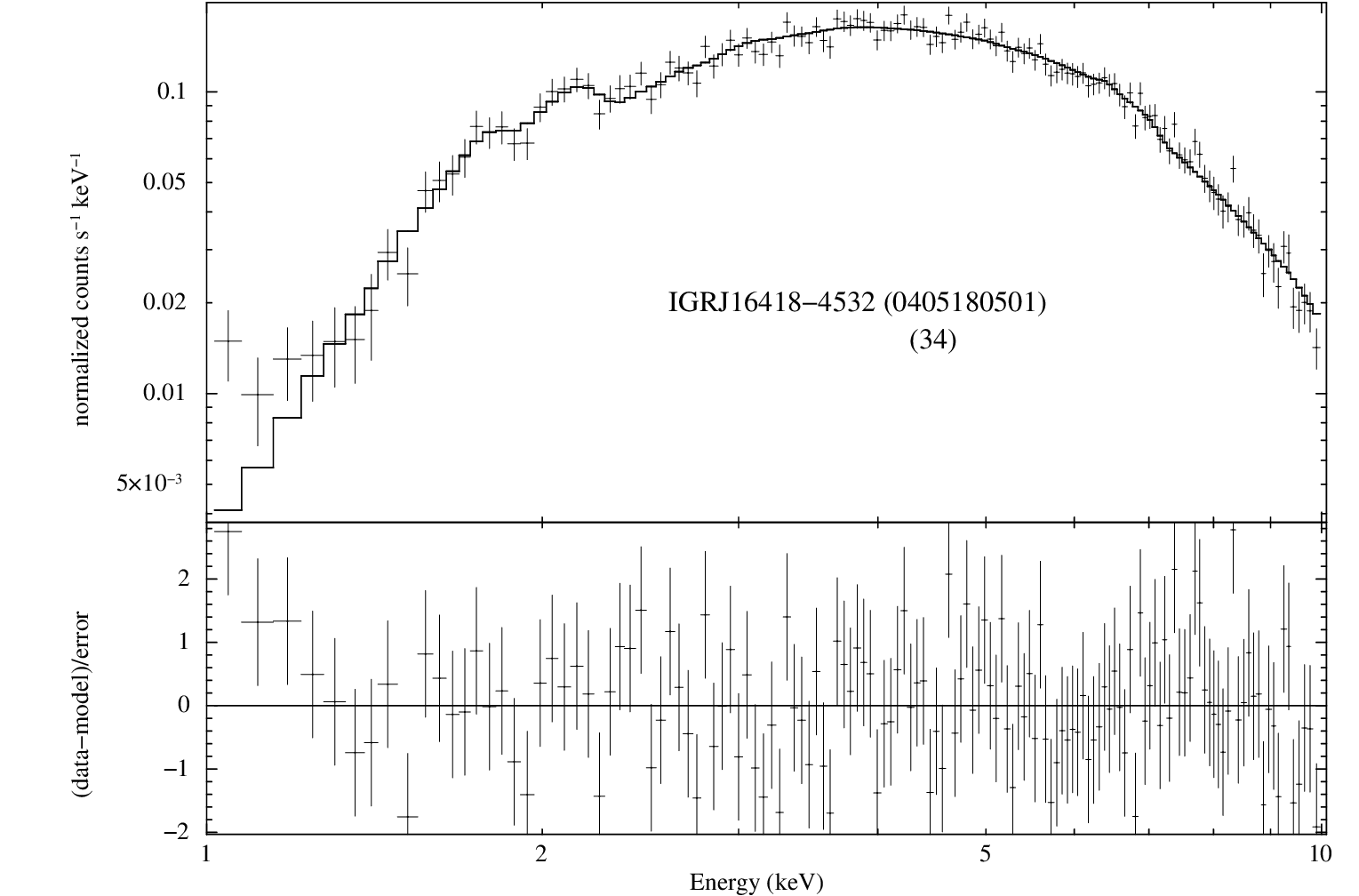}
\includegraphics[height=4.2cm,width=4.3cm,angle=-90]{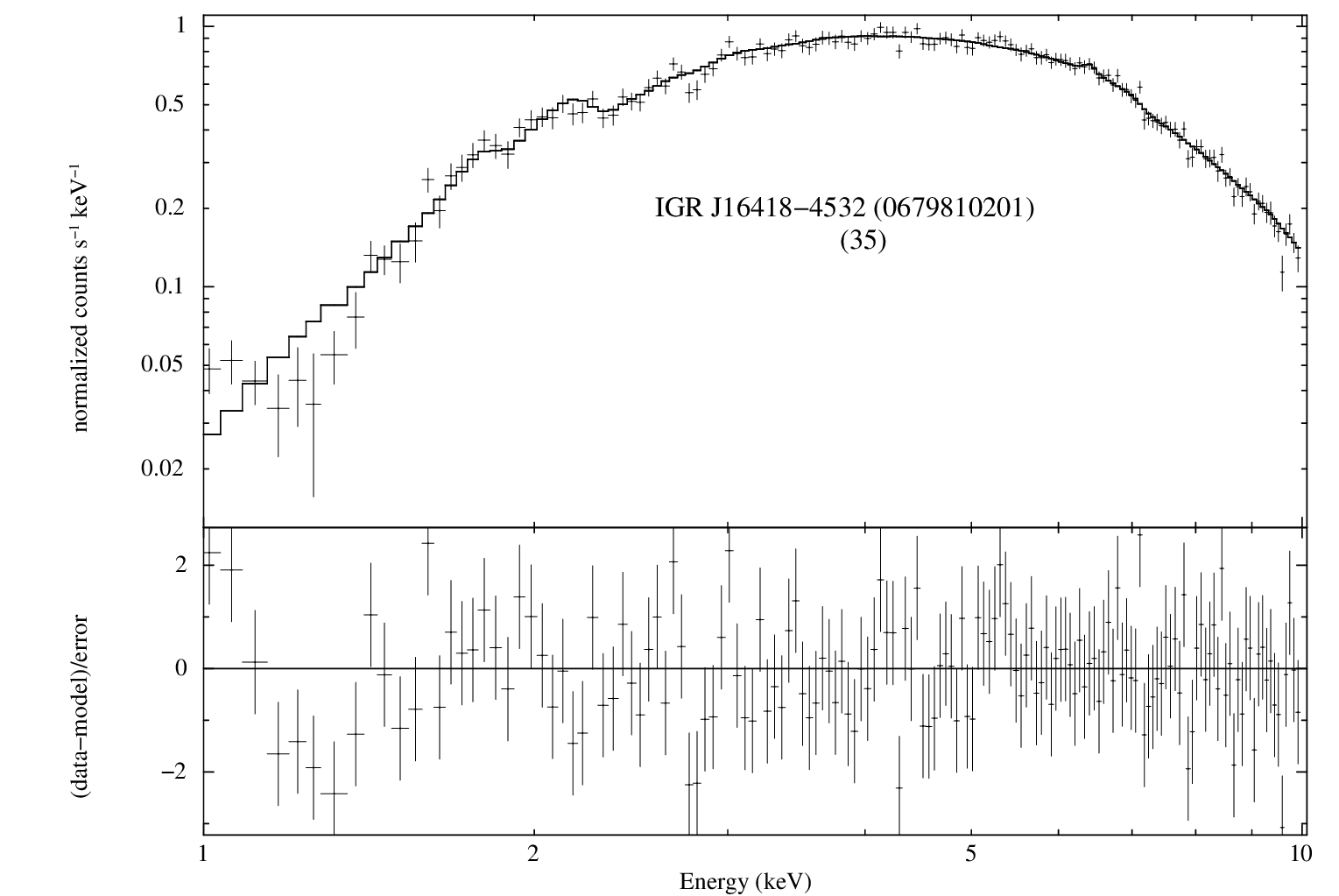}
\includegraphics[height=4.2cm,width=4.3cm,angle=-90]{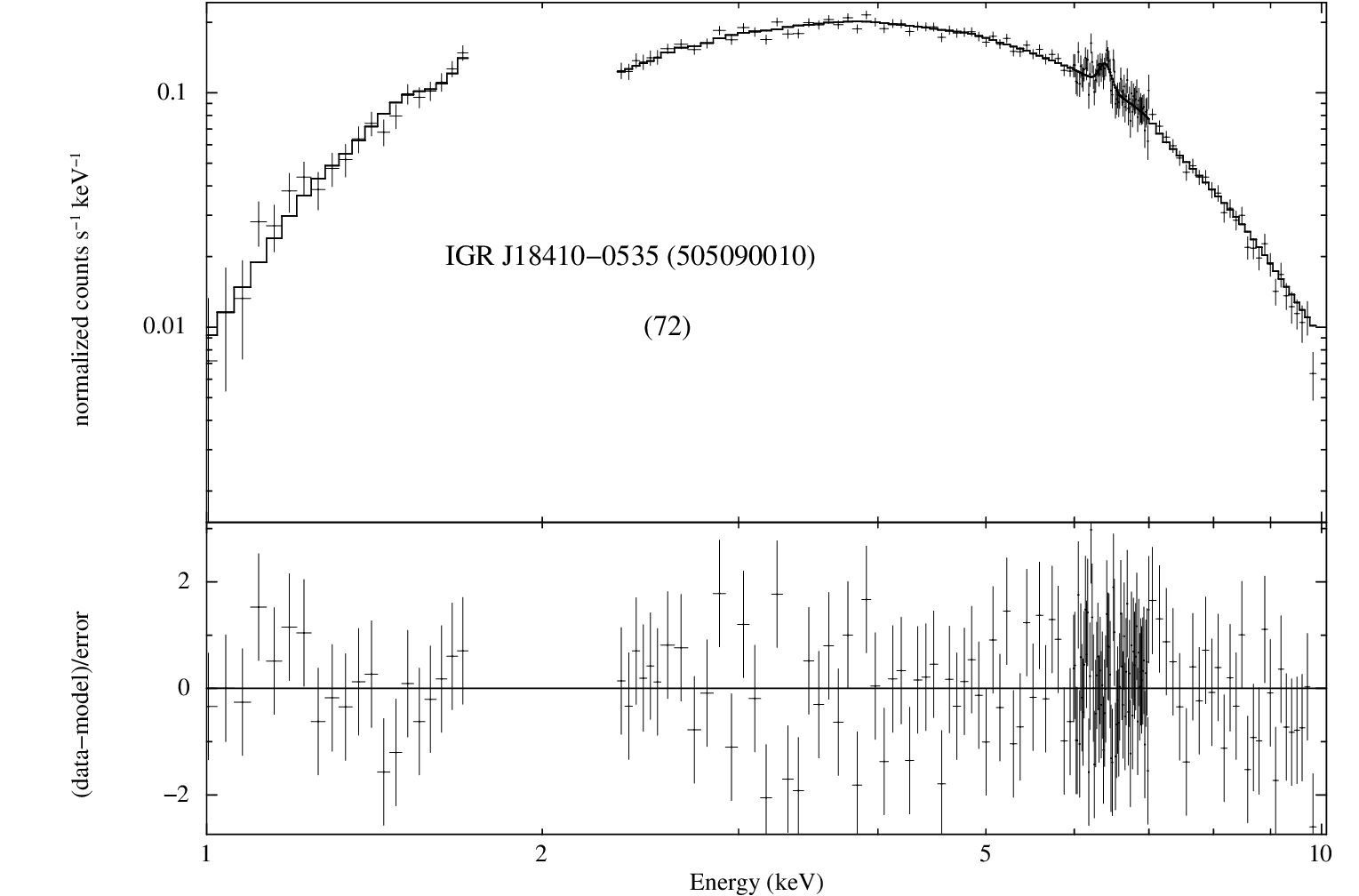}
\includegraphics[height=4.2cm,width=4.3cm,angle=-90]{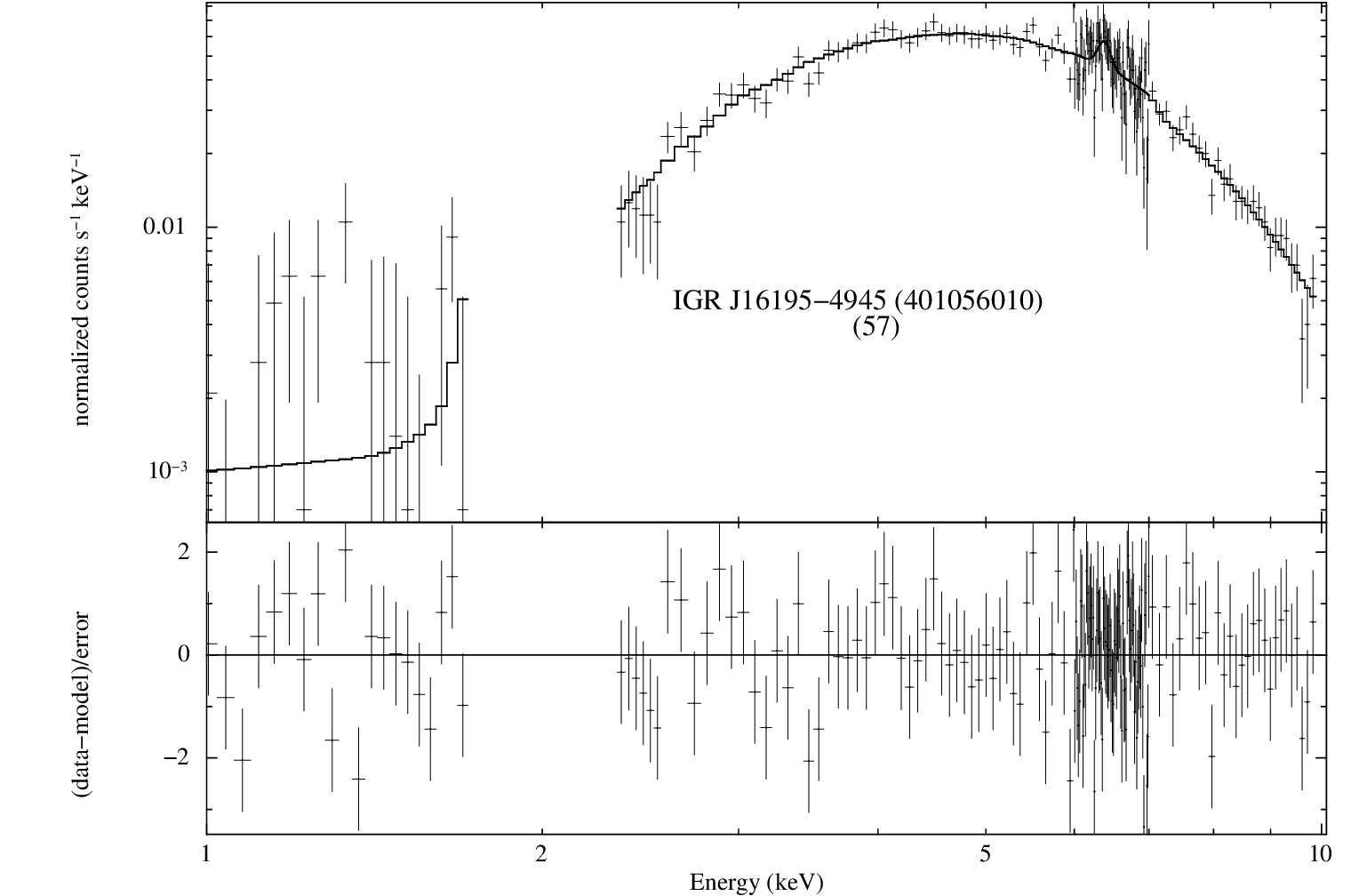}
\includegraphics[height=4.2cm,width=4.3cm,angle=-90]{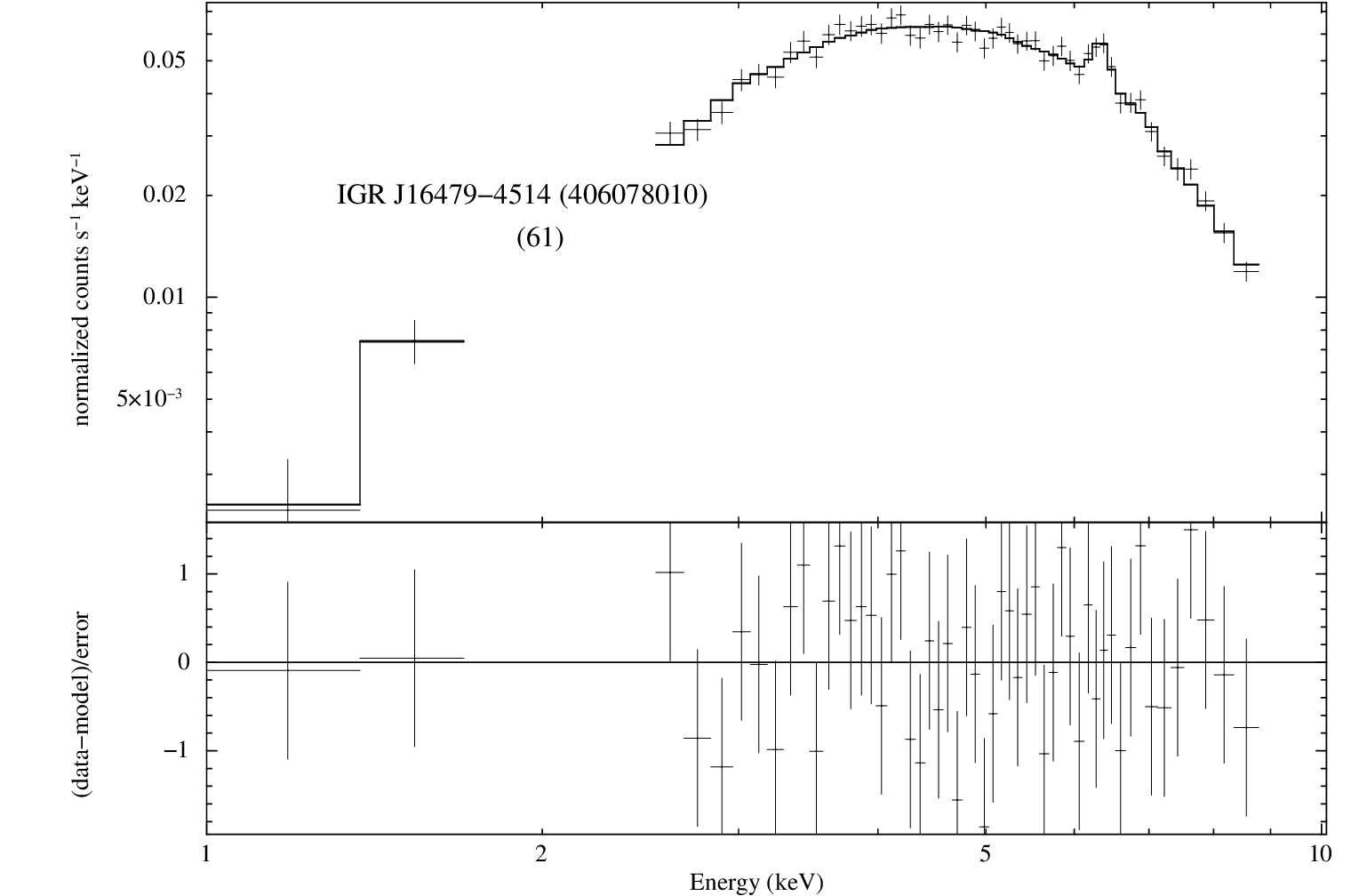}
\includegraphics[height=4.2cm,width=4.3cm,angle=-90]{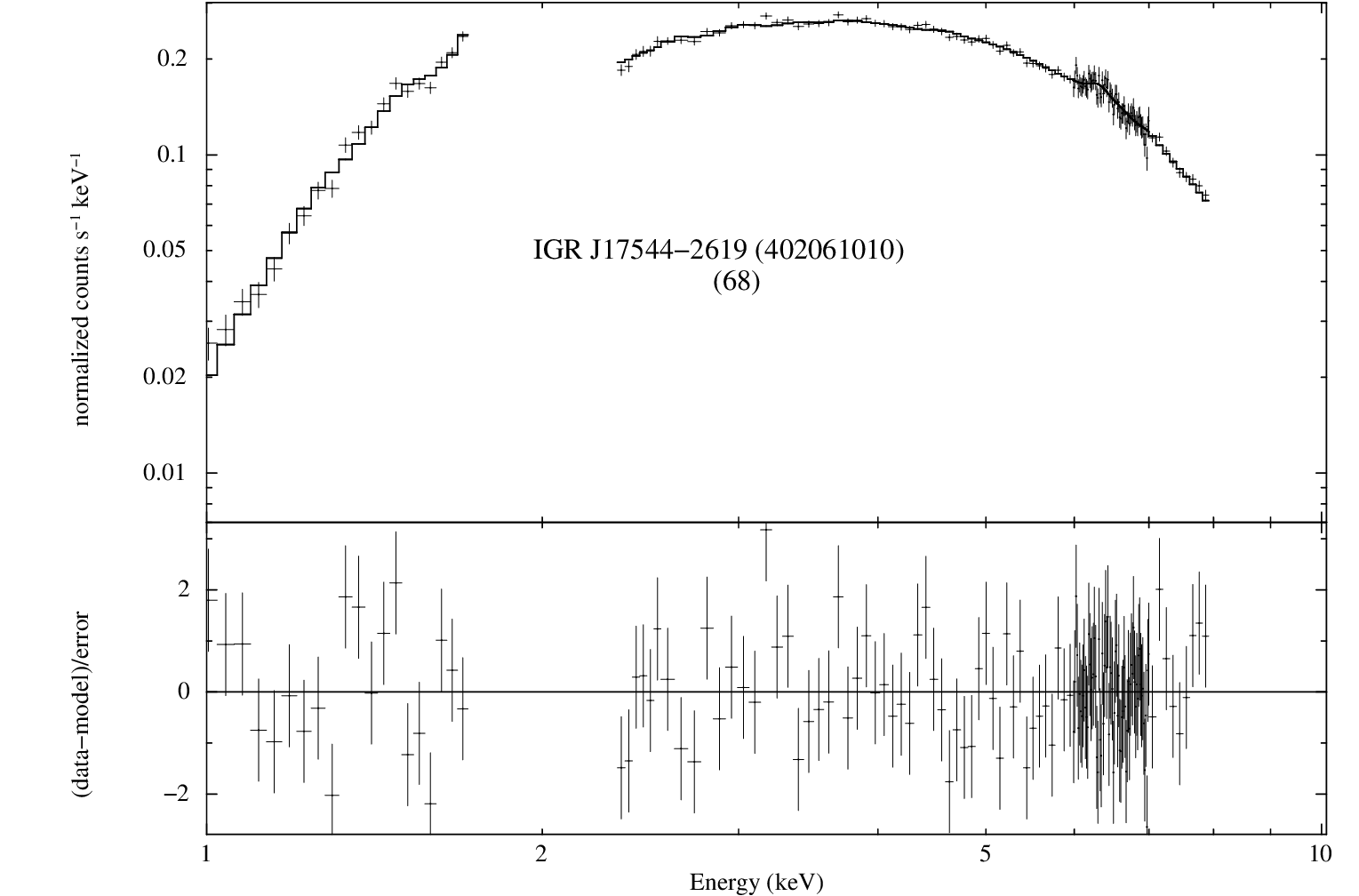}
\label{spectra2}
\end{figure*}

\begin{figure*}
\caption{\emph{XMM-Newton} spectra of all considered classical SgXBs. The source name and the observation ID is indicated in each figure.}
\includegraphics[height=4.2cm,width=4.3cm,angle=-90]{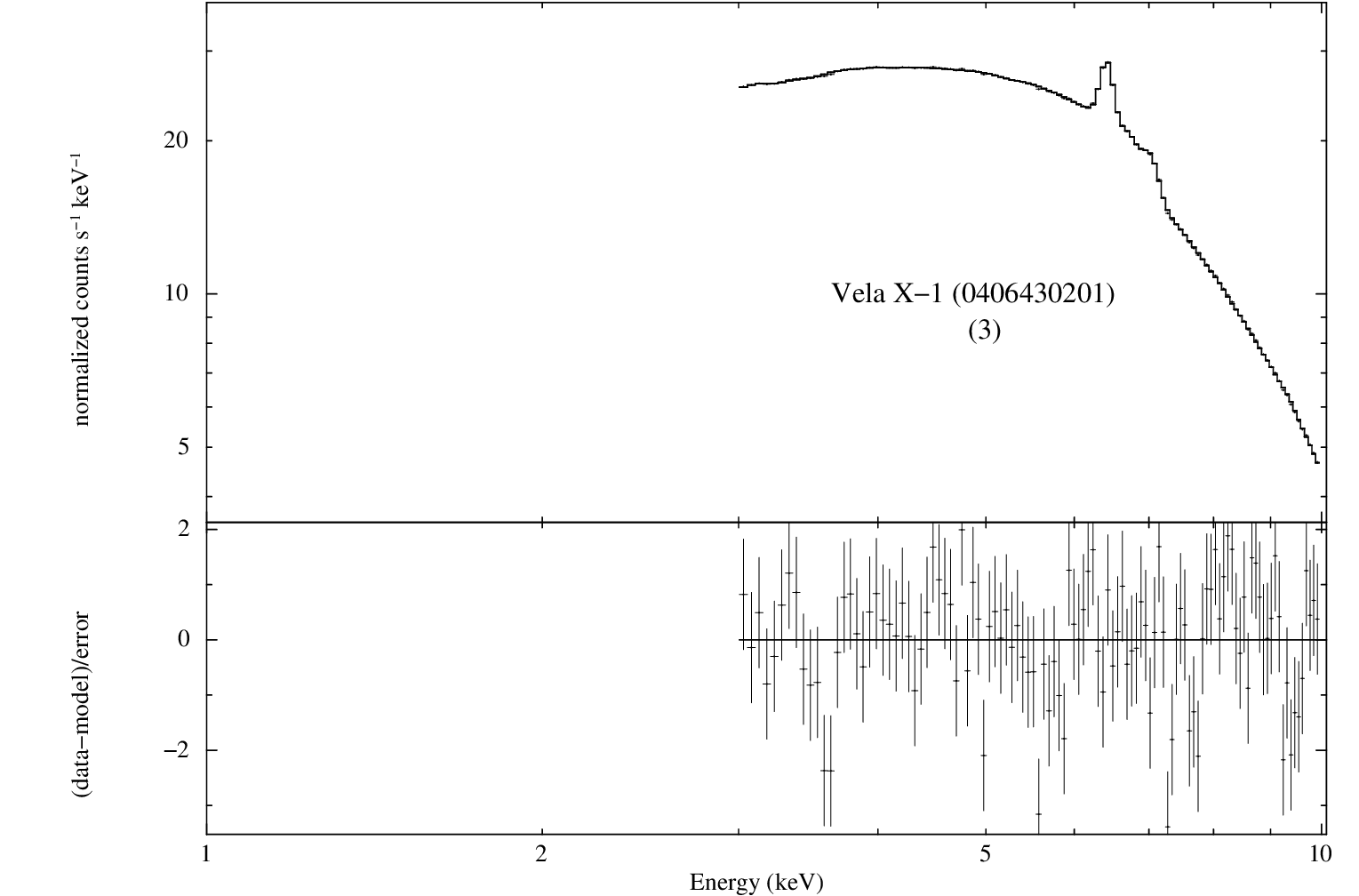}
\includegraphics[height=4.2cm,width=4.3cm,angle=-90]{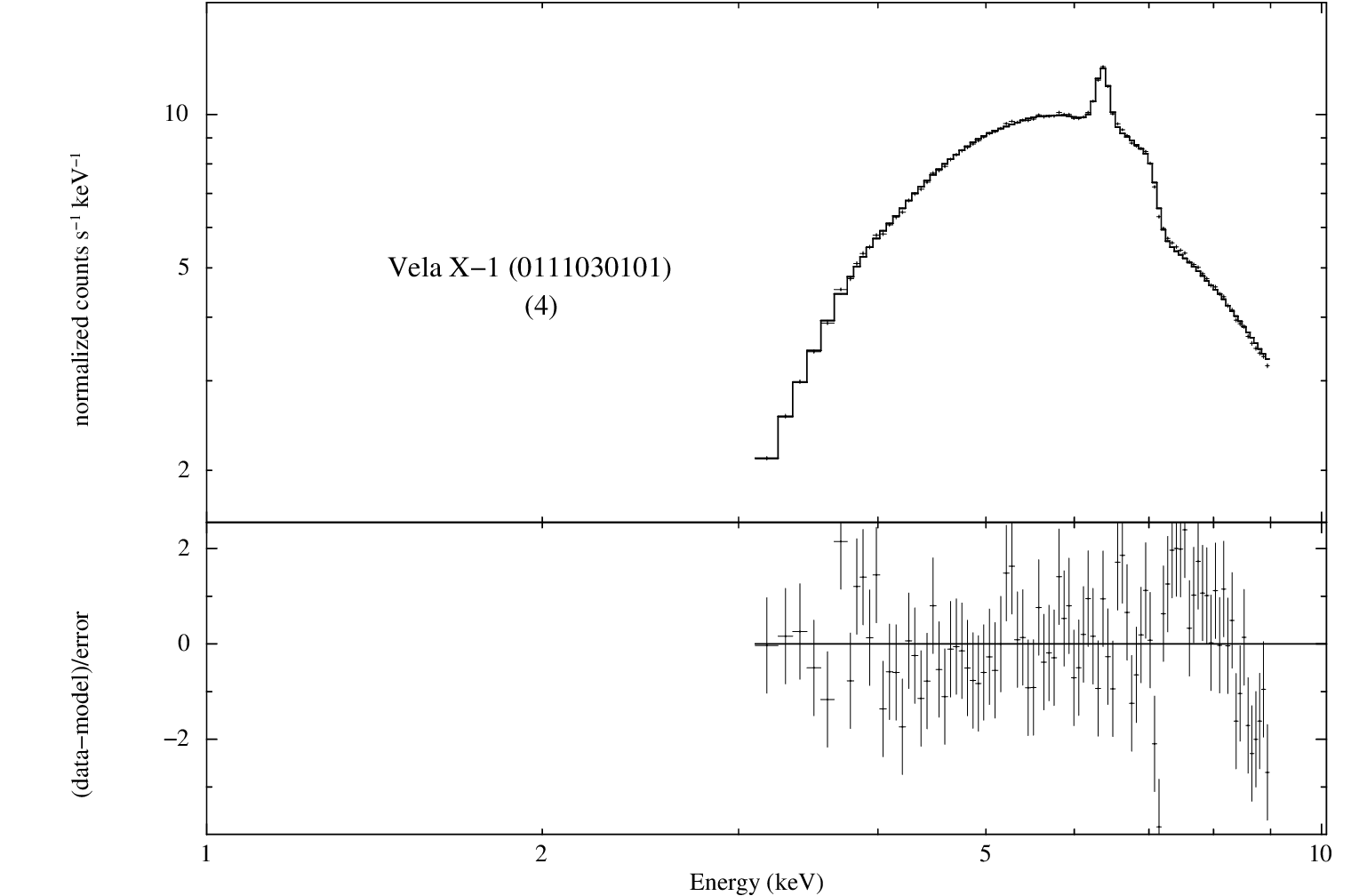}
\includegraphics[height=4.2cm,width=4.3cm,angle=-90]{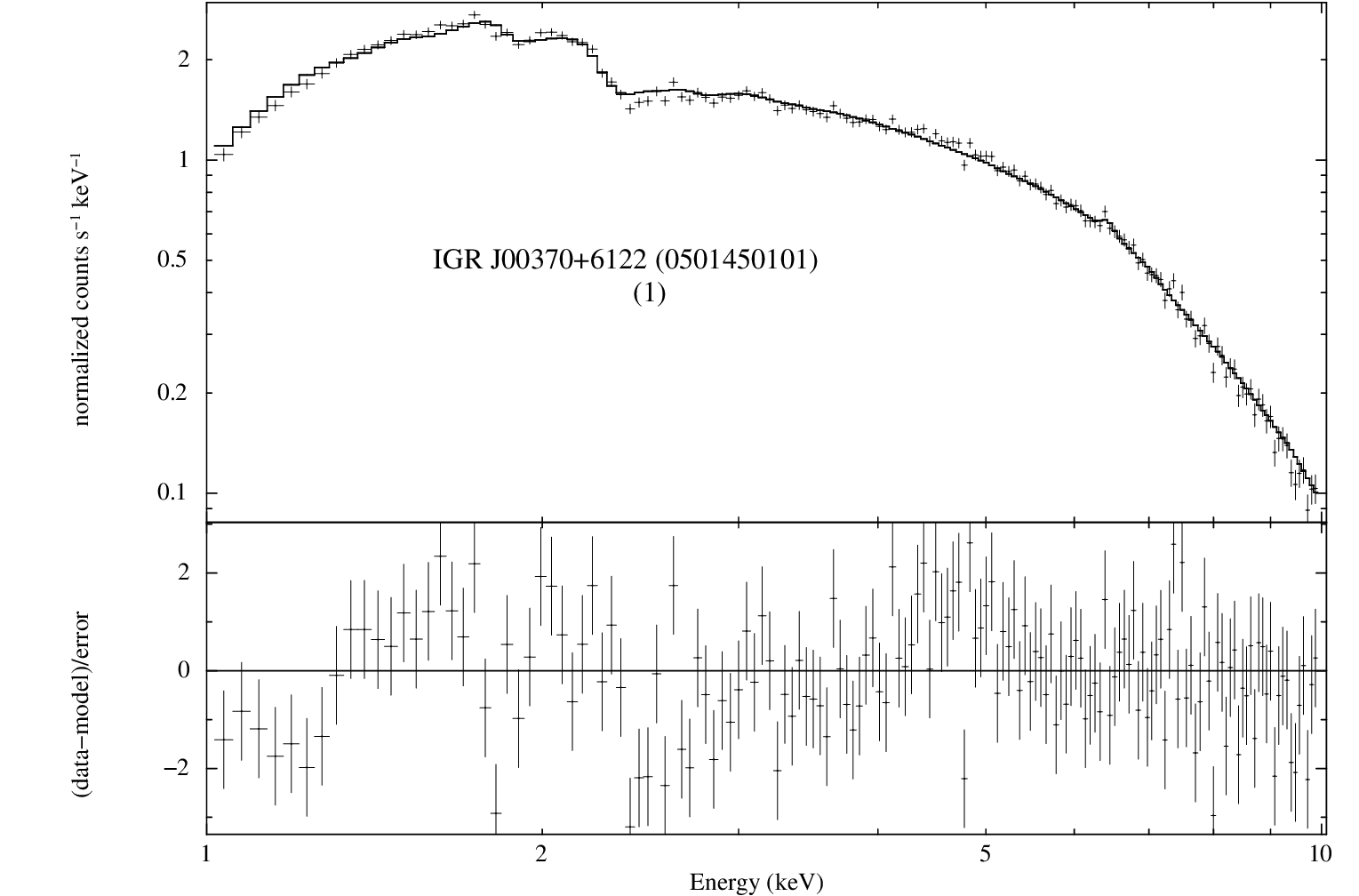}
\includegraphics[height=4.2cm,width=4.3cm,angle=-90]{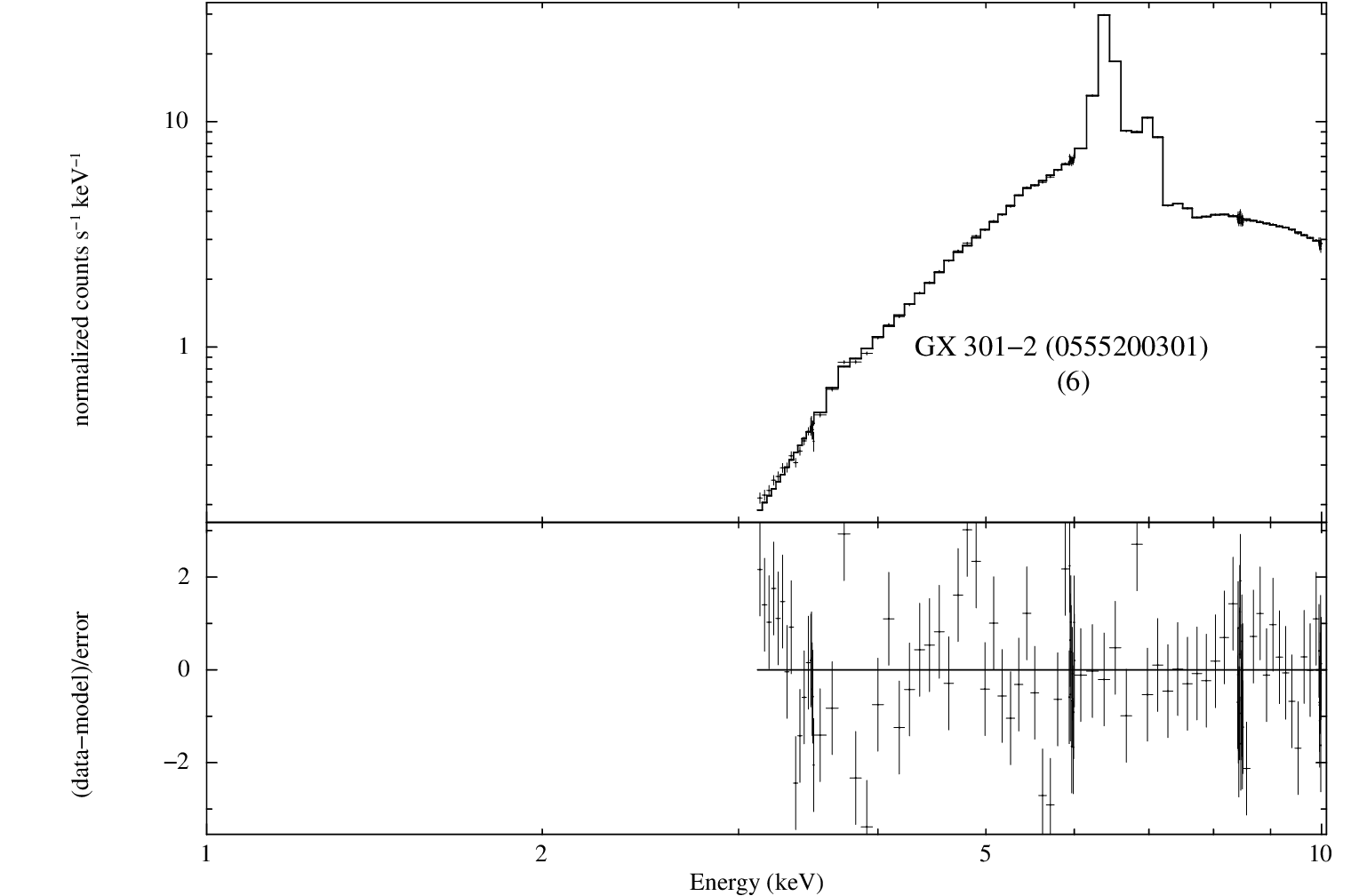}
\includegraphics[height=4.2cm,width=4.3cm,angle=-90]{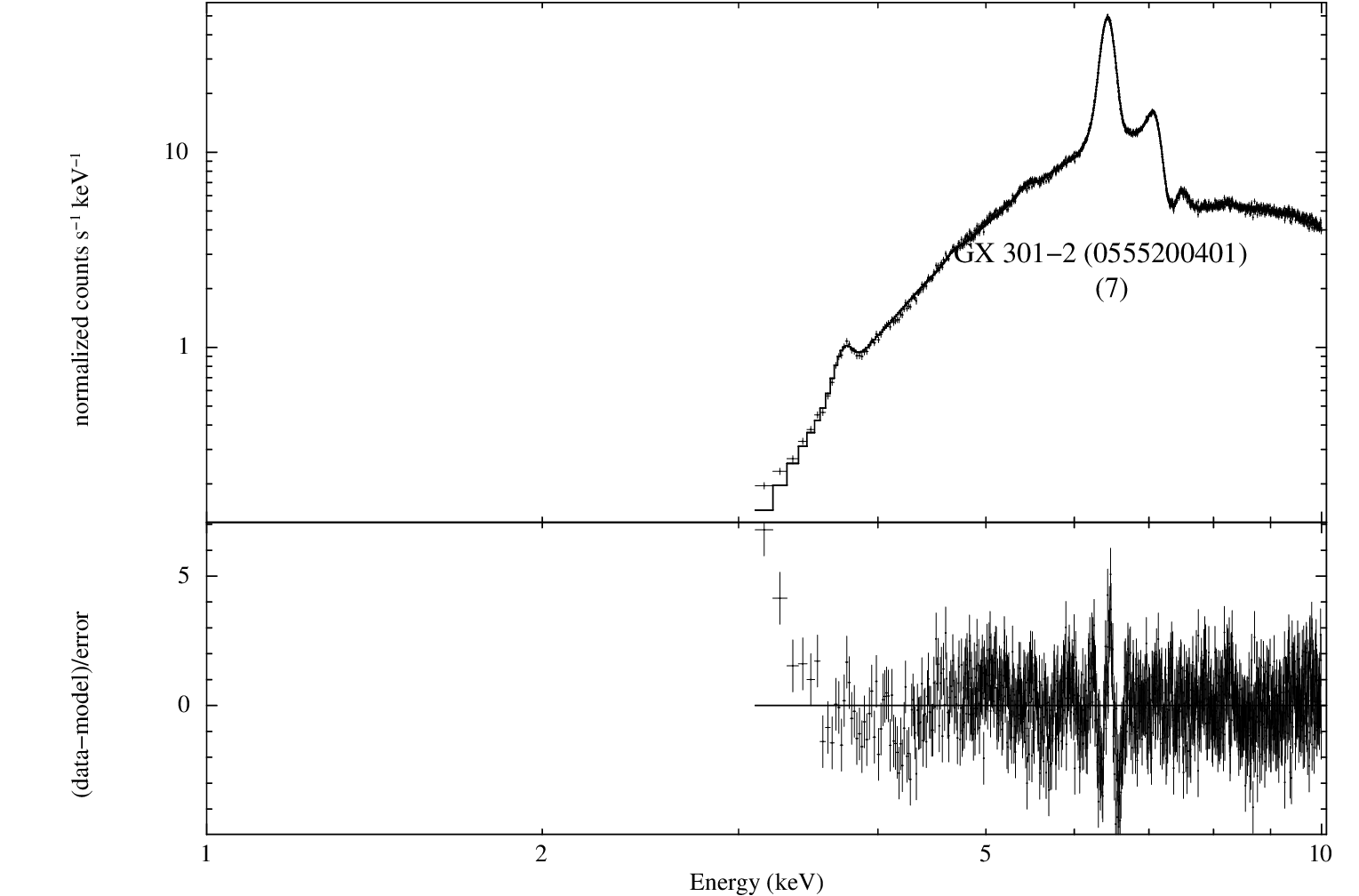}
\includegraphics[height=4.2cm,width=4.3cm,angle=-90]{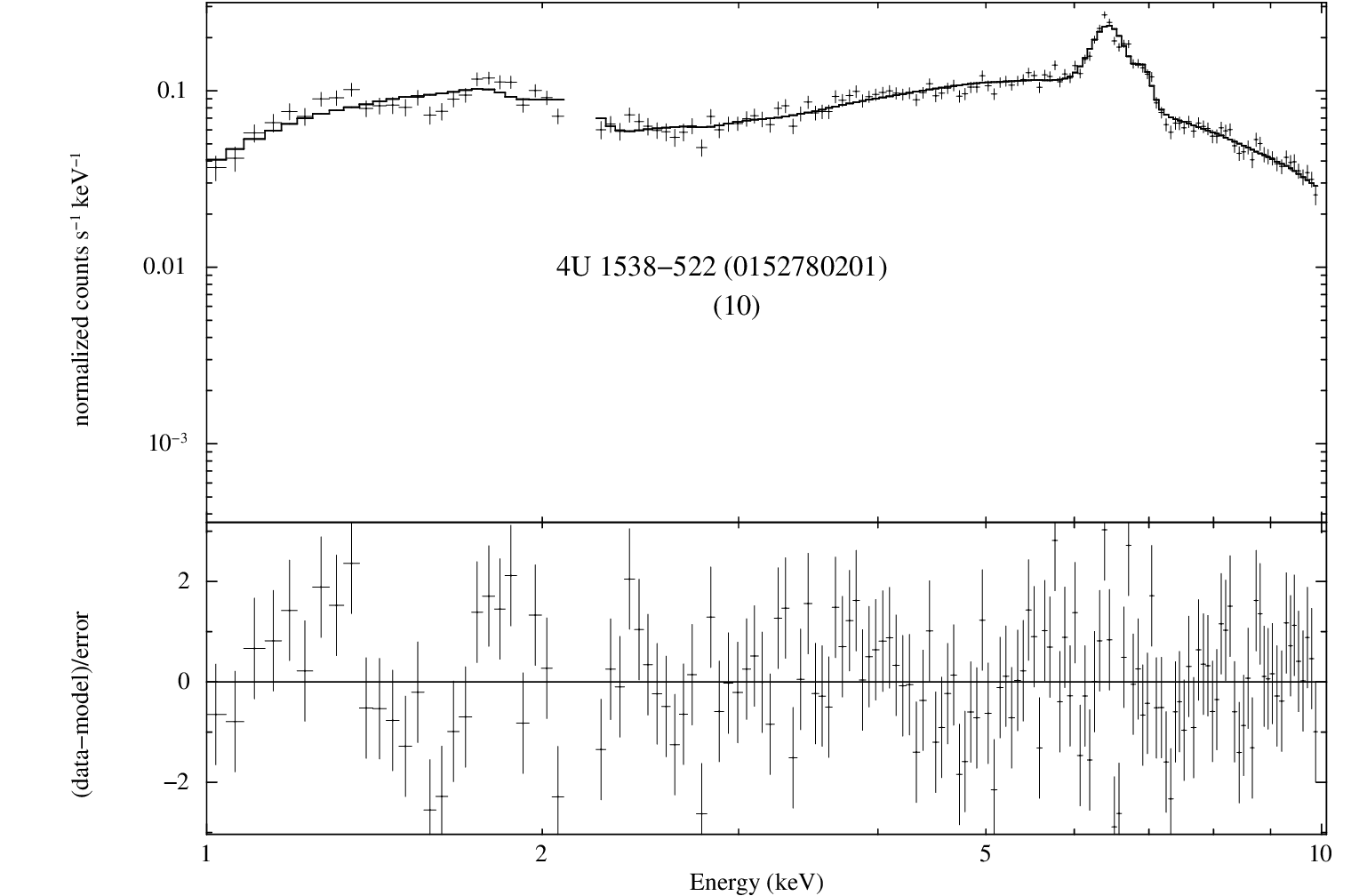}
\includegraphics[height=4.2cm,width=4.3cm,angle=-90]{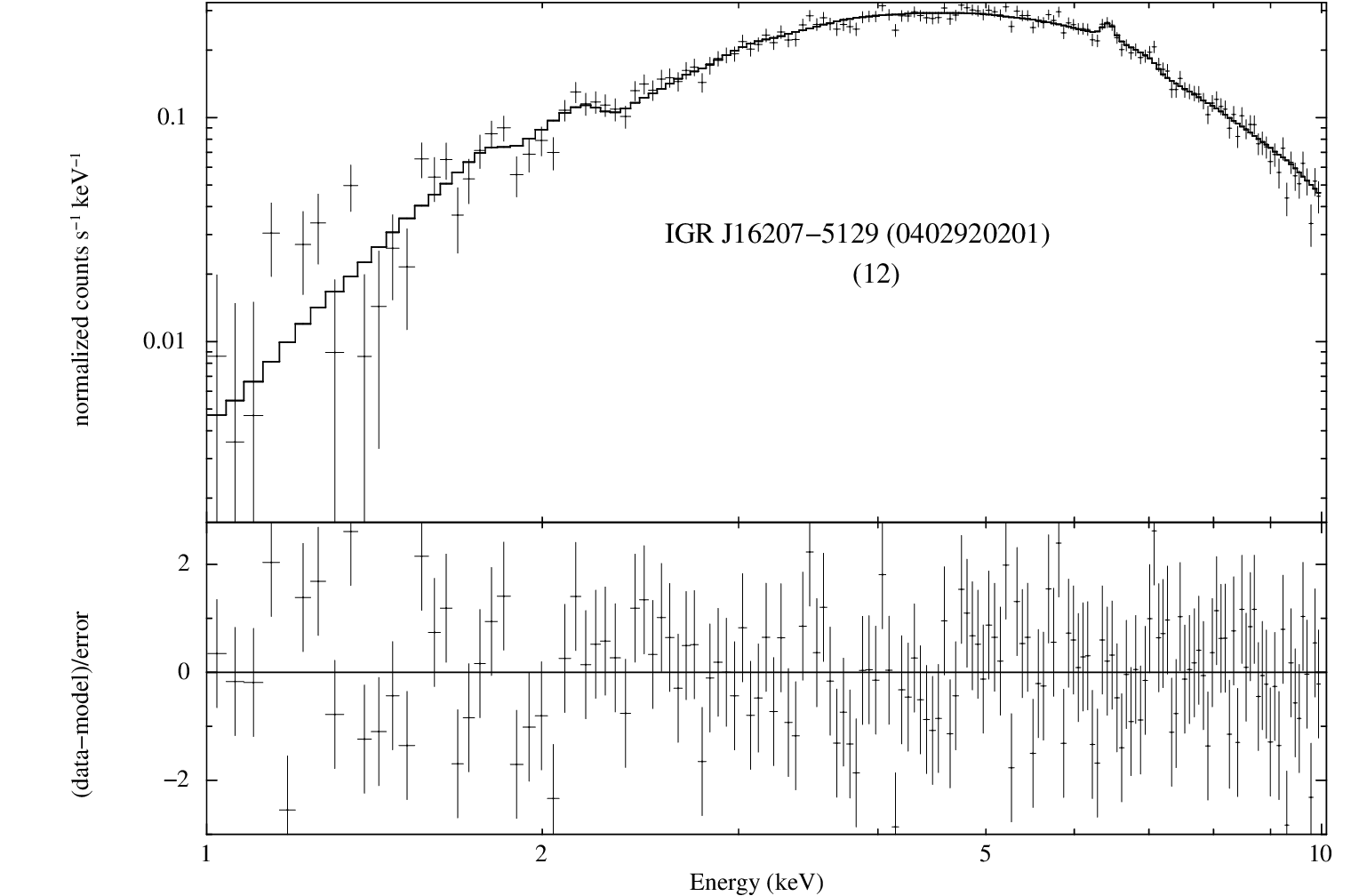}
\includegraphics[height=4.2cm,width=4.3cm,angle=-90]{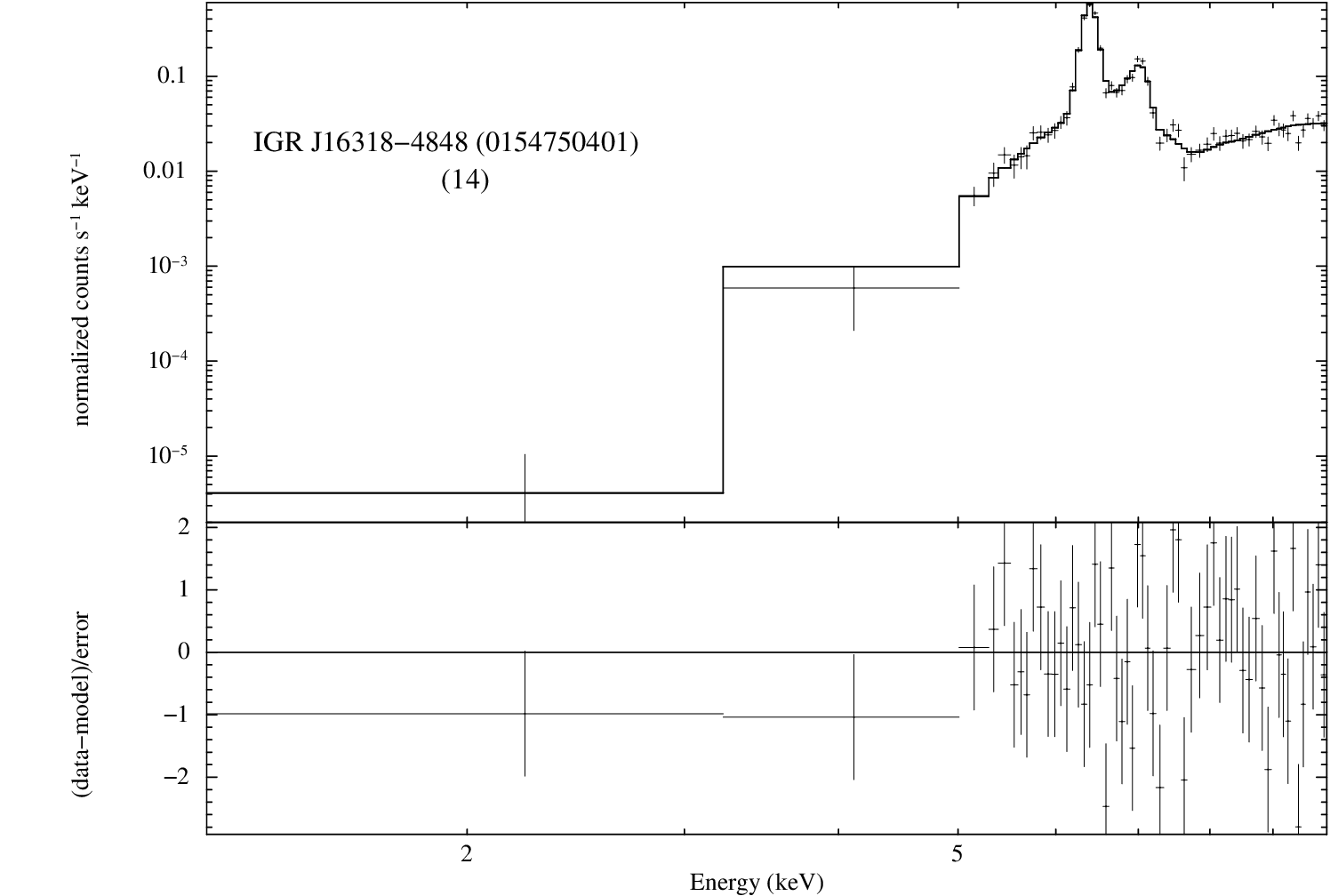}
\includegraphics[height=4.2cm,width=4.3cm,angle=-90]{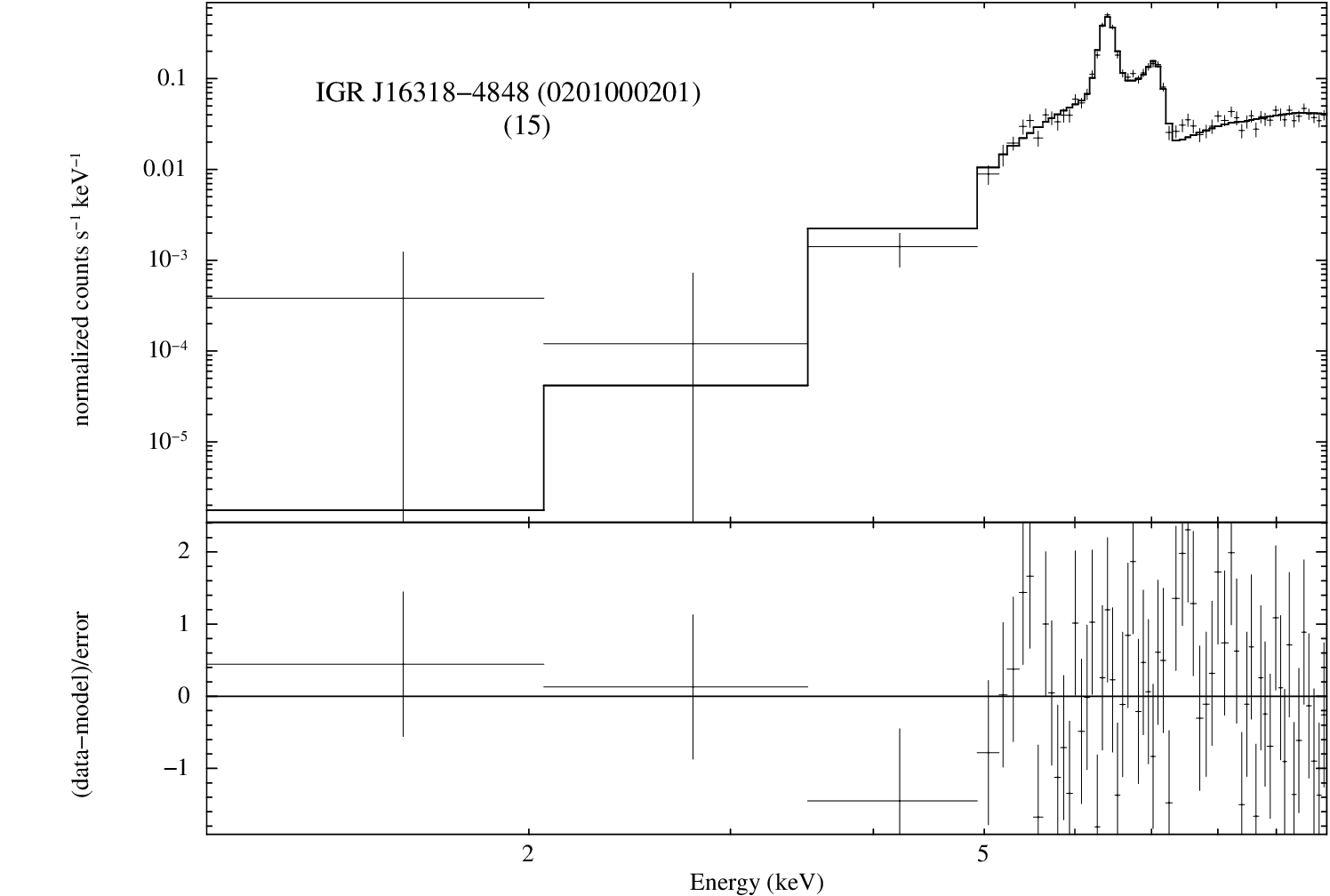}
\includegraphics[height=4.2cm,width=4.3cm,angle=-90]{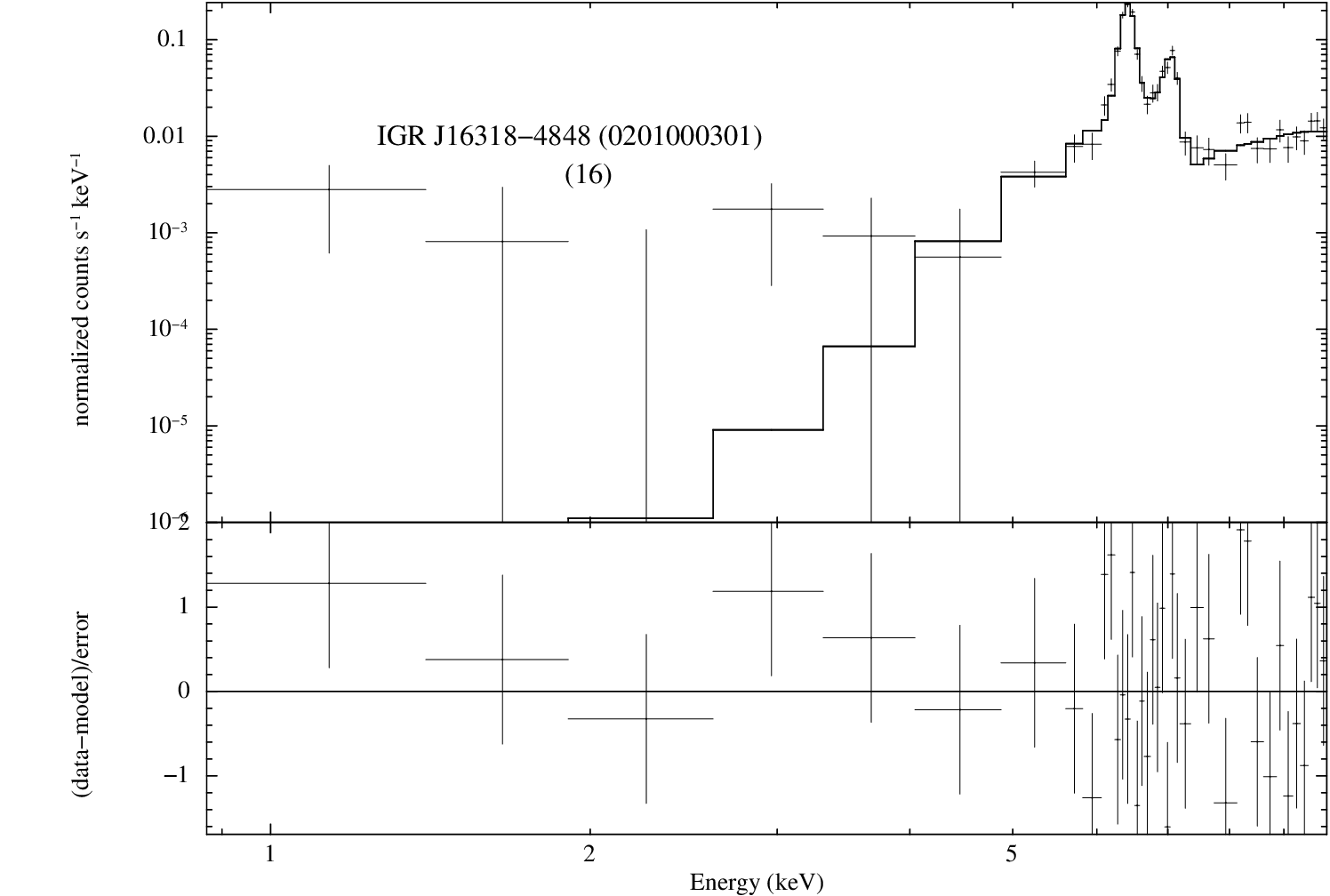}
\includegraphics[height=4.2cm,width=4.3cm,angle=-90]{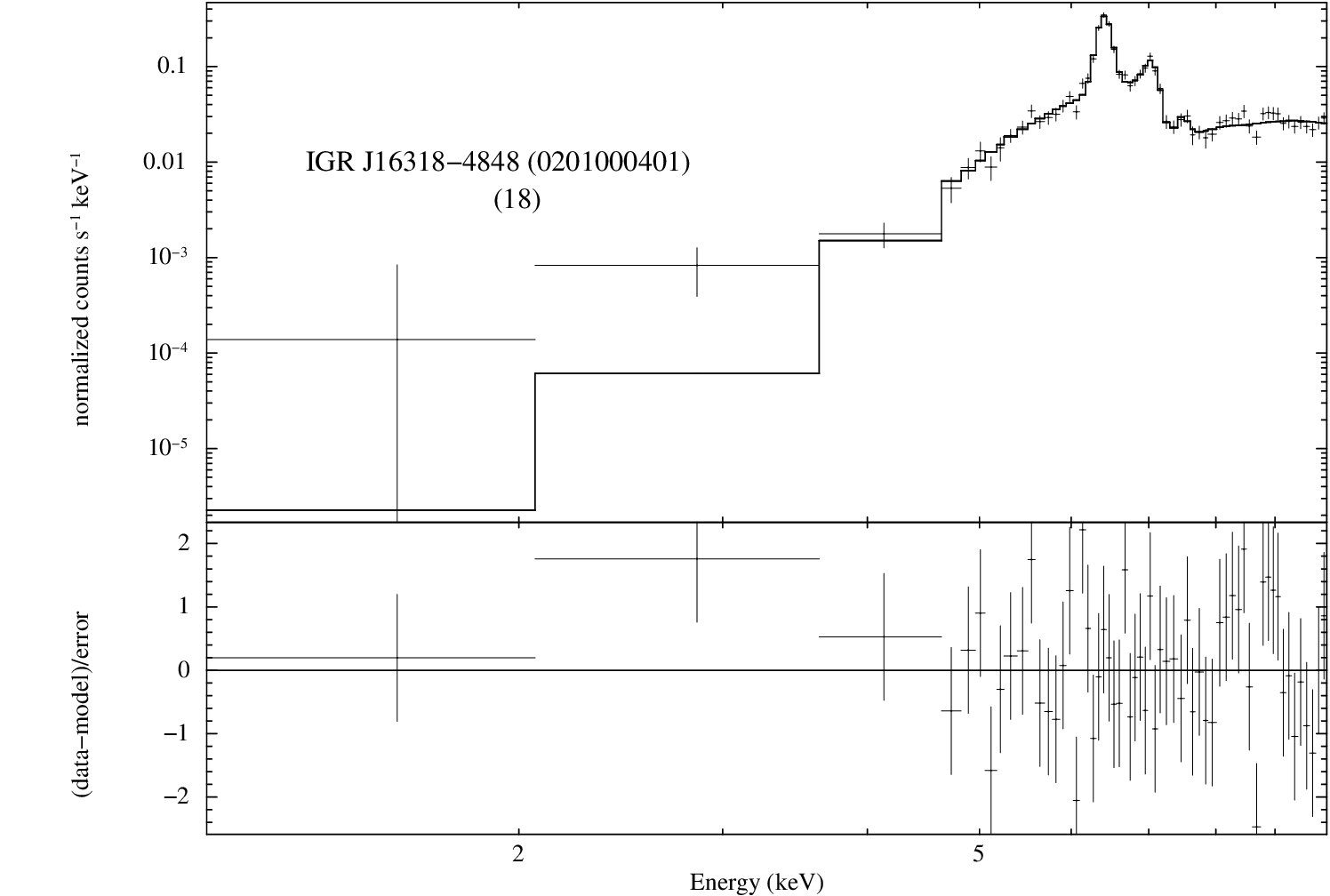}
\includegraphics[height=4.2cm,width=4.3cm,angle=-90]{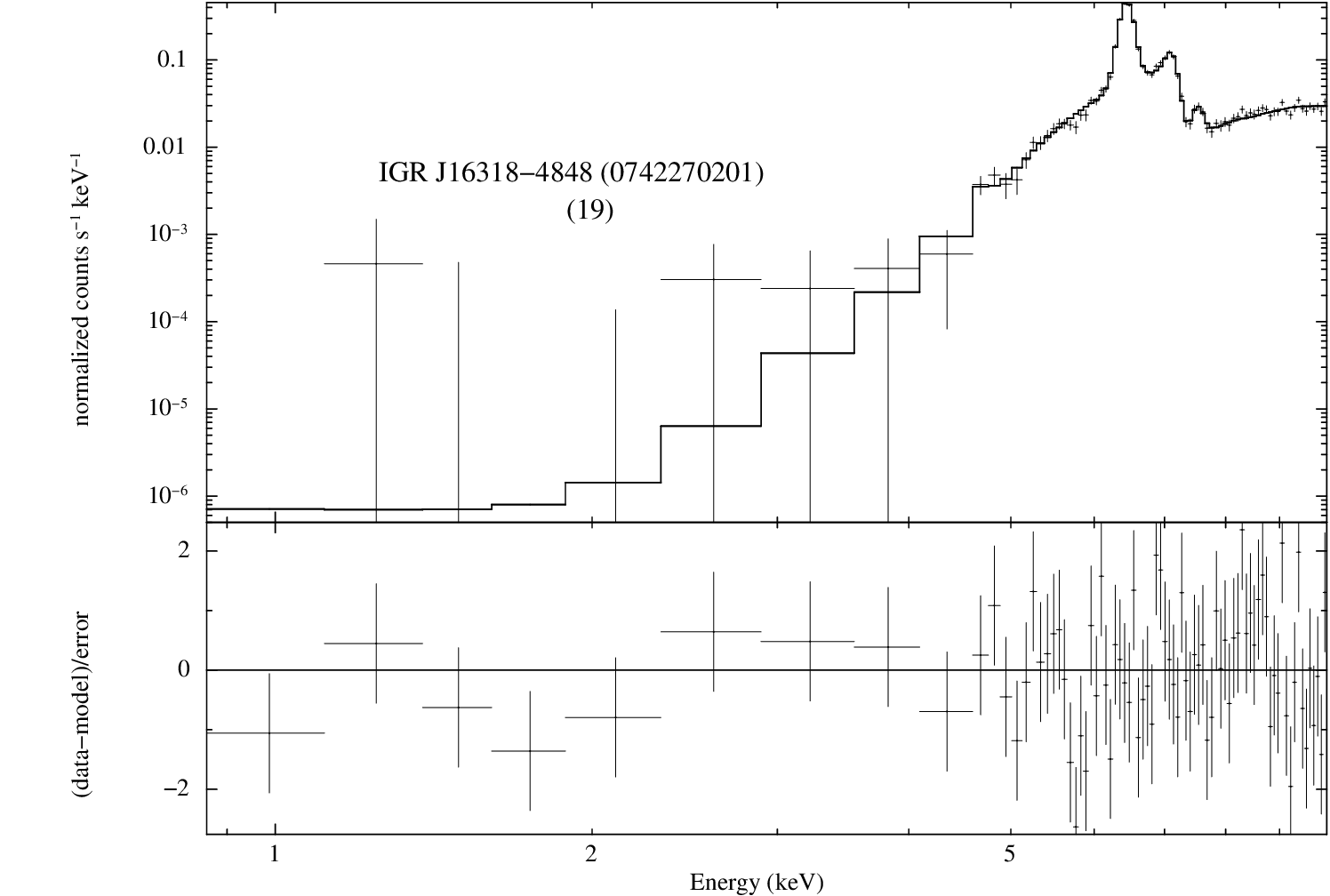} 
\includegraphics[height=4.2cm,width=4.3cm,angle=-90]{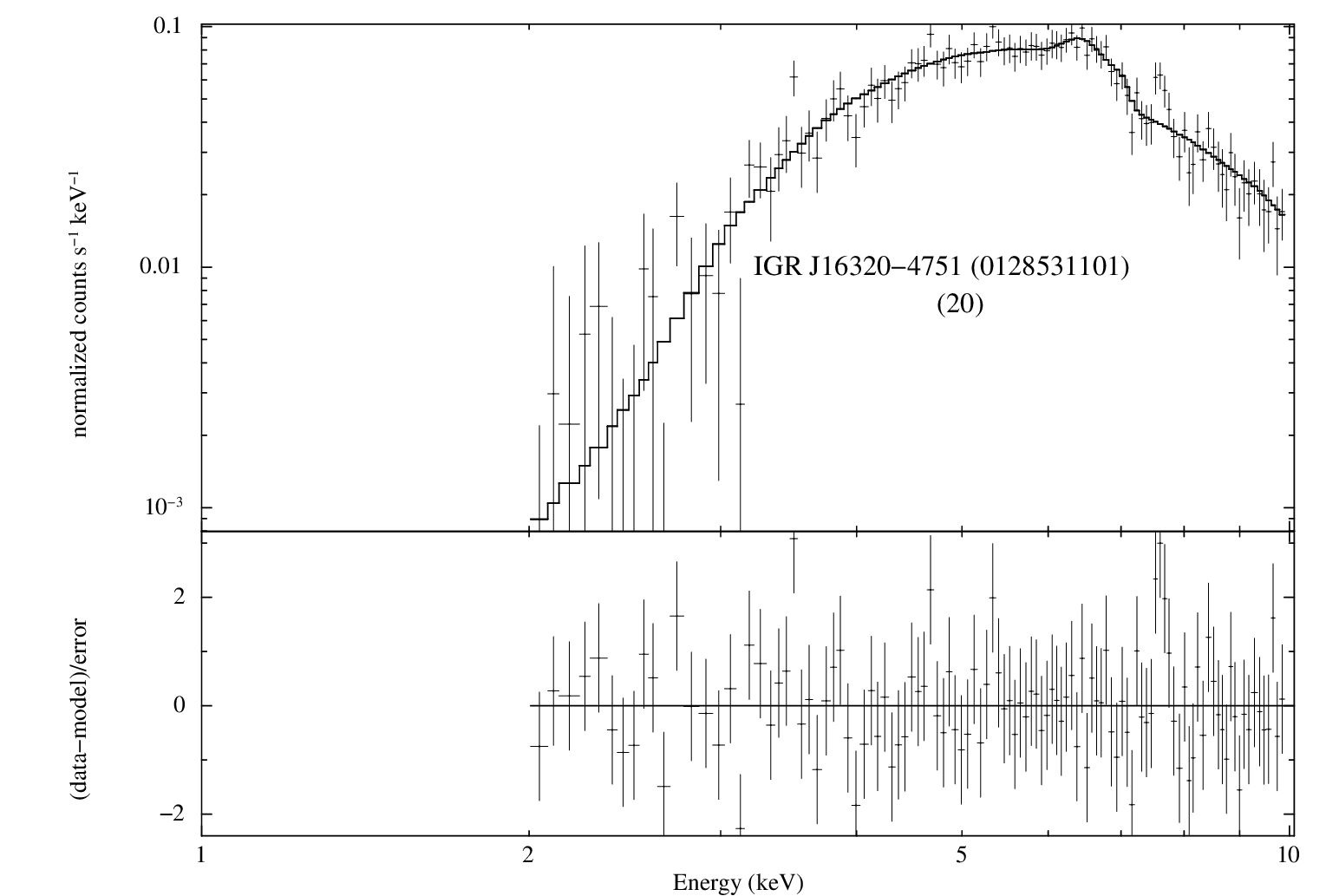}
\includegraphics[height=4.2cm,width=4.3cm,angle=-90]{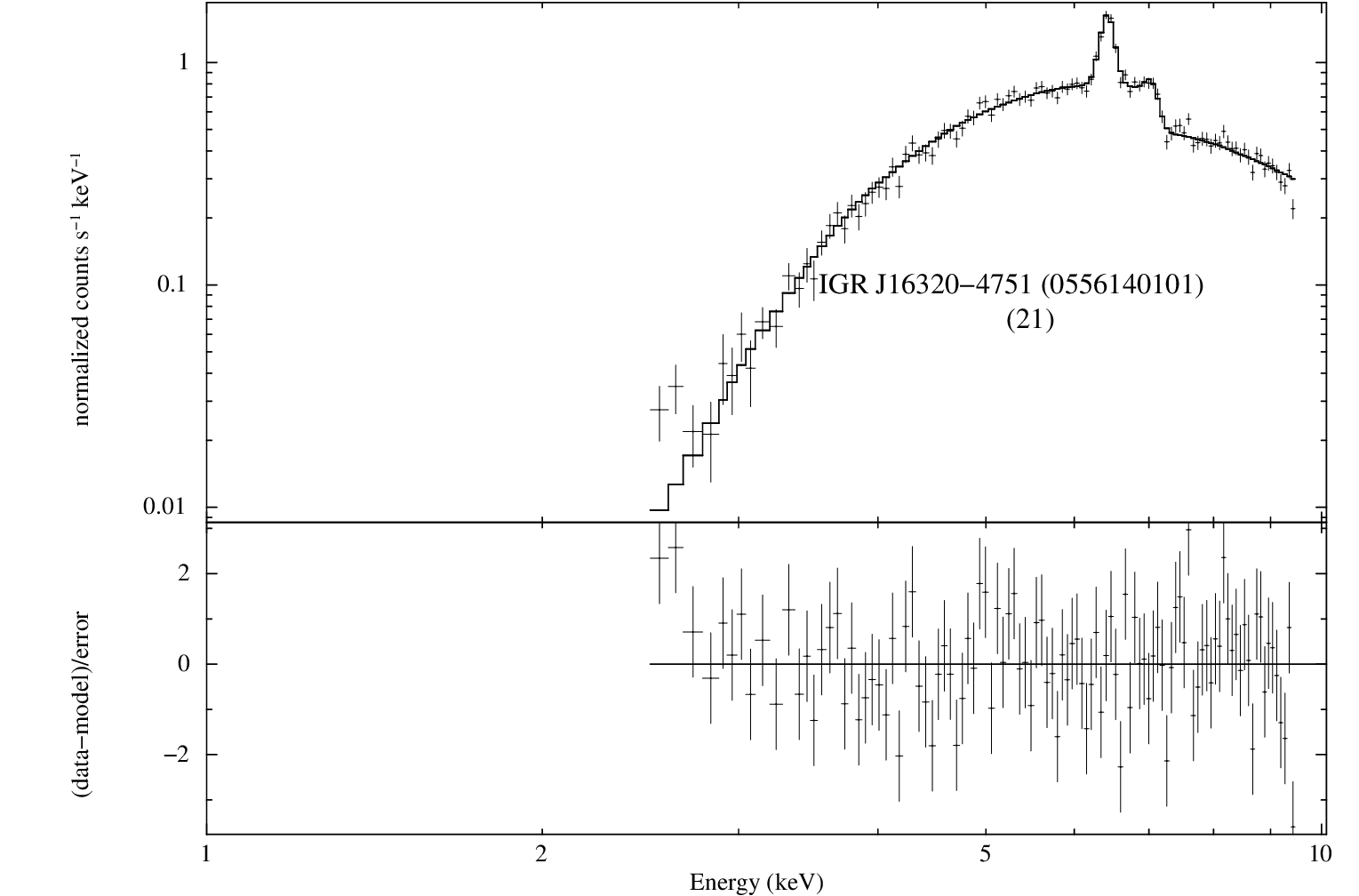}
\includegraphics[height=4.2cm,width=4.3cm,angle=-90]{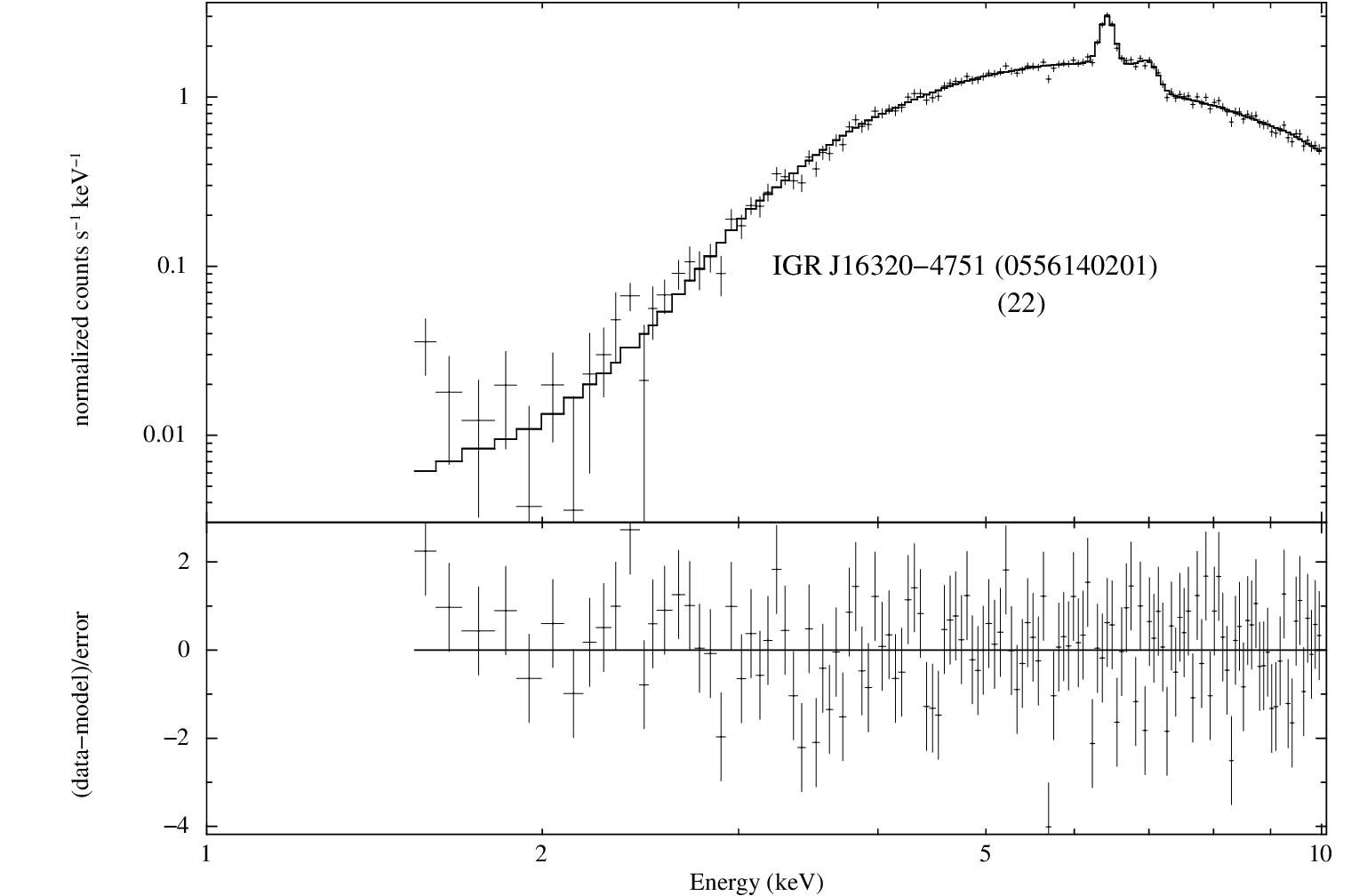}
\includegraphics[height=4.2cm,width=4.3cm,angle=-90]{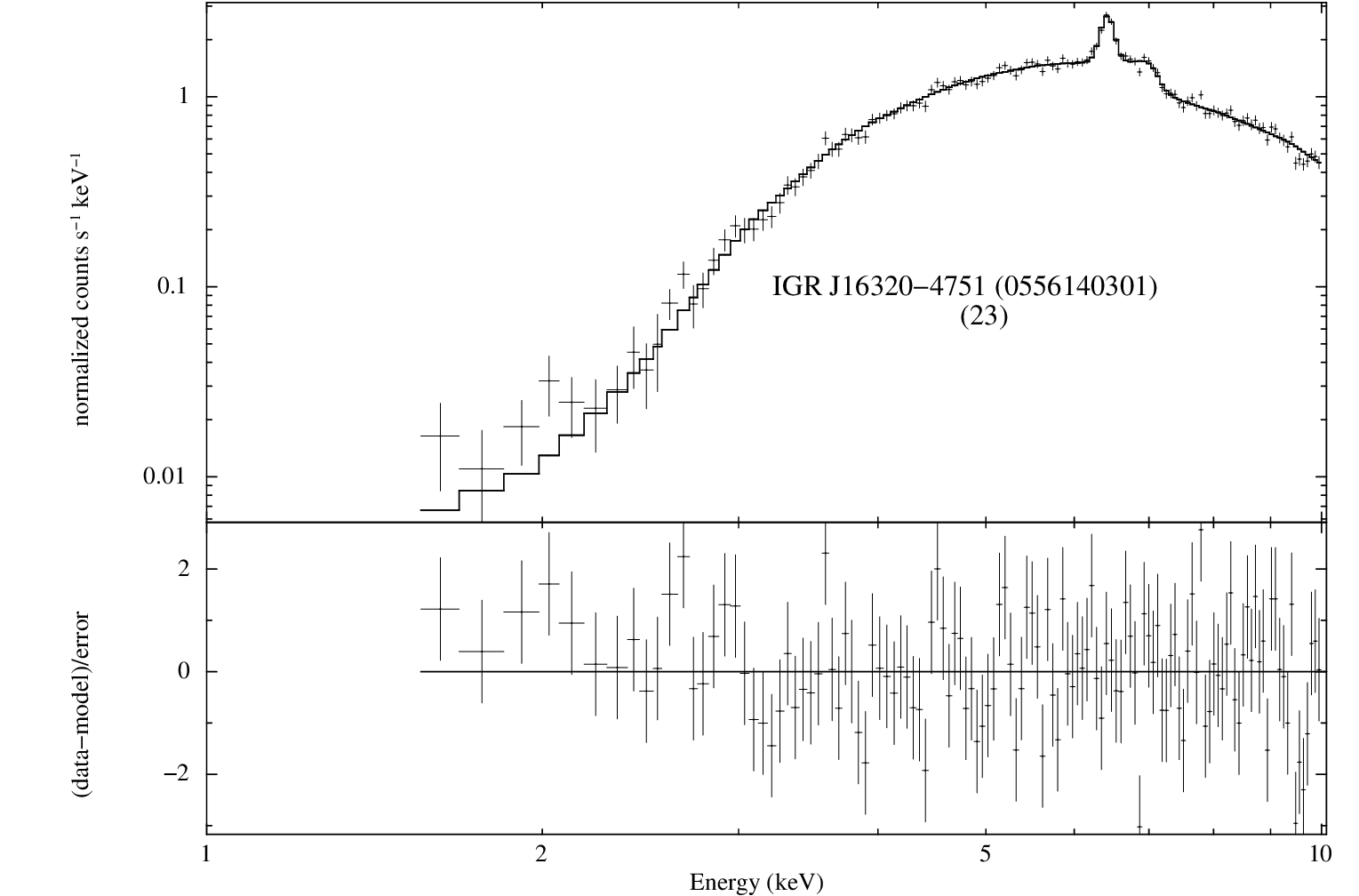}
\includegraphics[height=4.2cm,width=4.3cm,angle=-90]{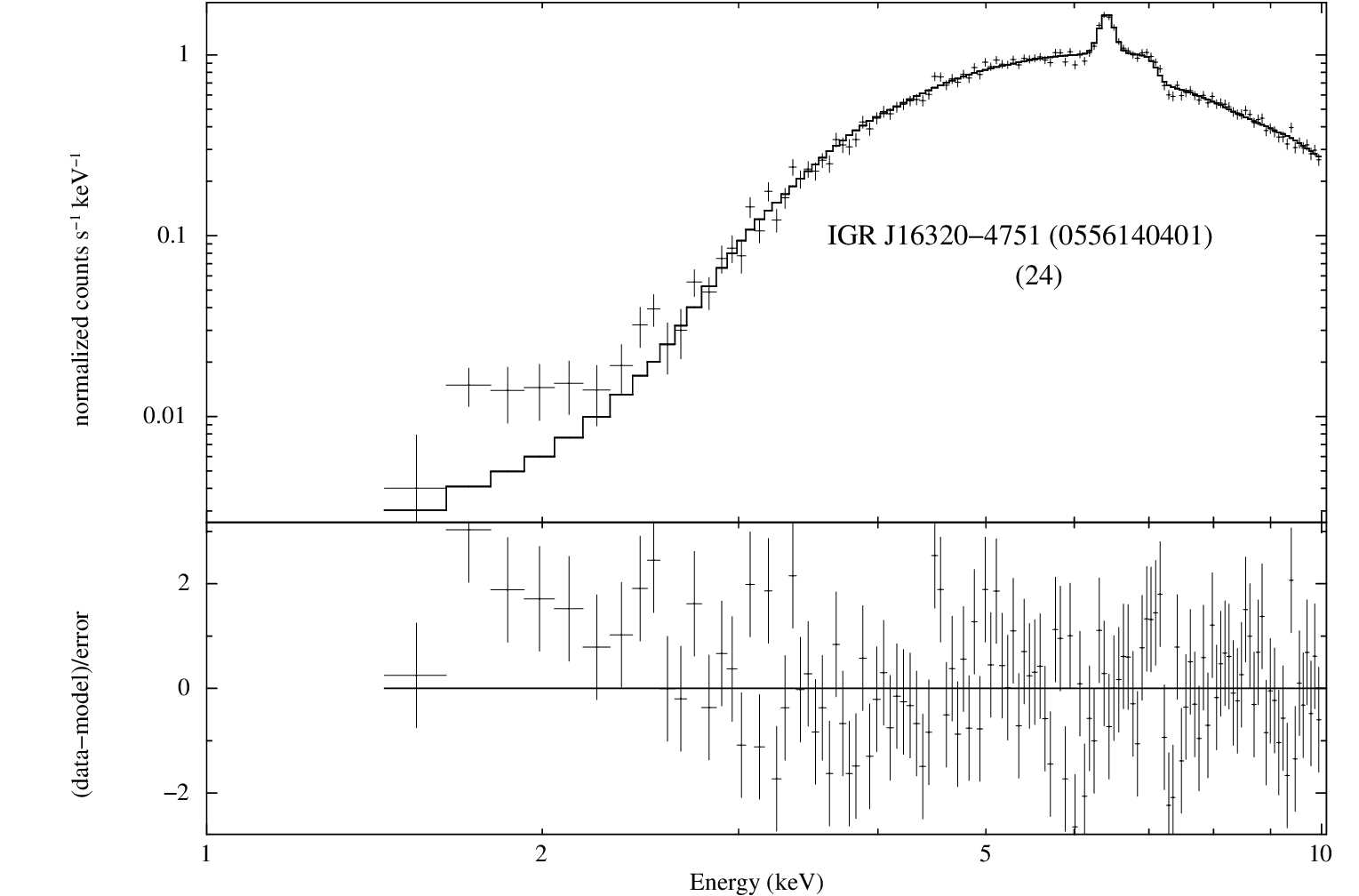}
\includegraphics[height=4.2cm,width=4.3cm,angle=-90]{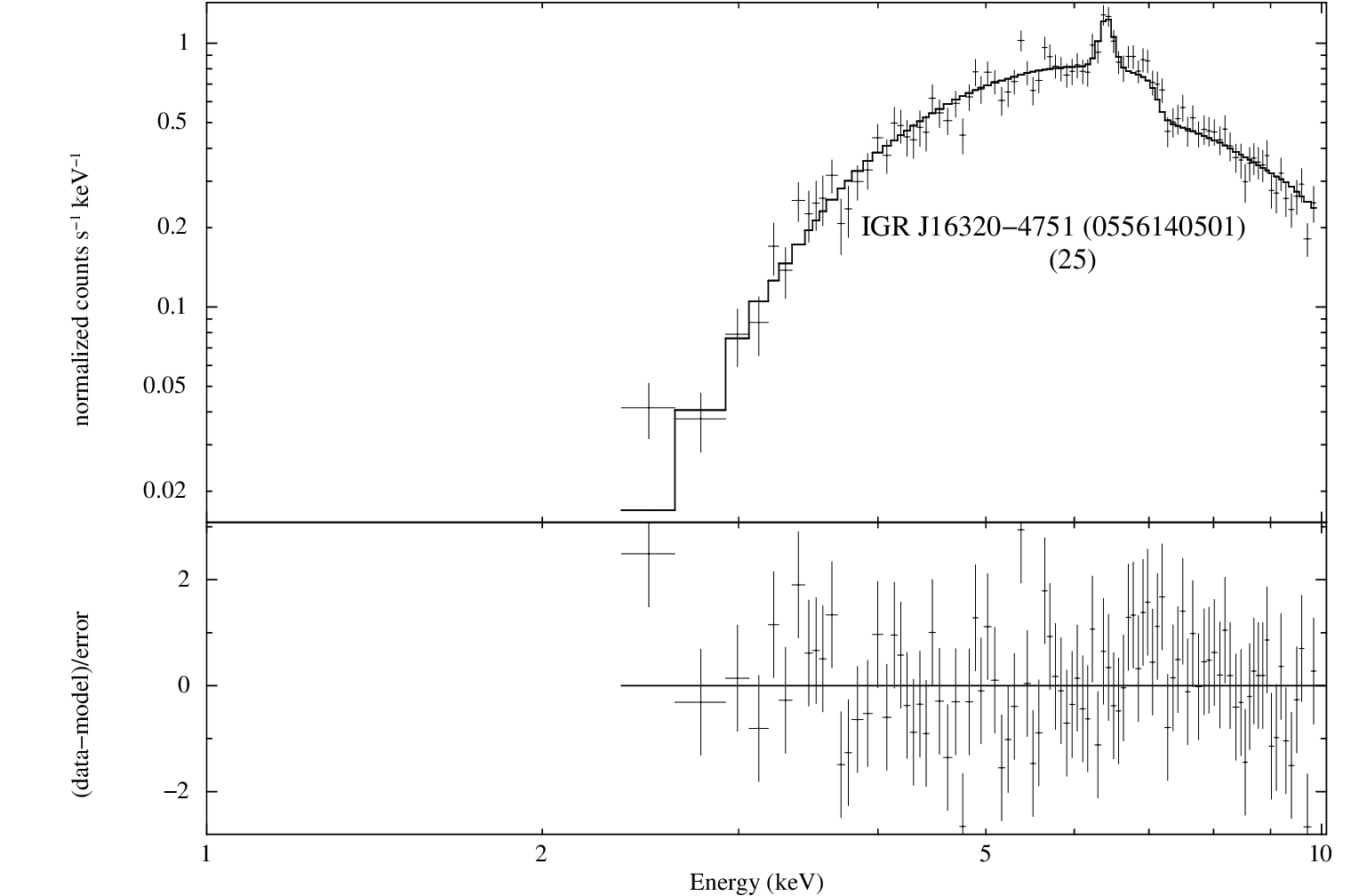}
\includegraphics[height=4.2cm,width=4.3cm,angle=-90]{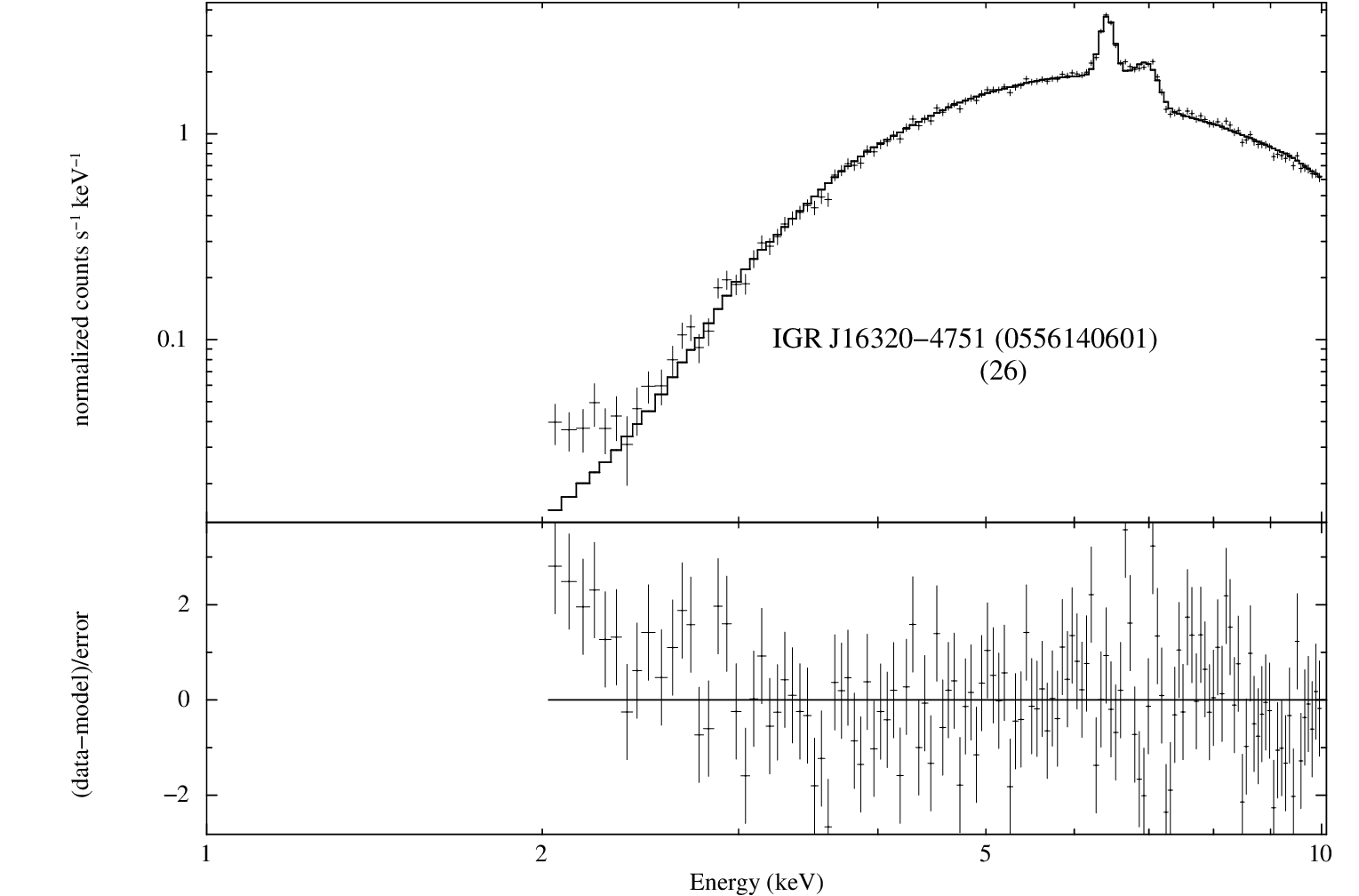}
\includegraphics[height=4.2cm,width=4.3cm,angle=-90]{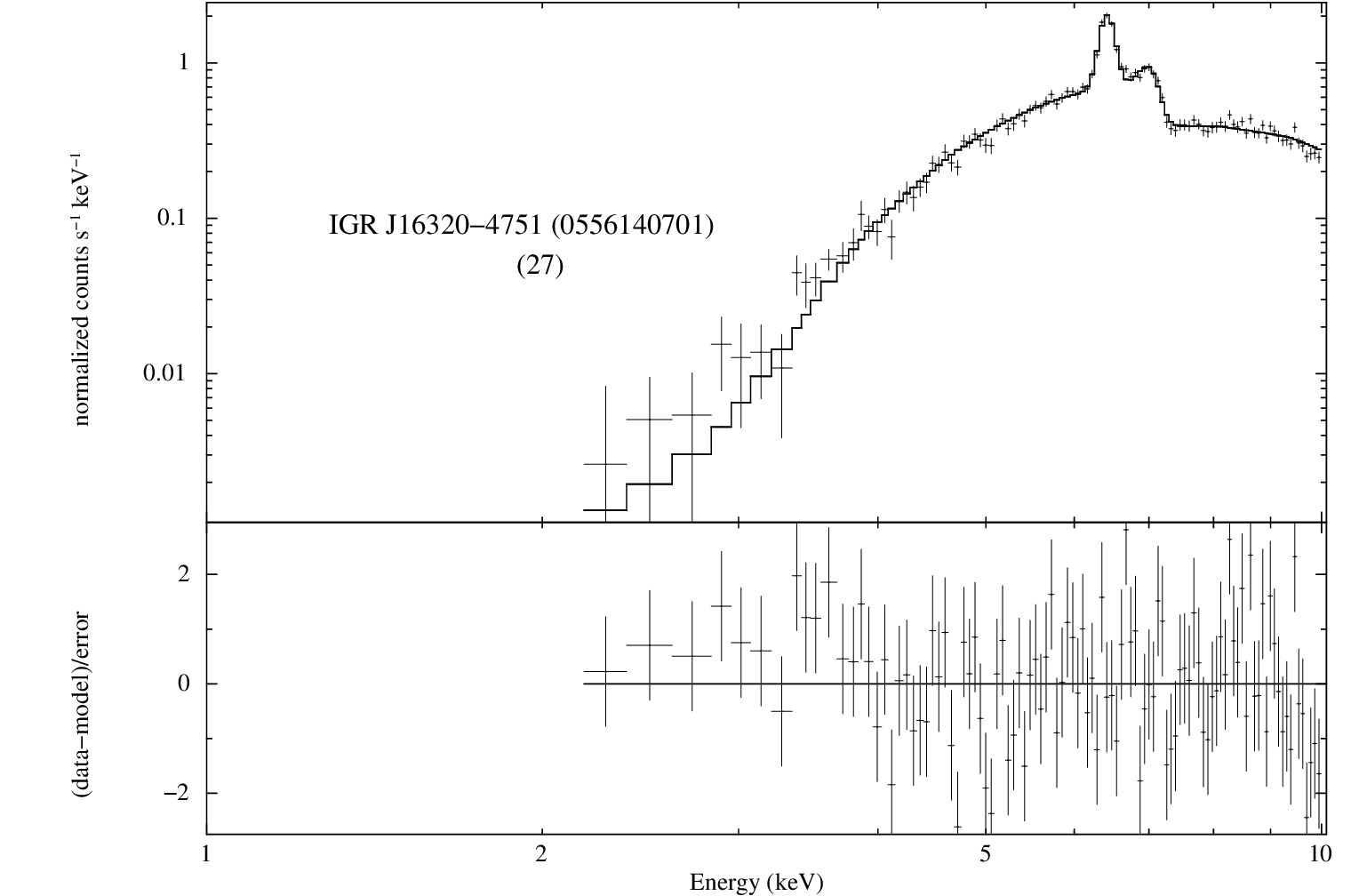}
\label{spectra3}
\end{figure*}

\begin{figure*}
\caption{\emph{XMM-Newton} spectra of all considered classical SgXBs. The source name and the observation ID is indicated in each figure.}
\includegraphics[height=4.2cm,width=4.3cm,angle=-90]{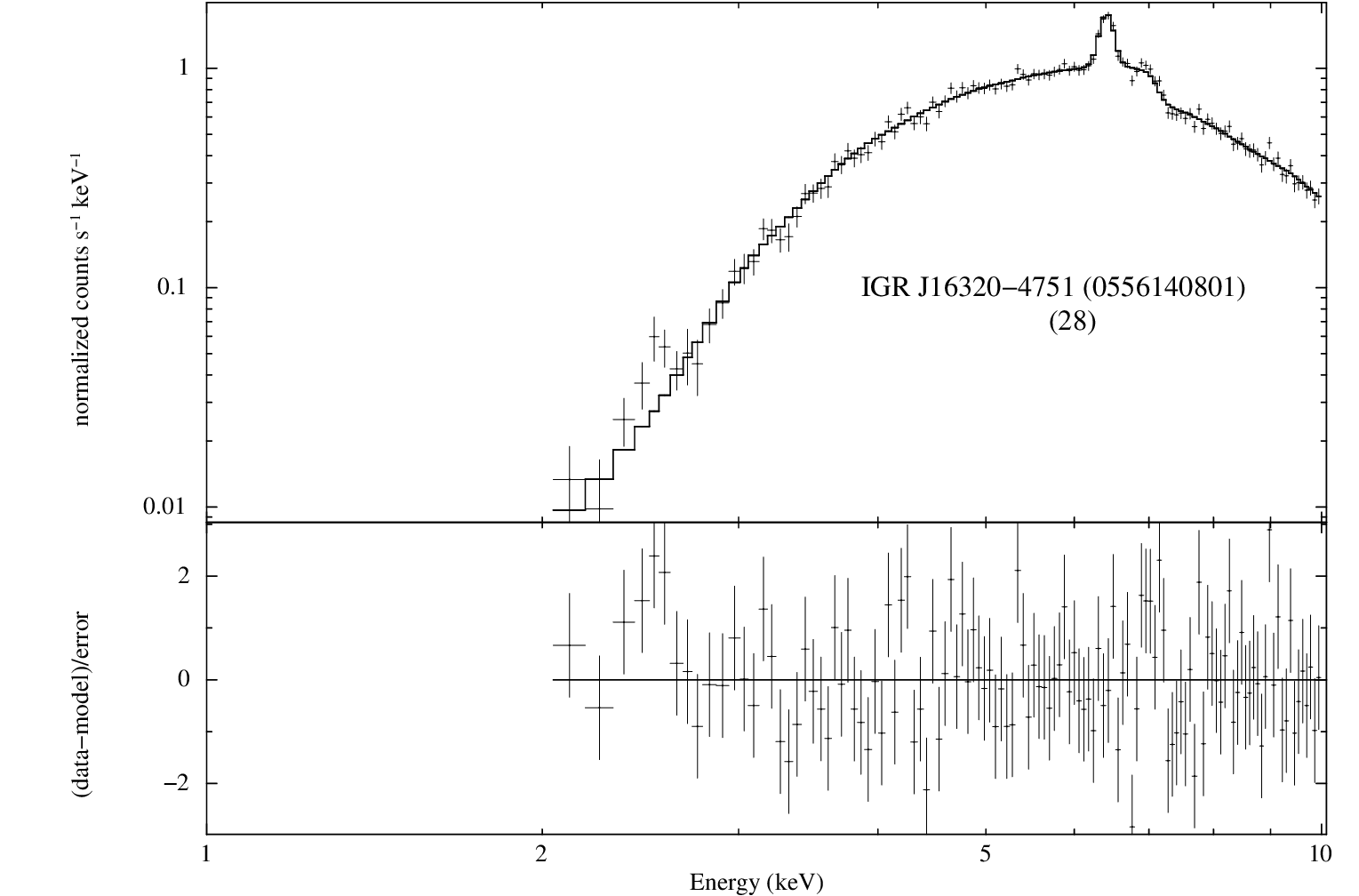}
\includegraphics[height=4.2cm,width=4.3cm,angle=-90]{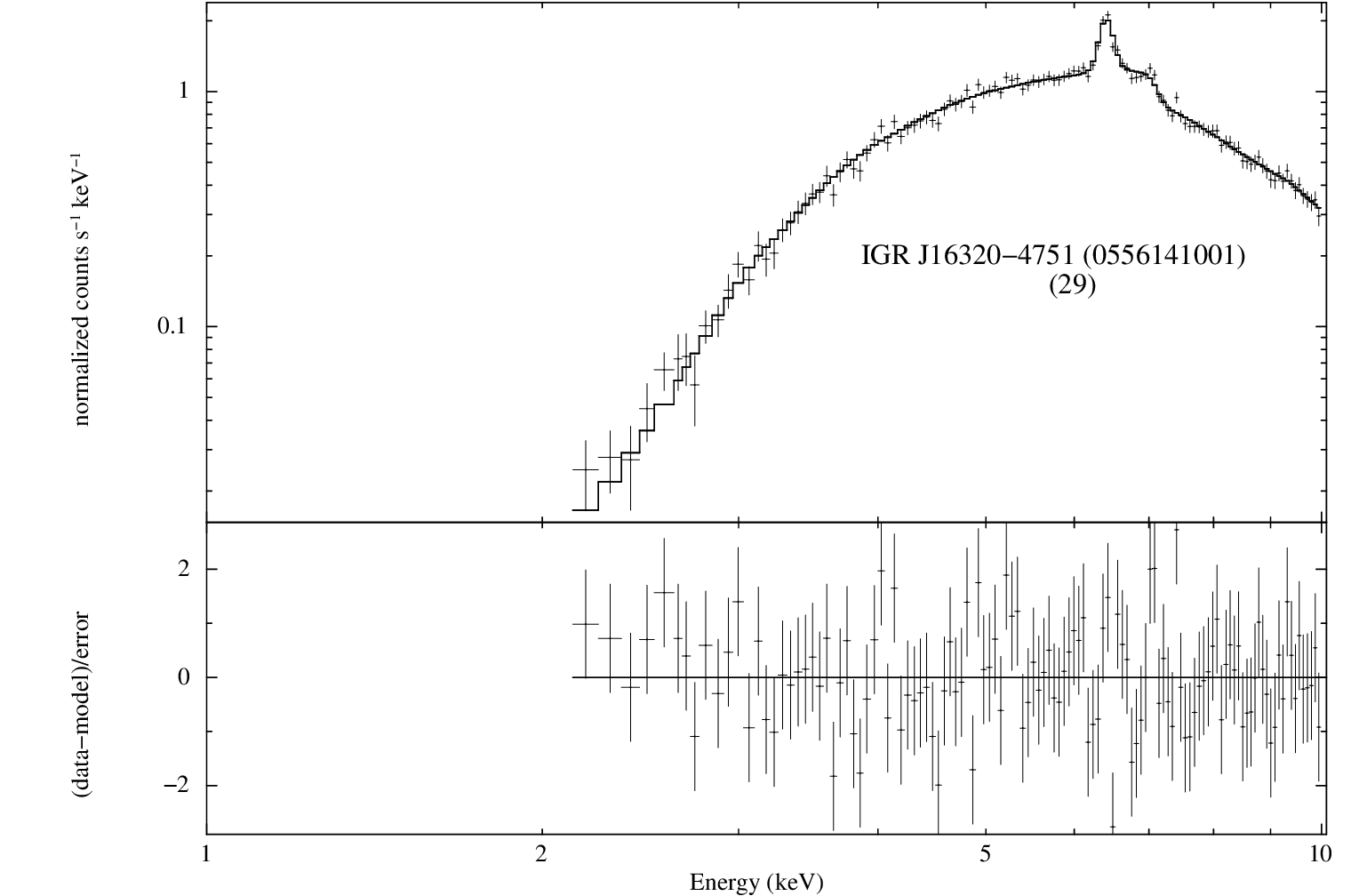}
\includegraphics[height=4.2cm,width=4.3cm,angle=-90]{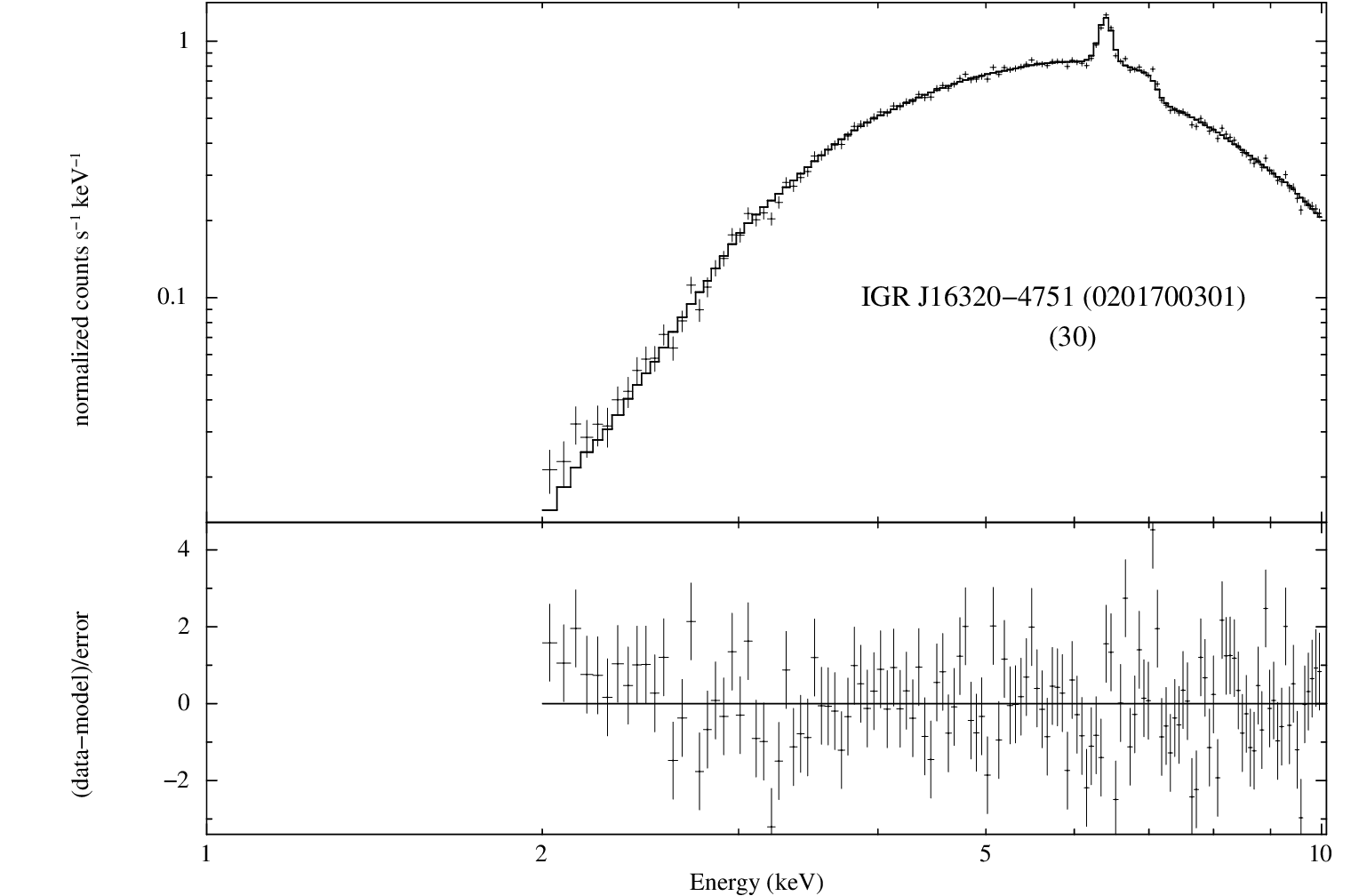}
\includegraphics[height=4.2cm,width=4.3cm,angle=-90]{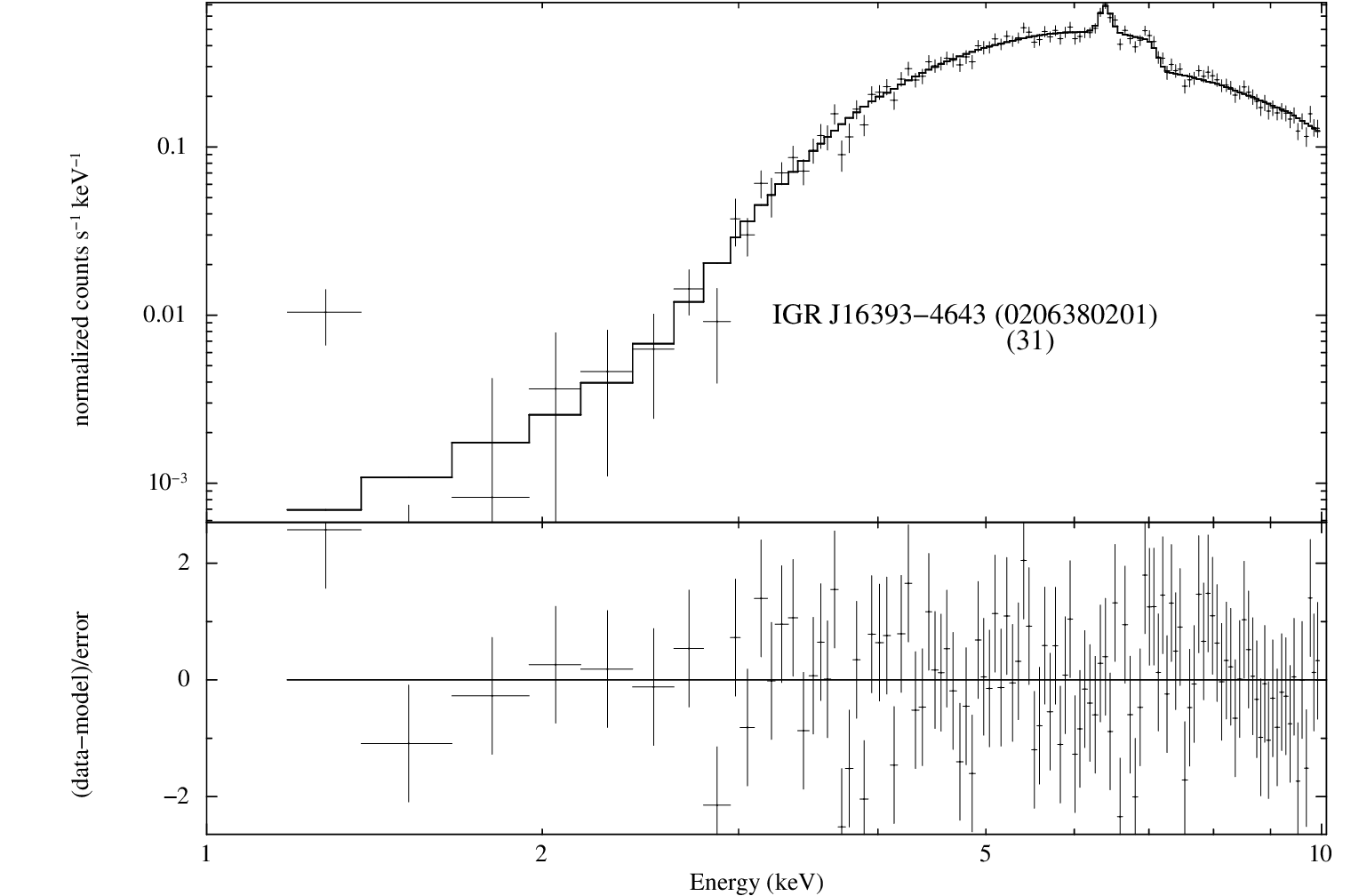}
\includegraphics[height=4.2cm,width=4.3cm,angle=-90]{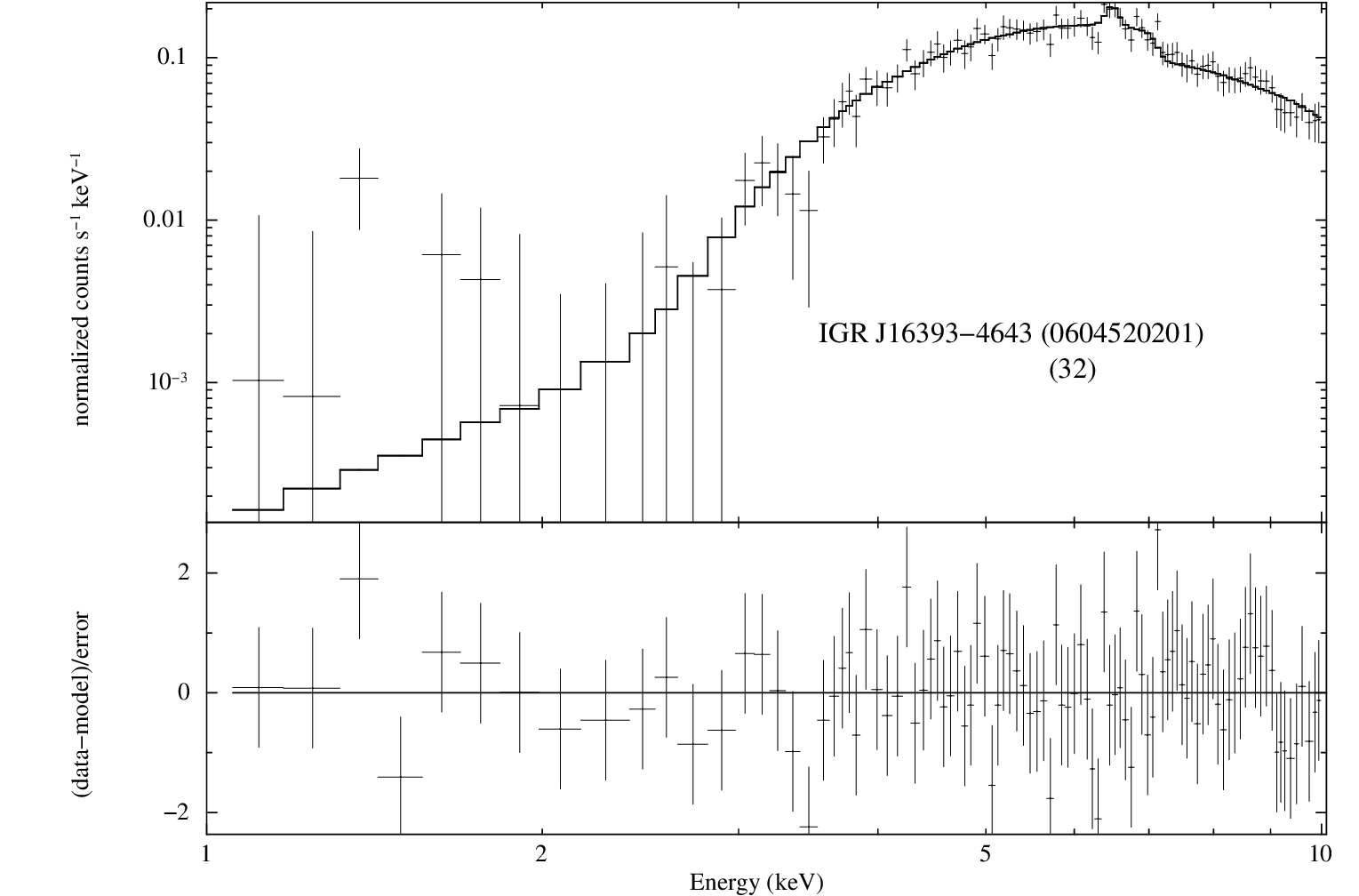}
\includegraphics[height=4.2cm,width=4.3cm,angle=-90]{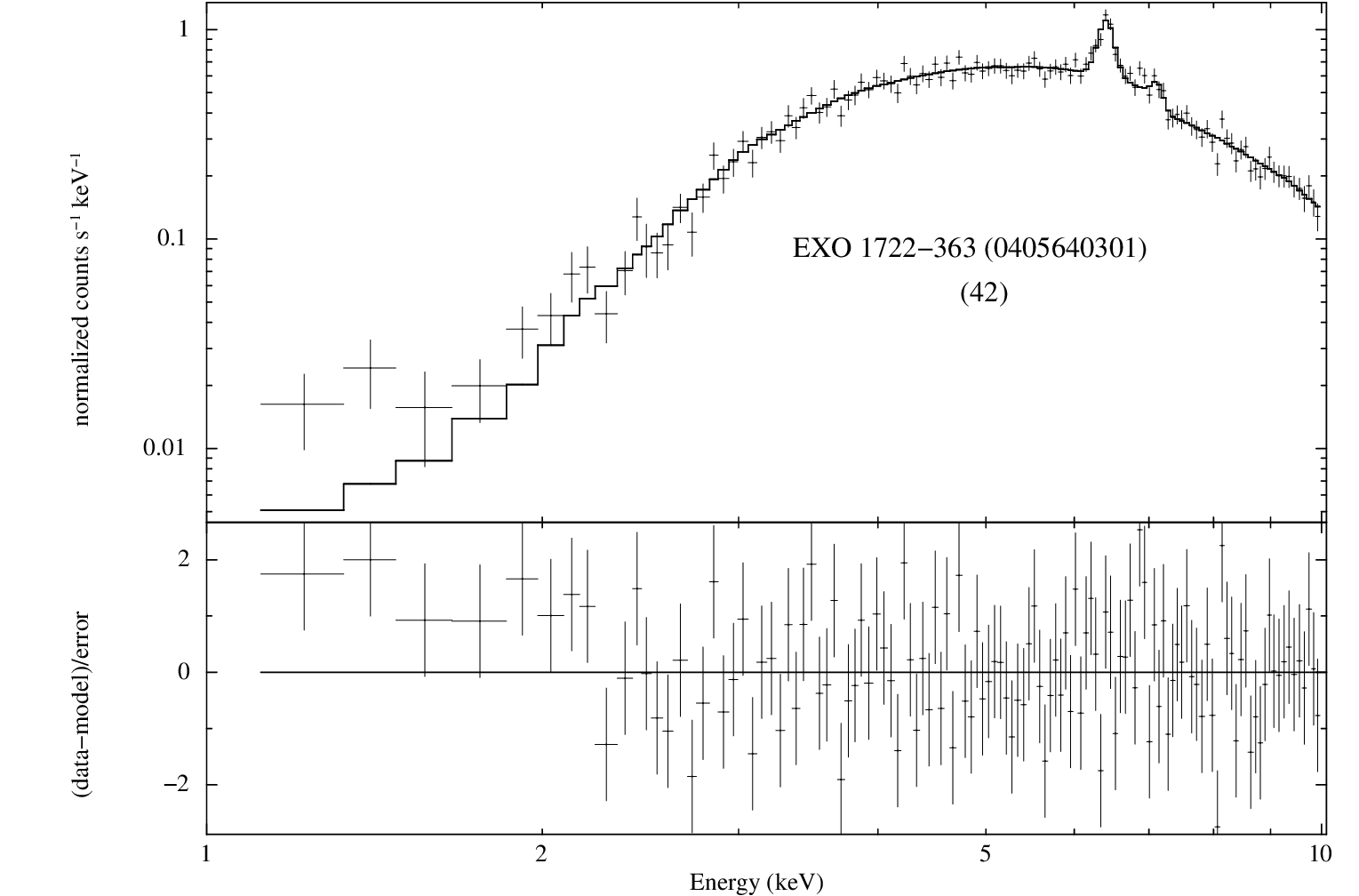}
\includegraphics[height=4.2cm,width=4.3cm,angle=-90]{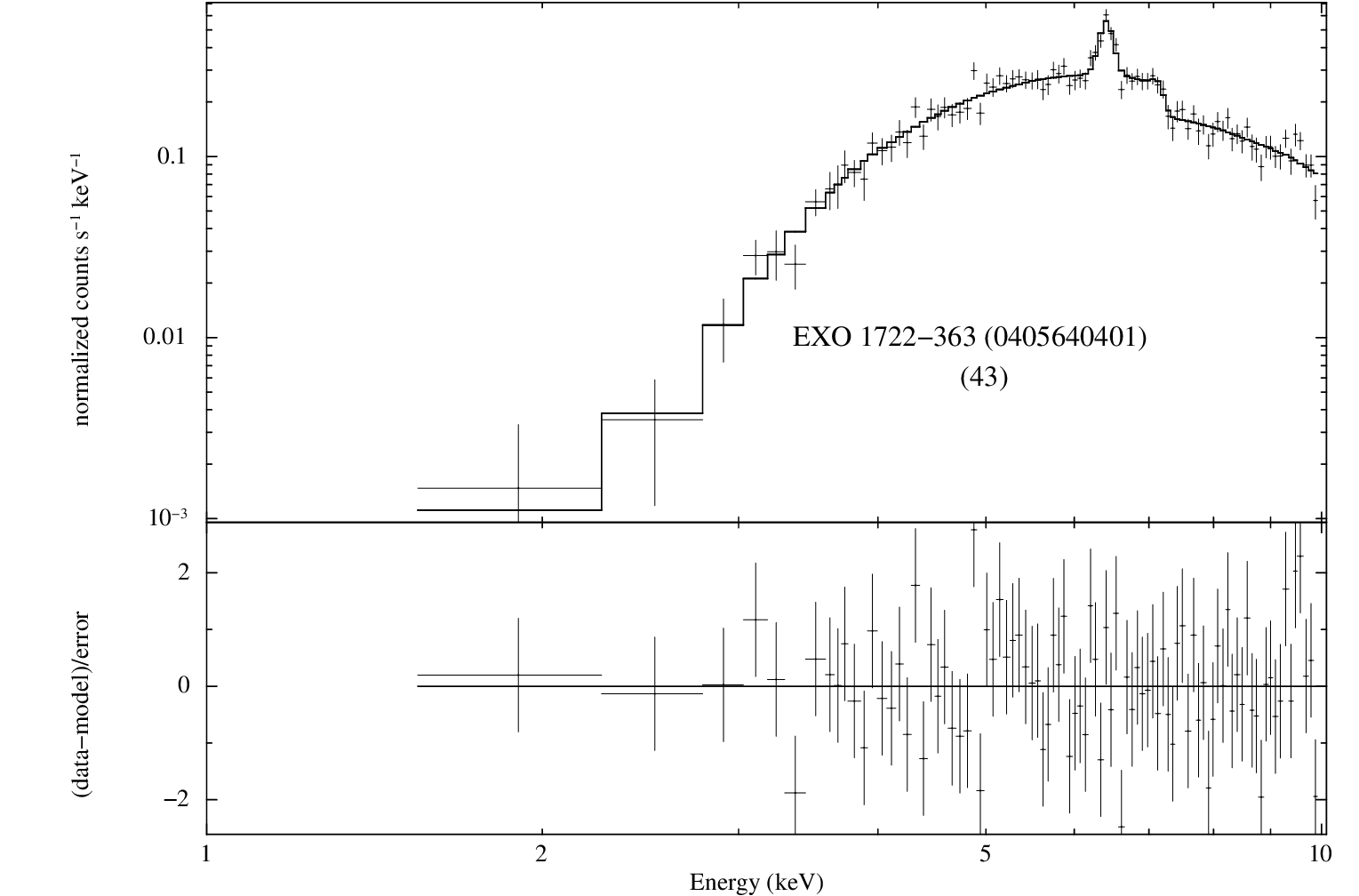}
\includegraphics[height=4.2cm,width=4.3cm,angle=-90]{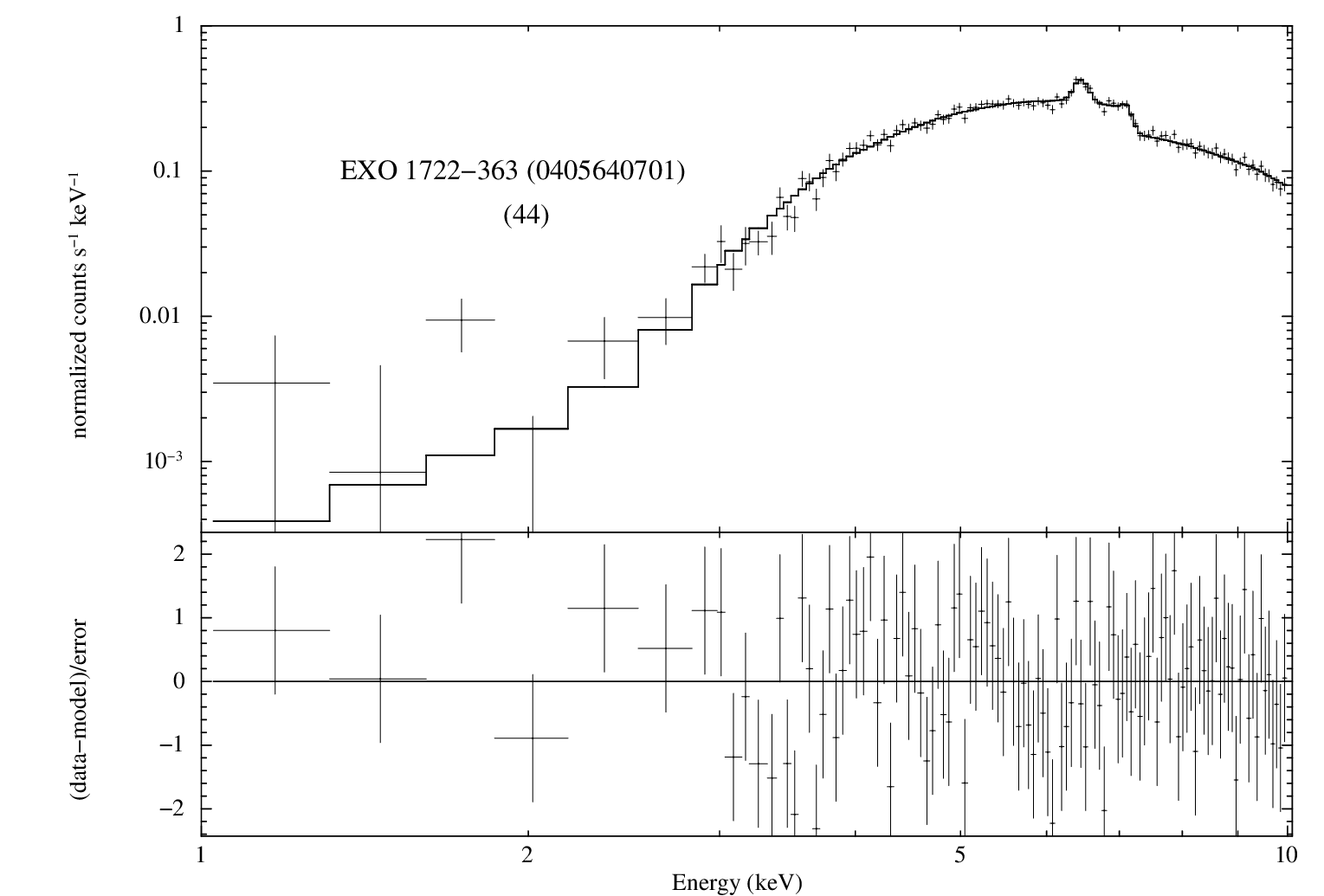}
\includegraphics[height=4.2cm,width=4.3cm,angle=-90]{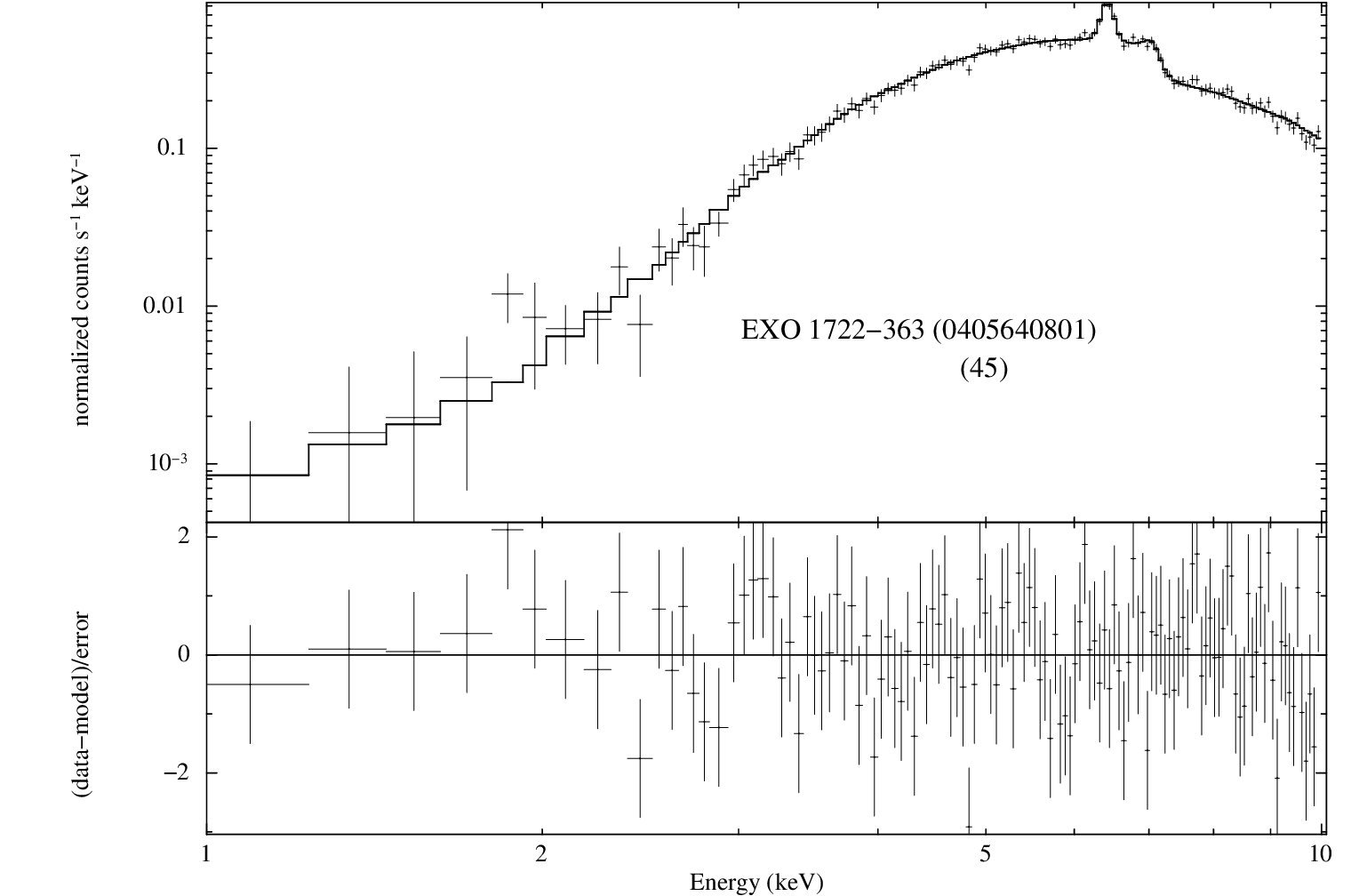}
\includegraphics[height=4.2cm,width=4.3cm,angle=-90]{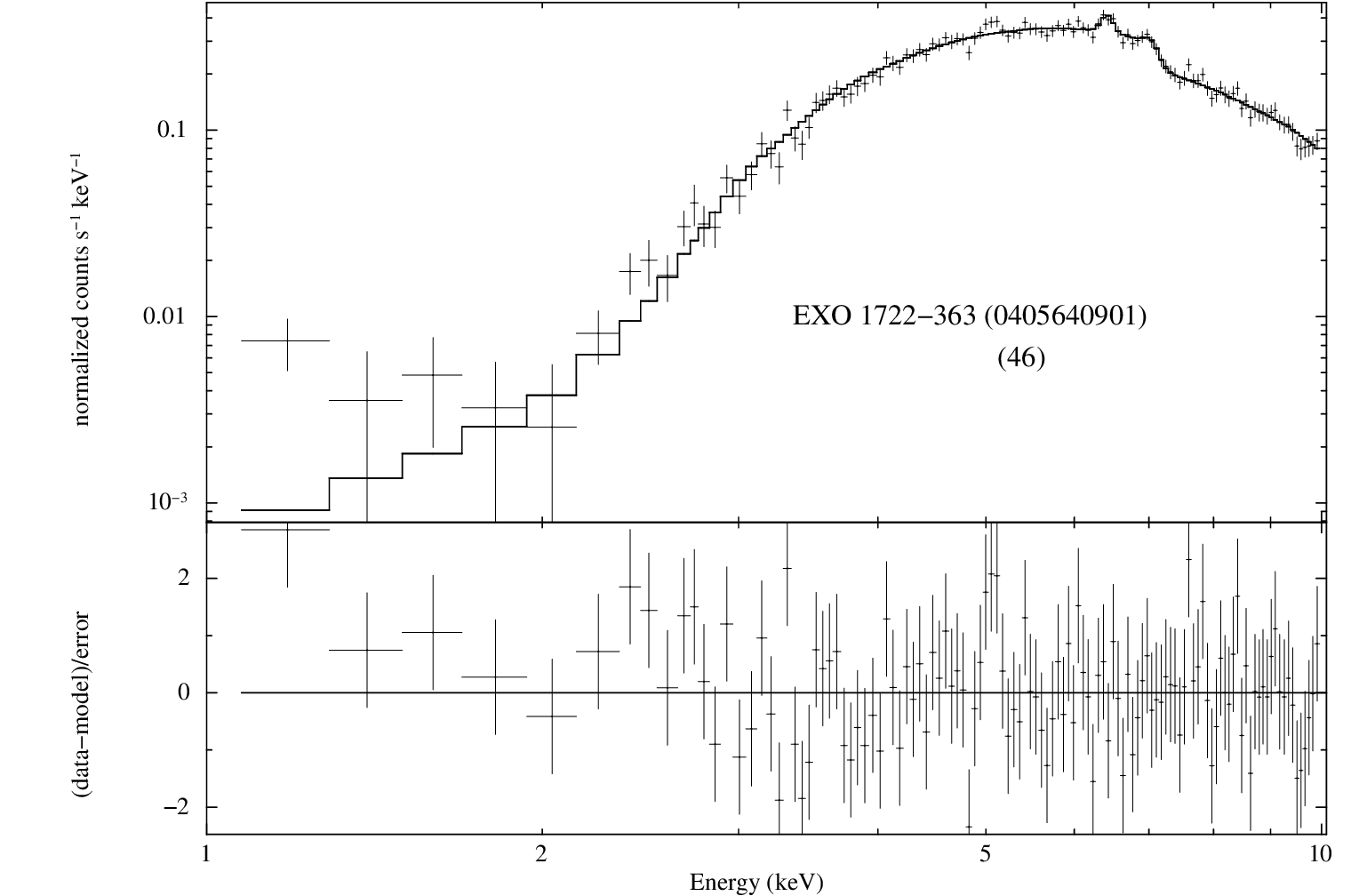}
\includegraphics[height=4.2cm,width=4.3cm,angle=-90]{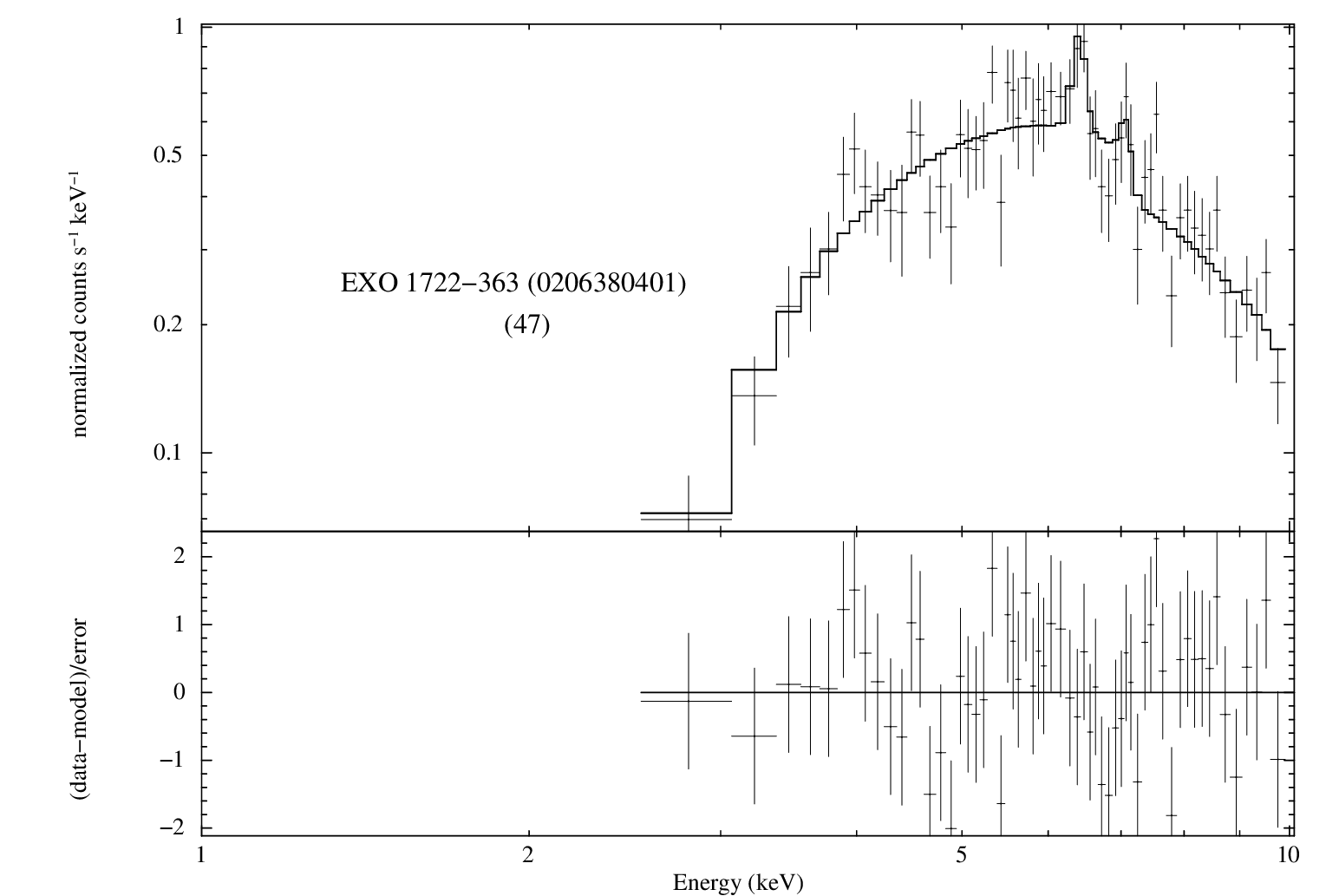}
\includegraphics[height=4.2cm,width=4.3cm,angle=-90]{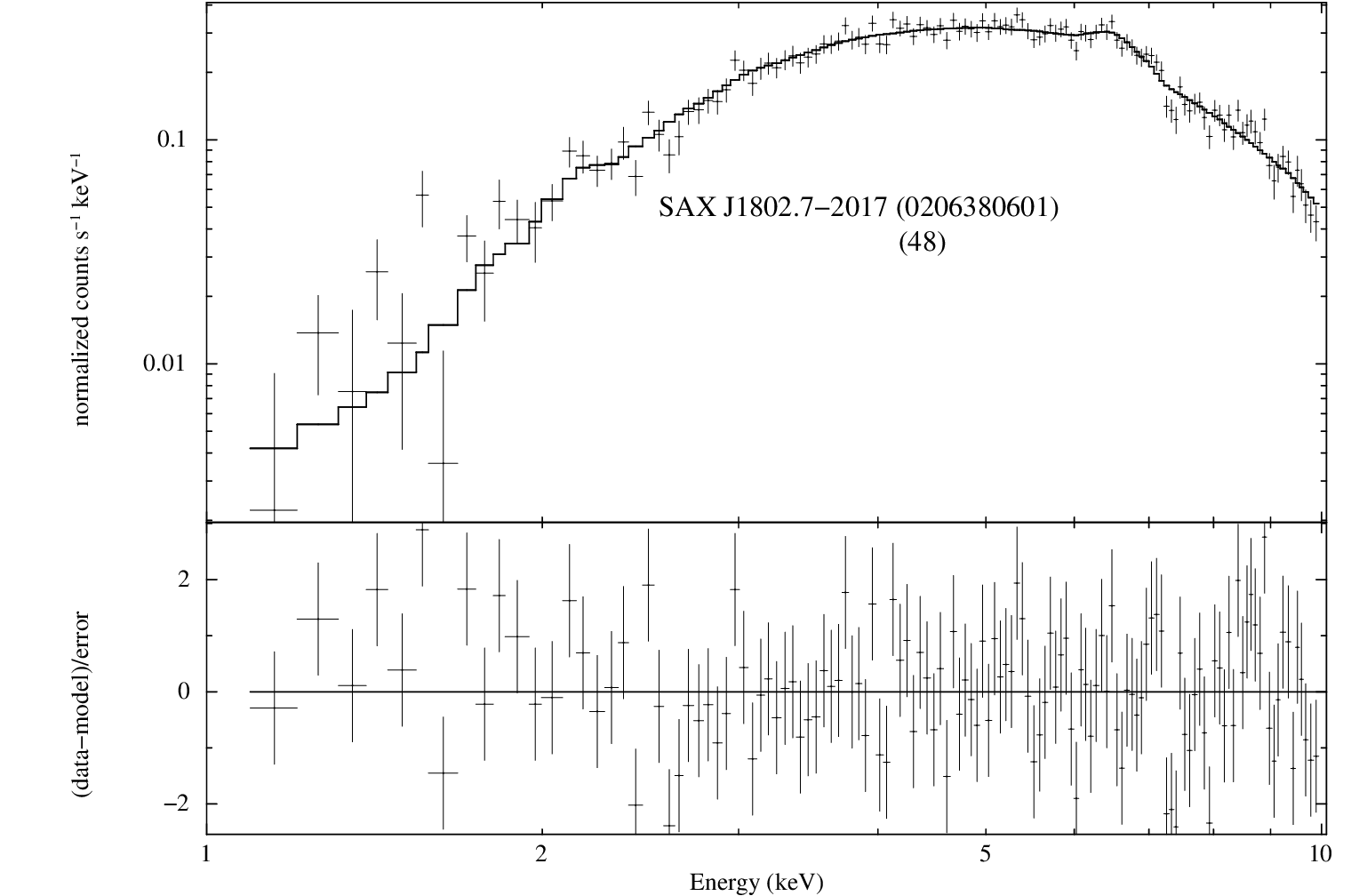}
\includegraphics[height=4.2cm,width=4.3cm,angle=-90]{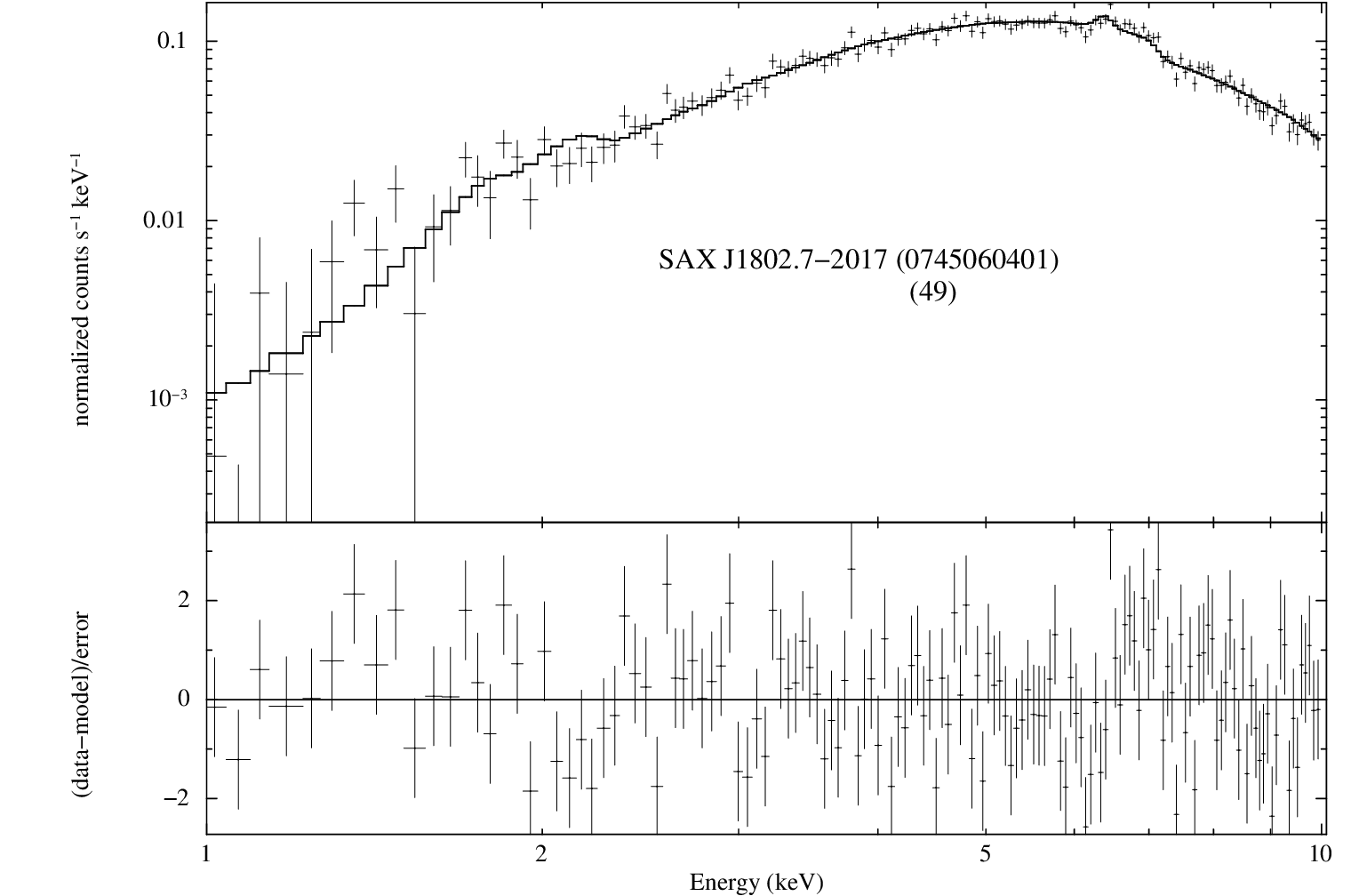}
\includegraphics[height=4.2cm,width=4.3cm,angle=-90]{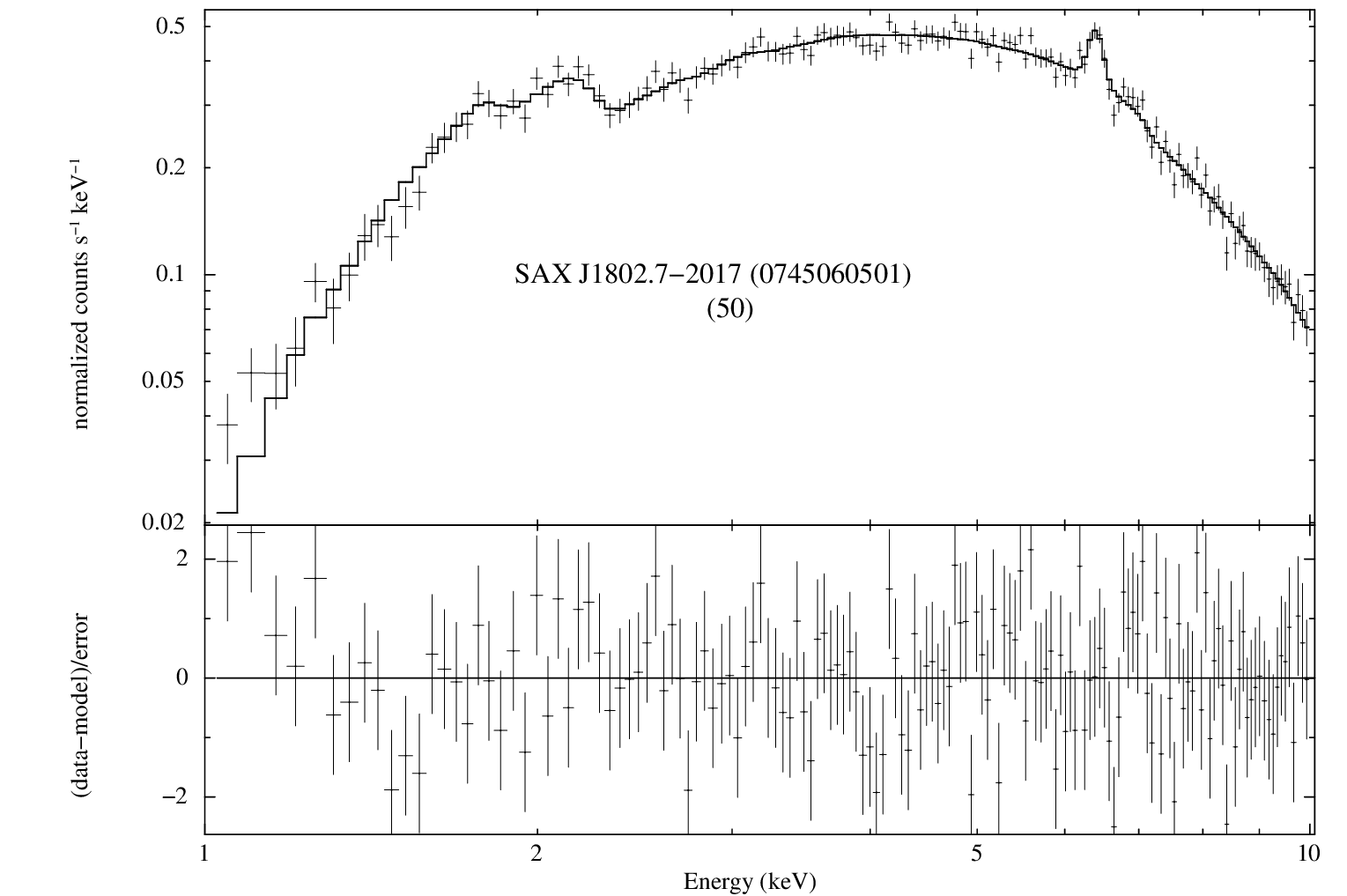}
\includegraphics[height=4.2cm,width=4.3cm,angle=-90]{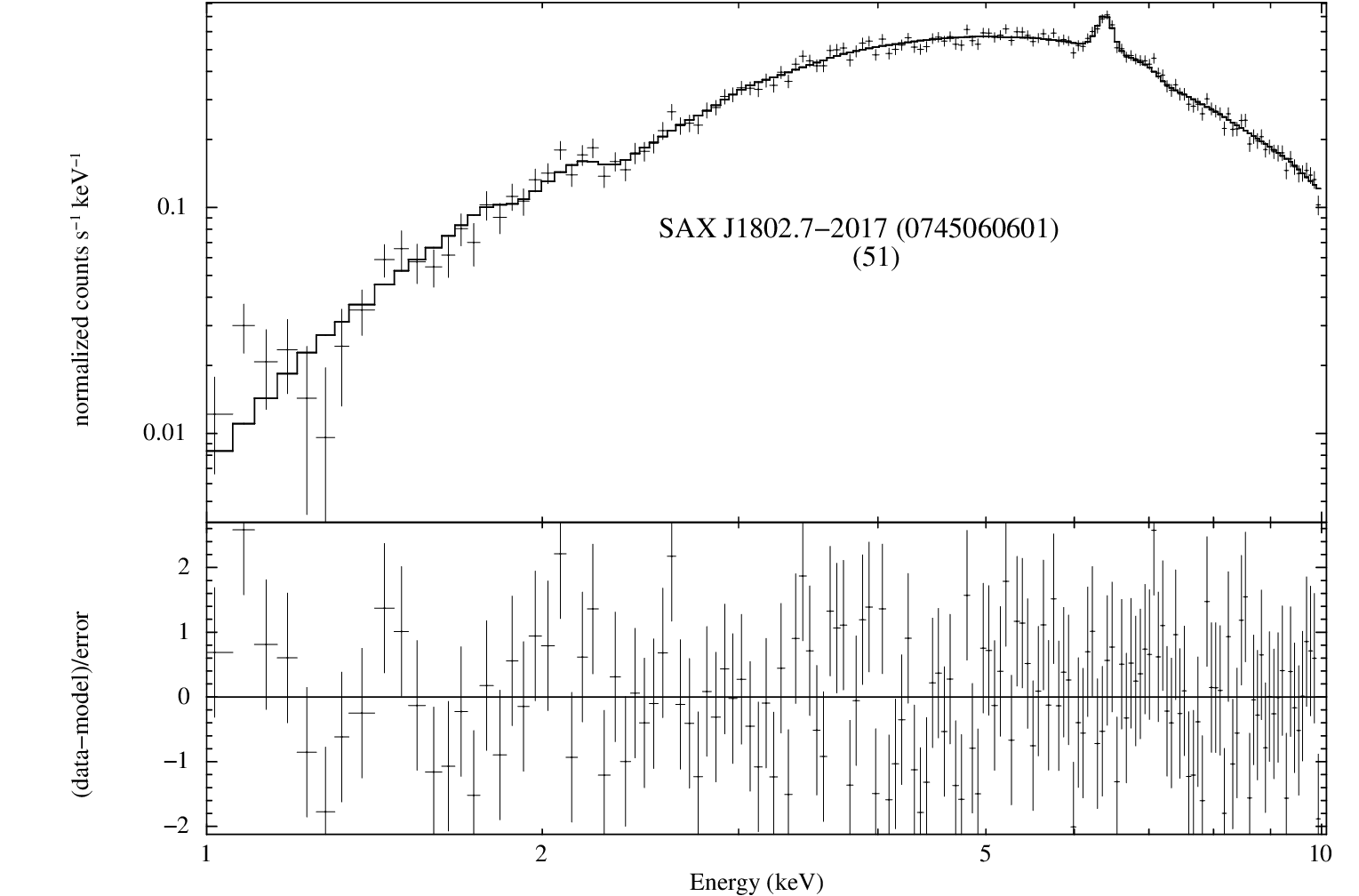}
\includegraphics[height=4.2cm,width=4.3cm,angle=-90]{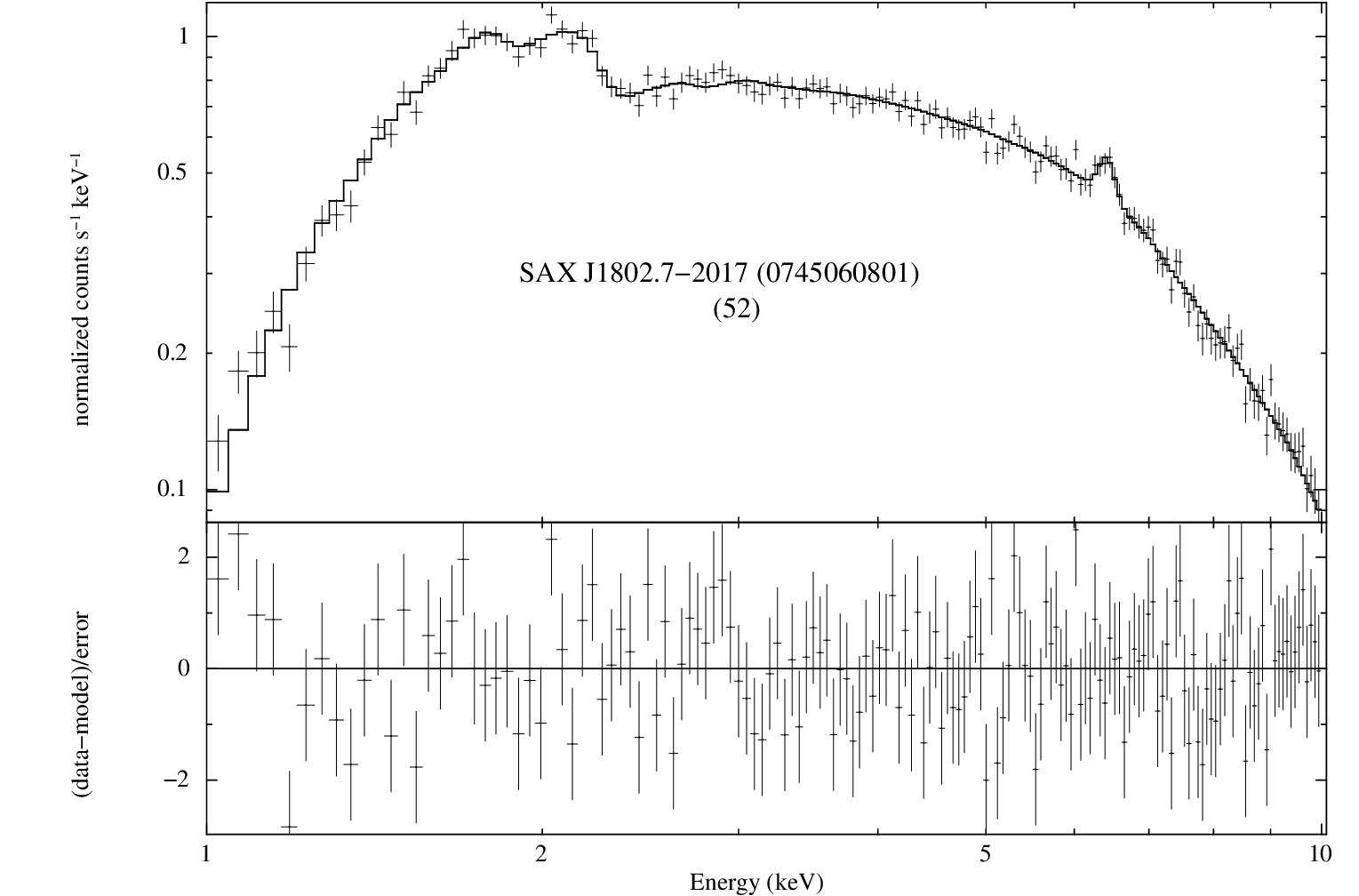}
\includegraphics[height=4.2cm,width=4.3cm,angle=-90]{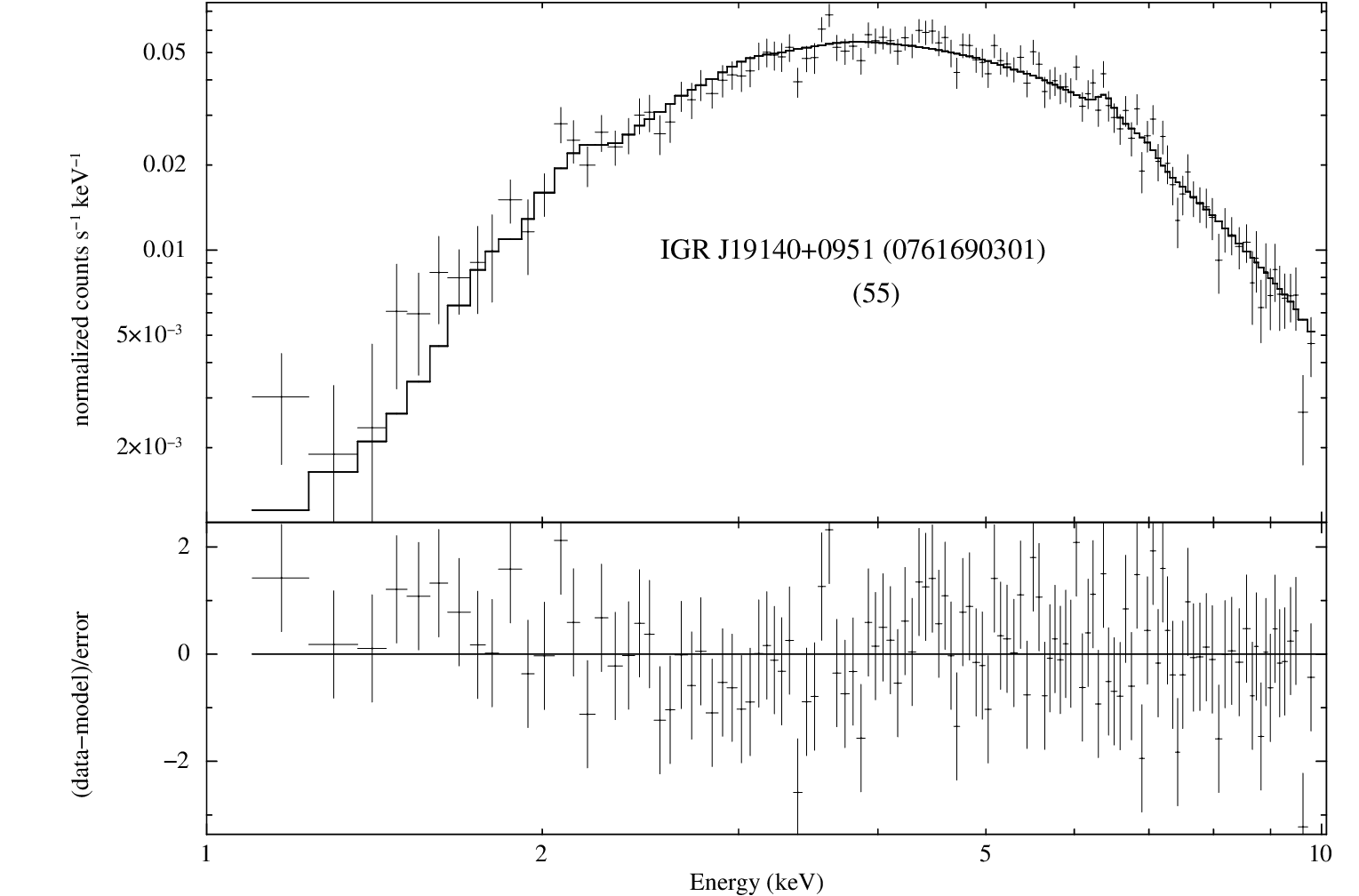}
\label{spectra4}
\end{figure*}

\begin{figure*}
\caption{\emph{XMM-Newton} spectra of all considered SFXTs. The source name and the observation ID is indicated in each figure.}
\includegraphics[height=4.2cm,width=4.3cm,angle=-90]{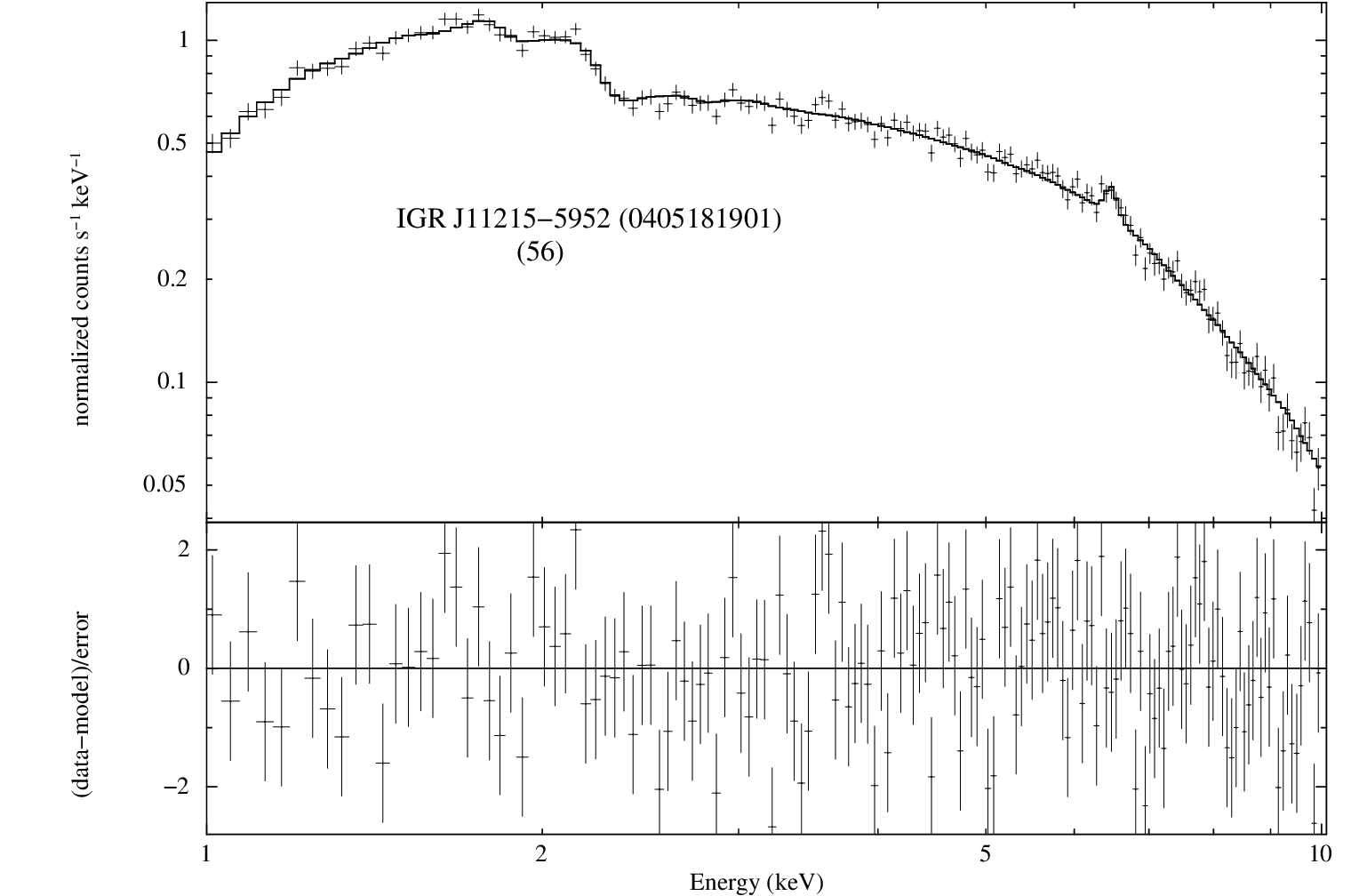}
\includegraphics[height=4.2cm,width=4.3cm,angle=-90]{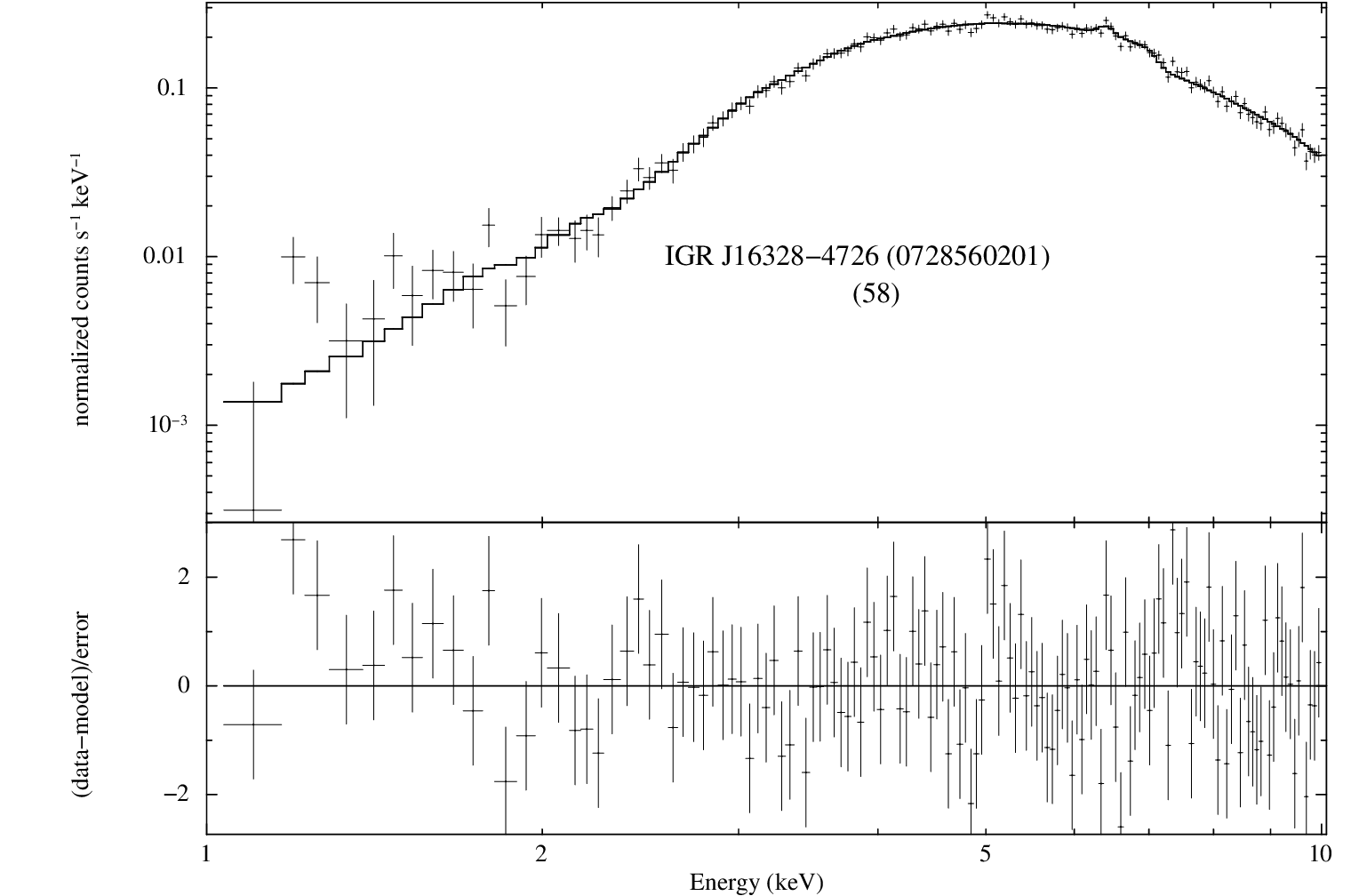}
\includegraphics[height=4.2cm,width=4.3cm,angle=-90]{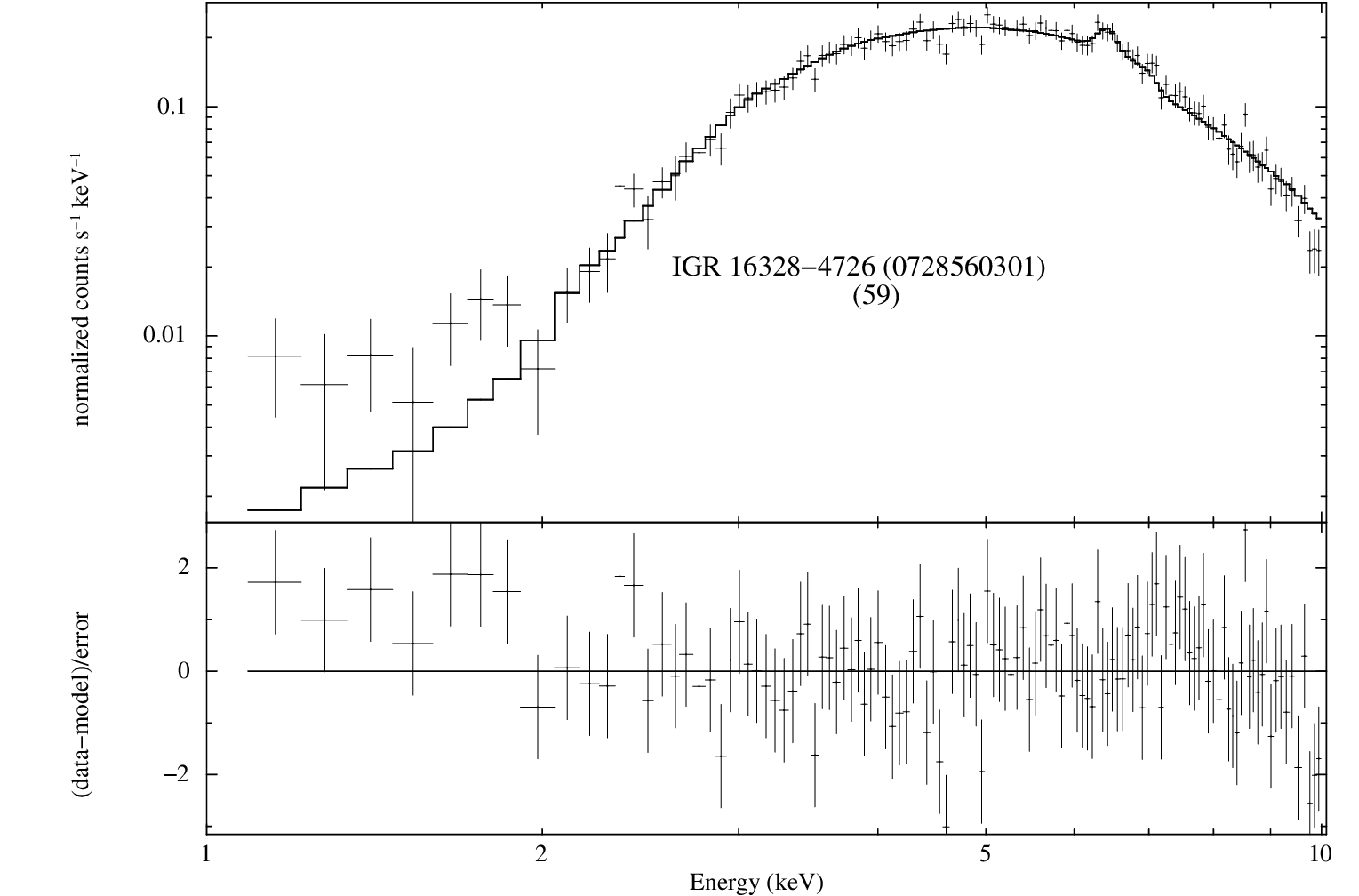}
\includegraphics[height=4.2cm,width=4.3cm,angle=-90]{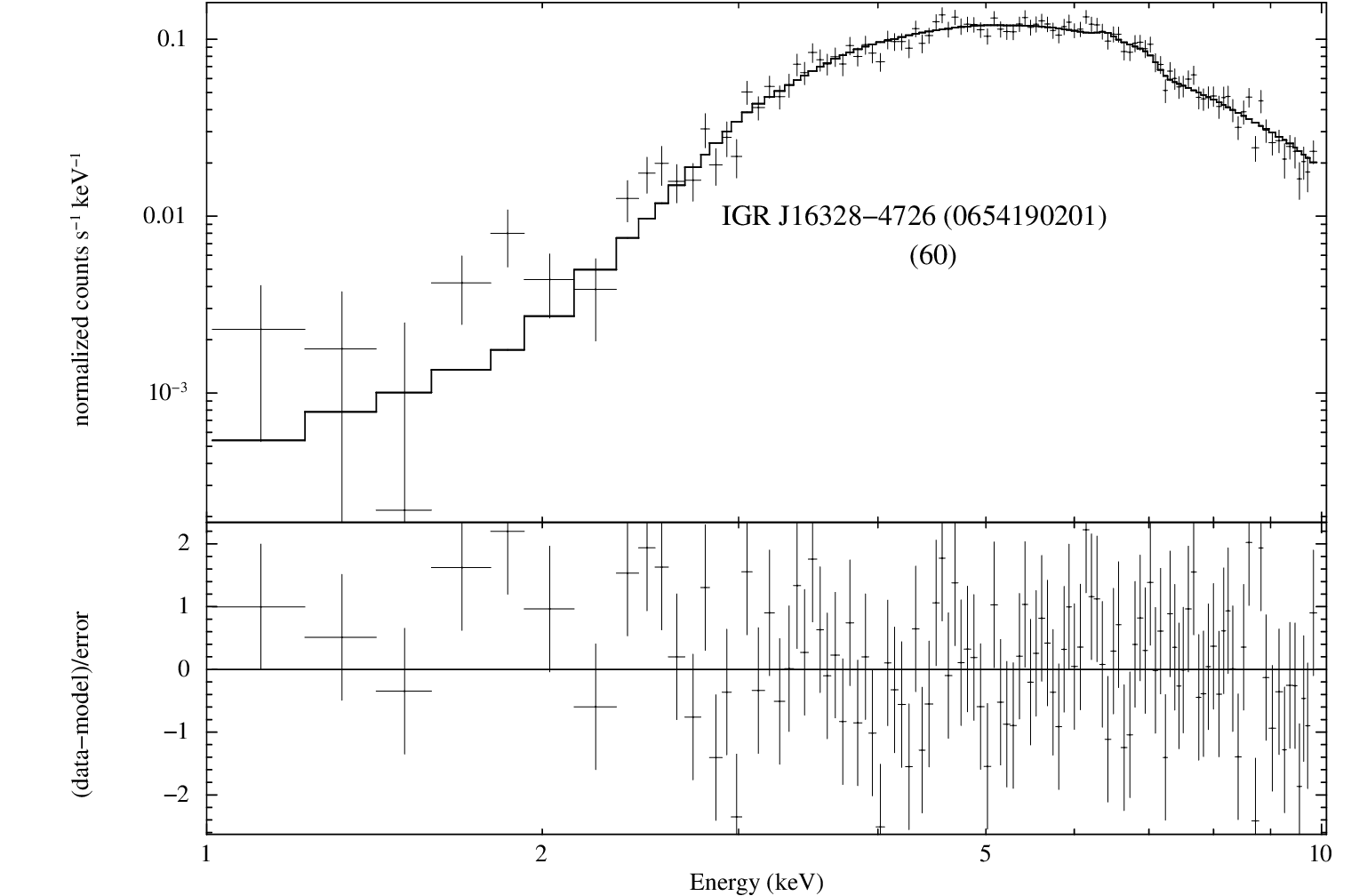}
\includegraphics[height=4.2cm,width=4.3cm,angle=-90]{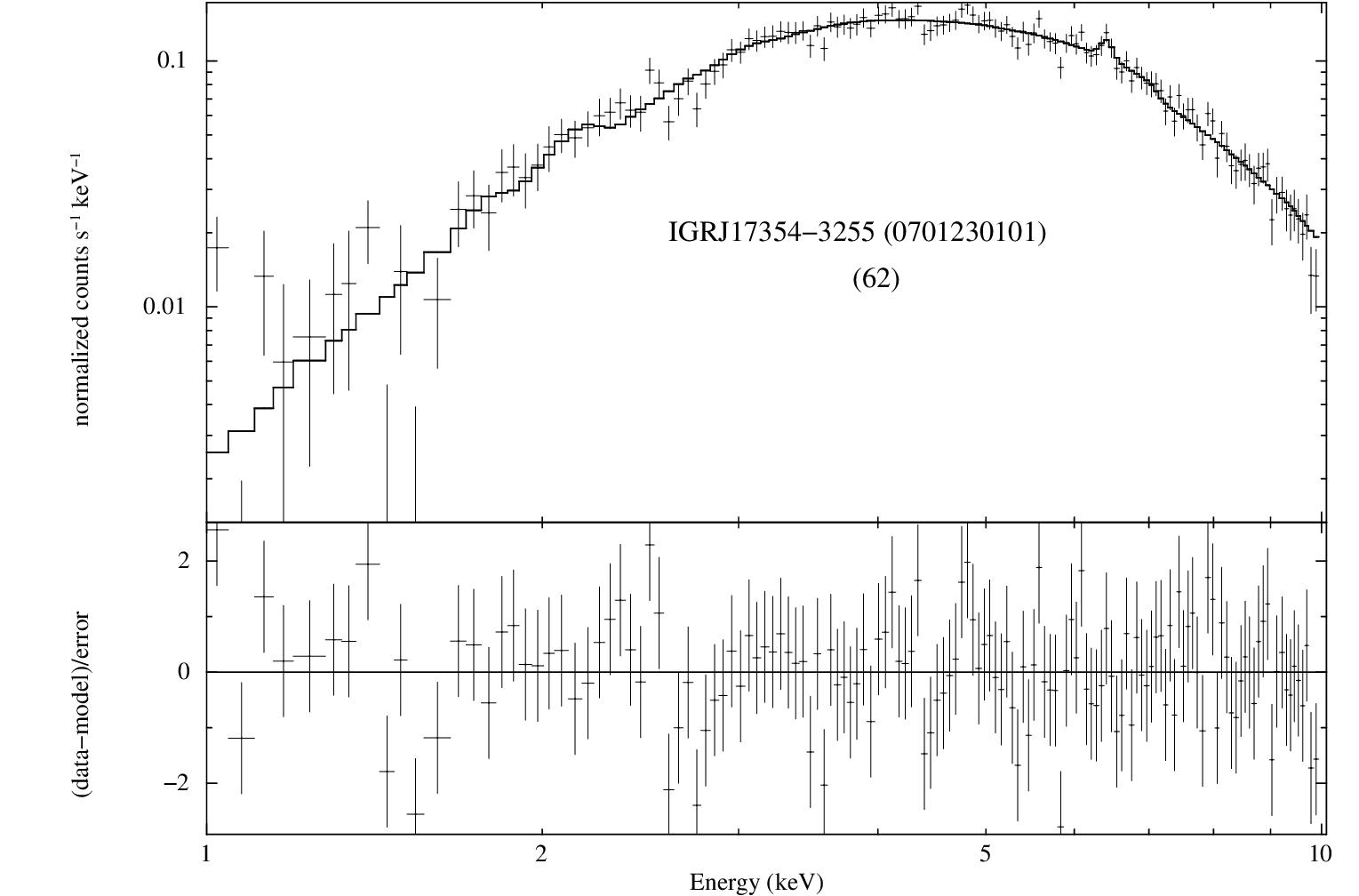}
\includegraphics[height=4.2cm,width=4.3cm,angle=-90]{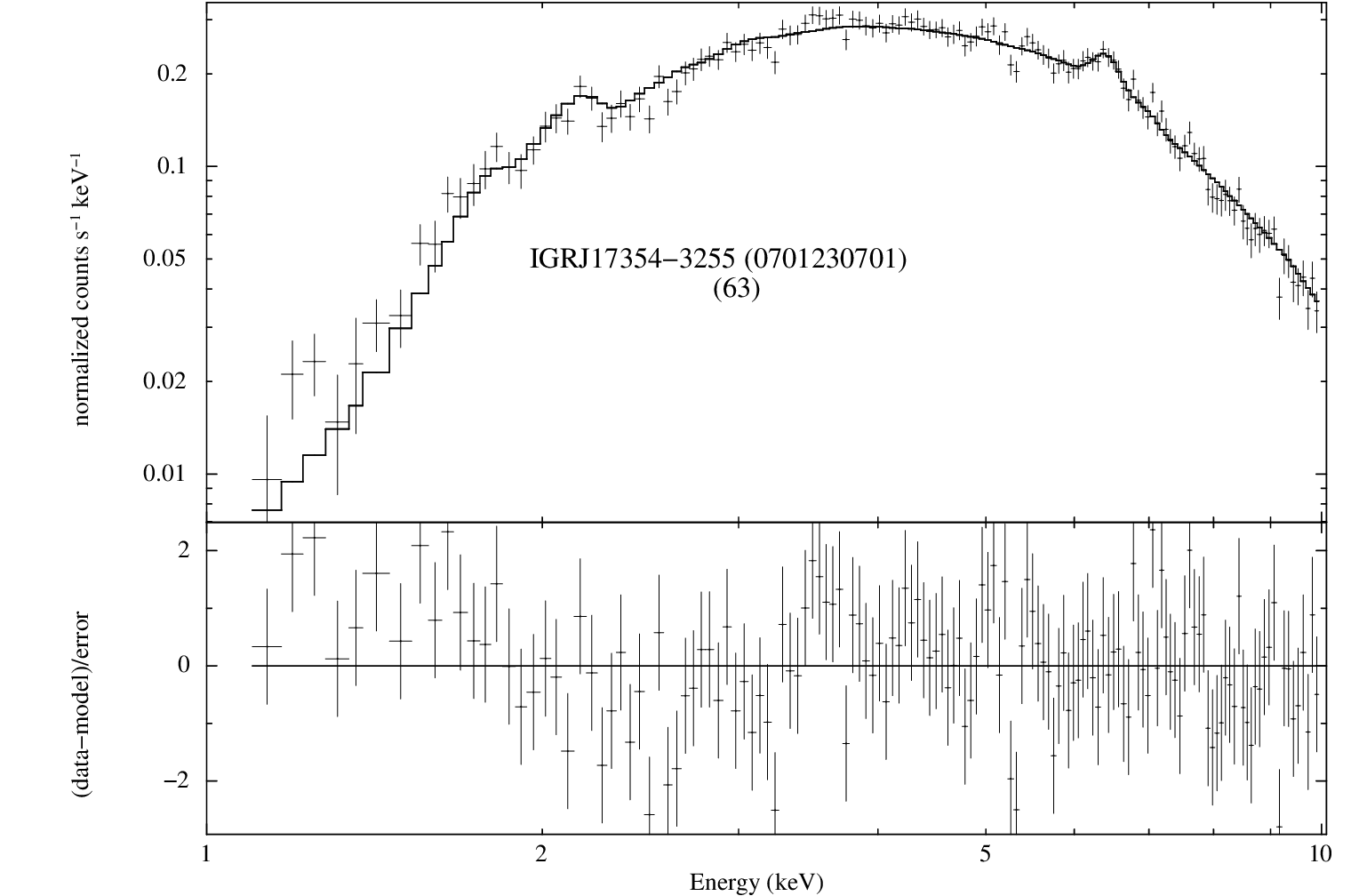}
\includegraphics[height=4.2cm,width=4.3cm,angle=-90]{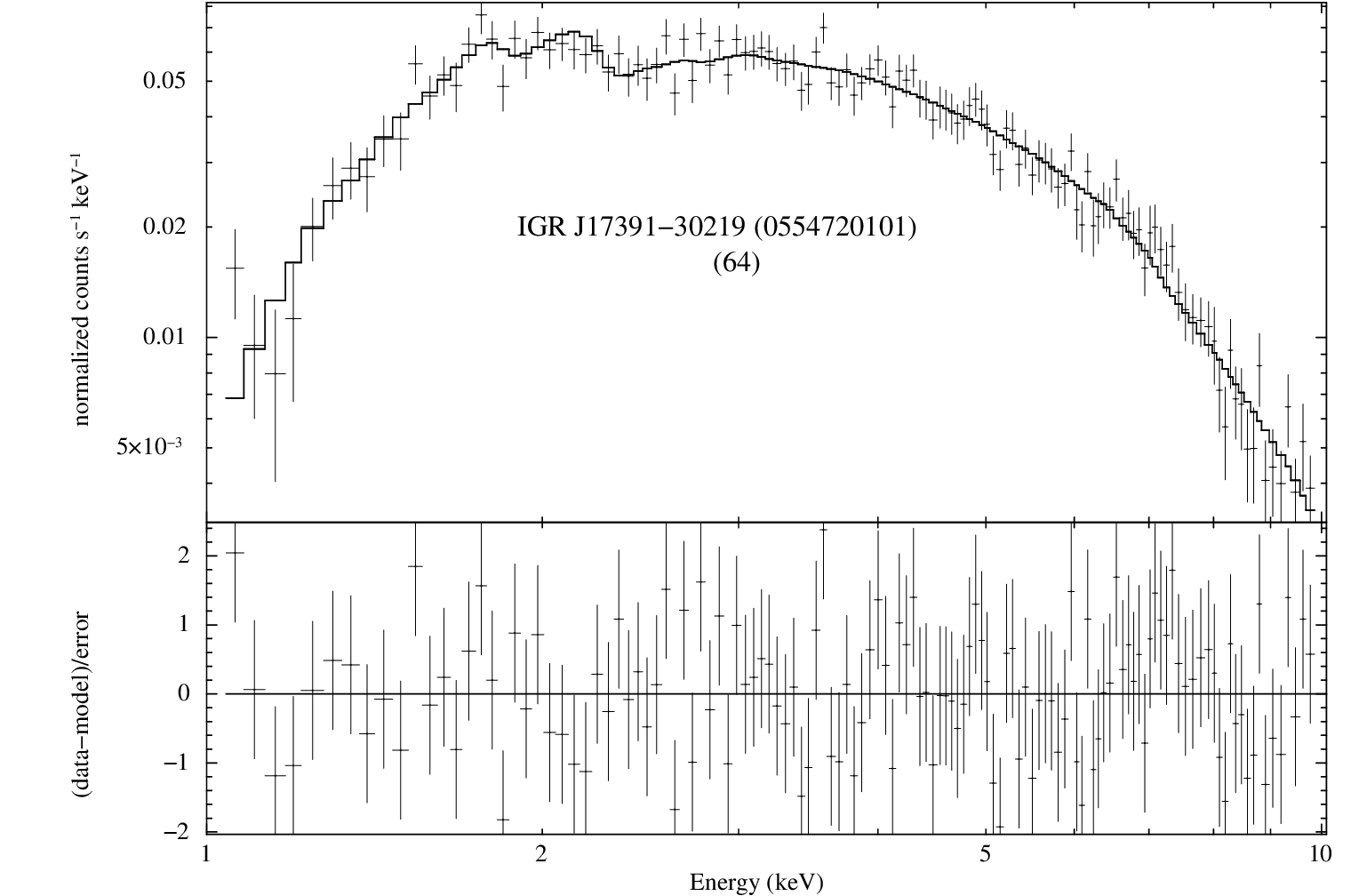}
\includegraphics[height=4.2cm,width=4.3cm,angle=-90]{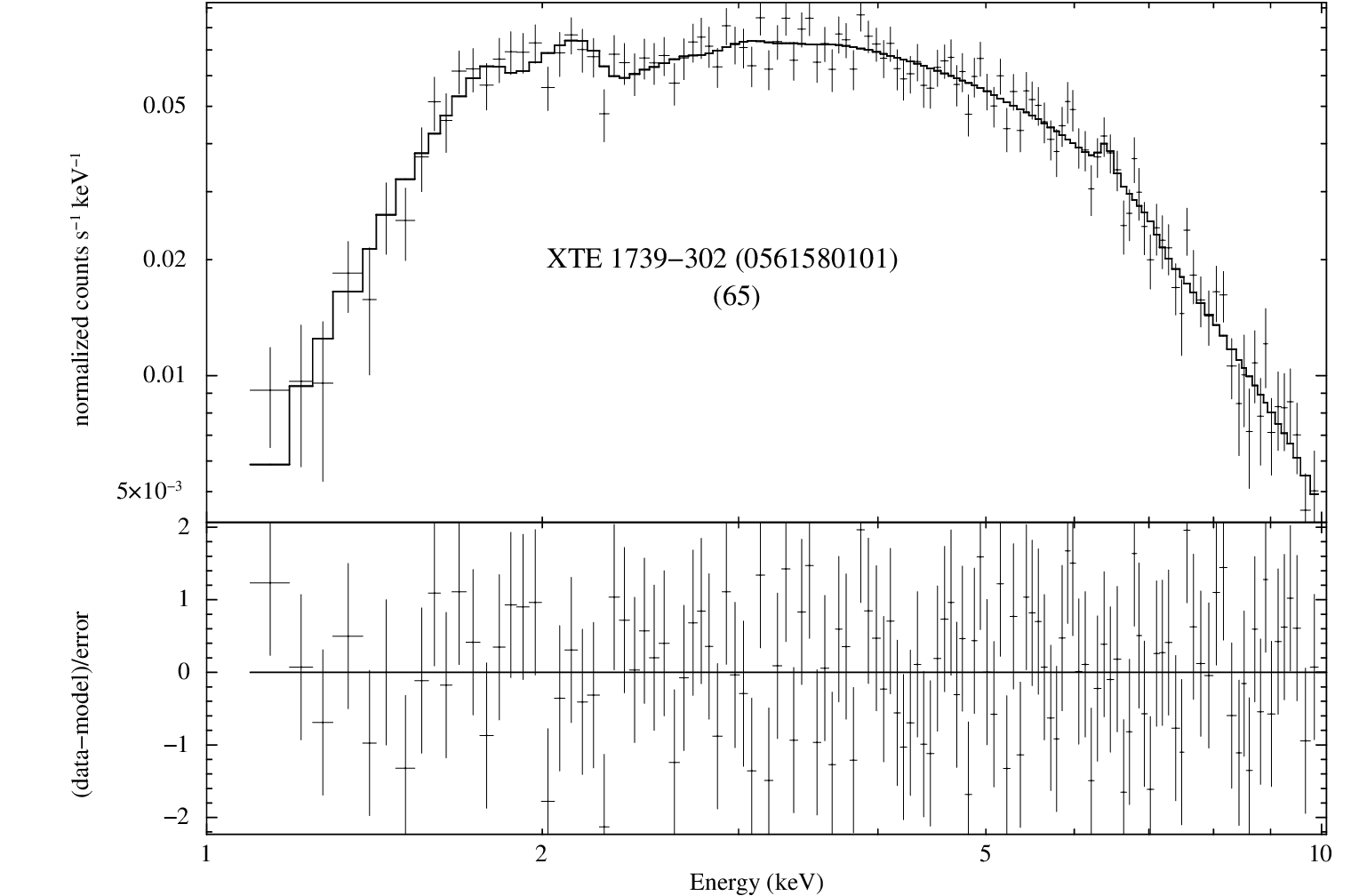}
\includegraphics[height=4.2cm,width=4.3cm,angle=-90]{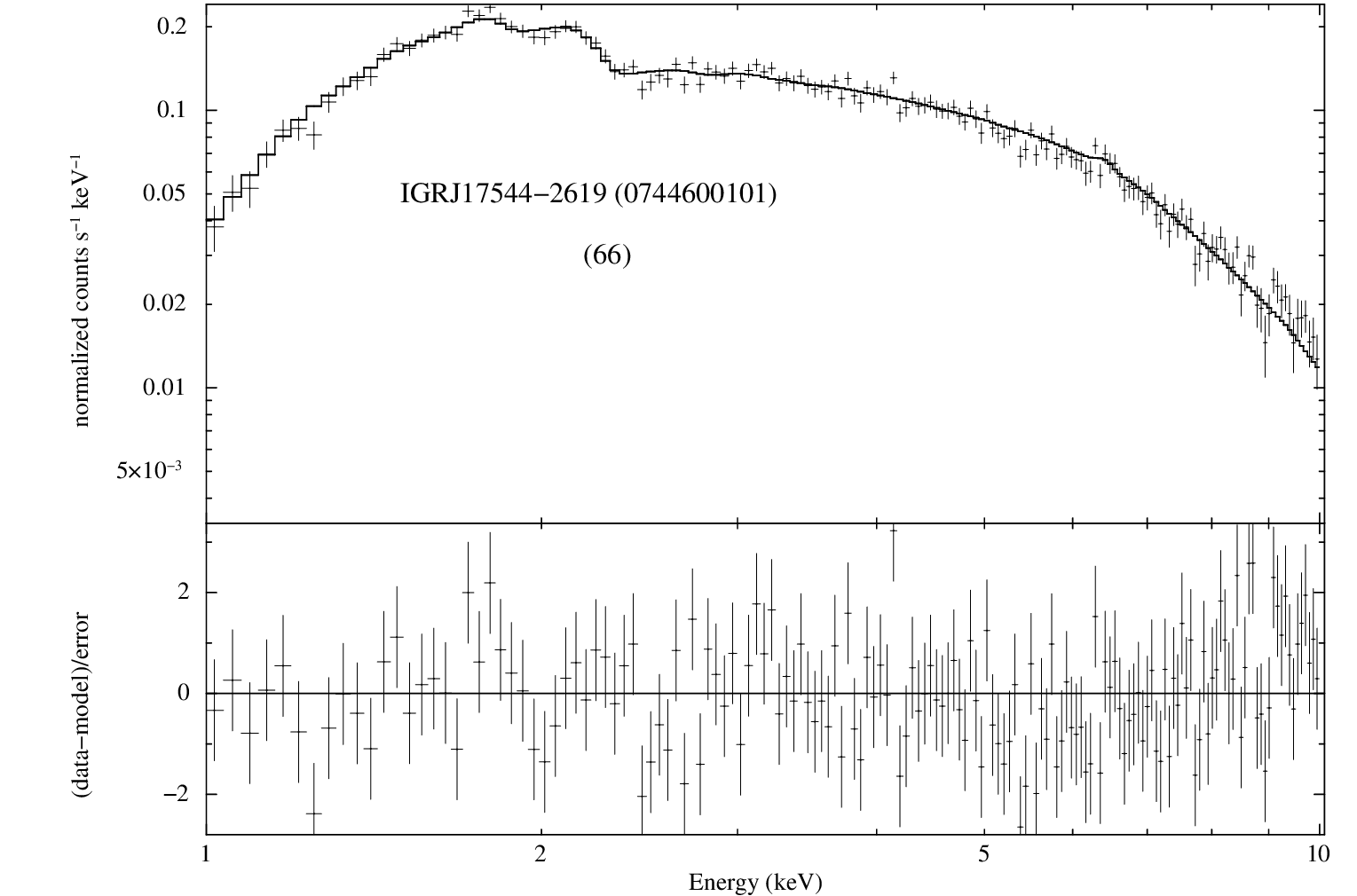}
\includegraphics[height=4.2cm,width=4.3cm,angle=-90]{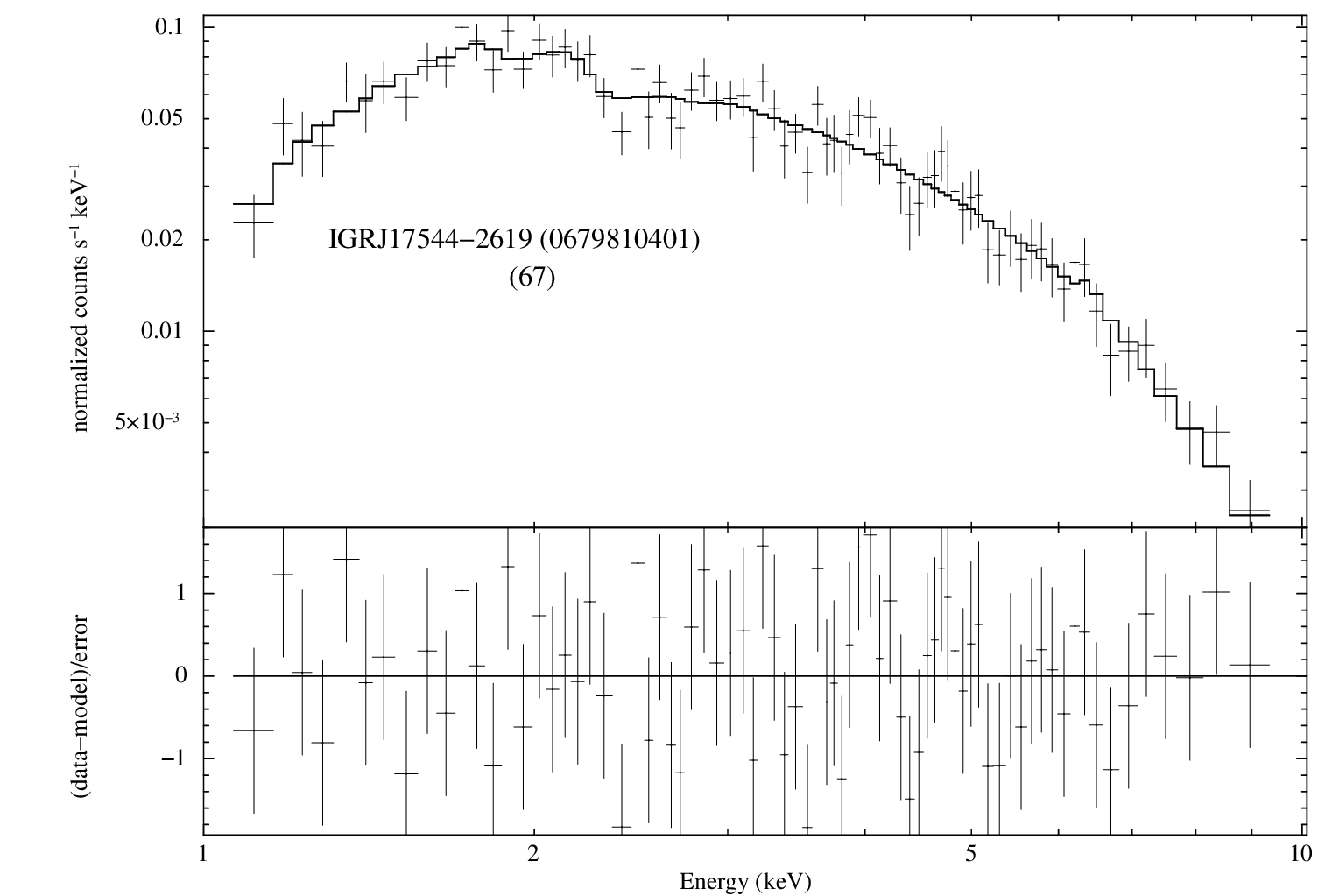}
\includegraphics[height=4.2cm,width=4.3cm,angle=-90]{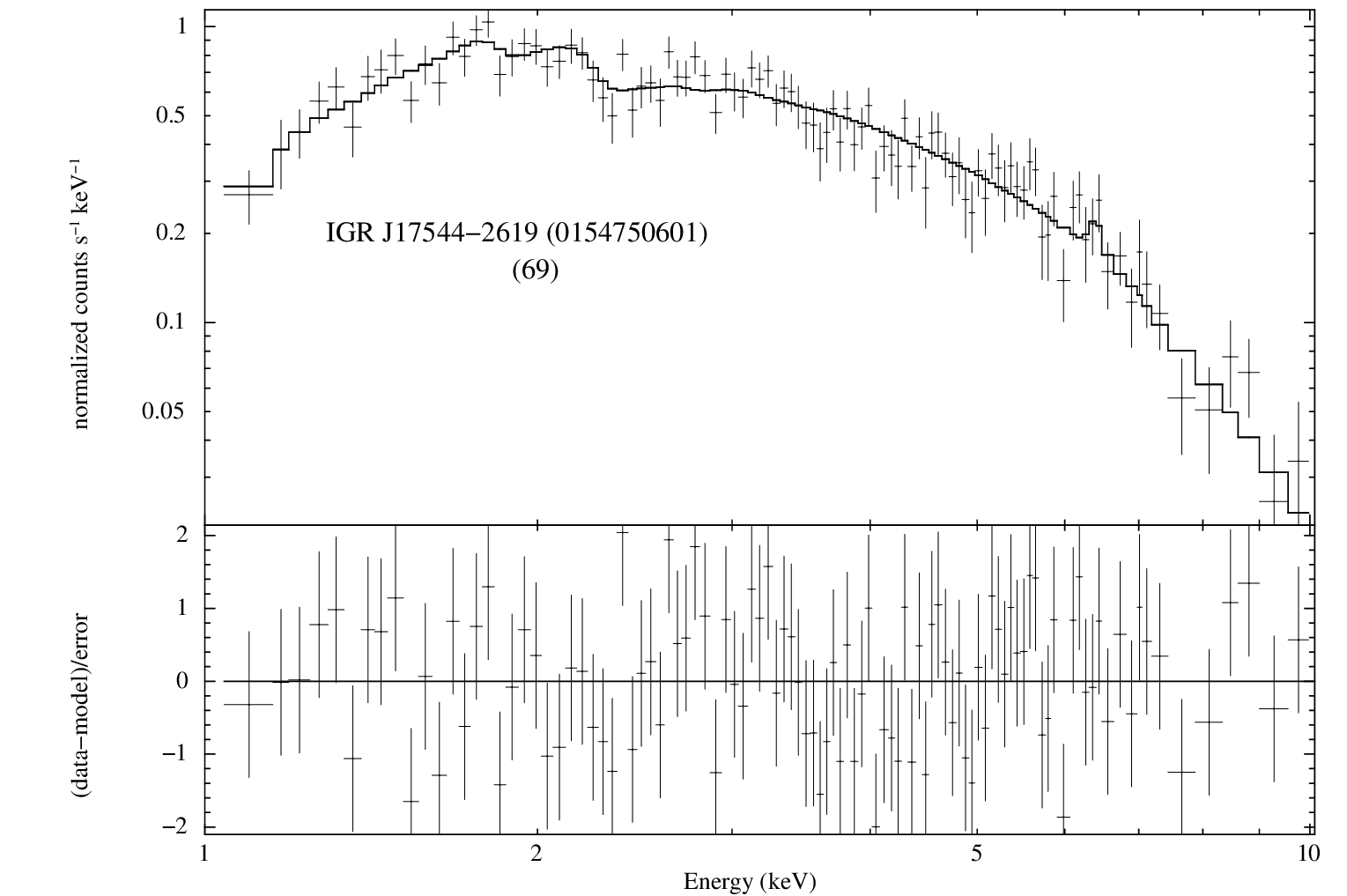}
\includegraphics[height=4.2cm,width=4.3cm,angle=-90]{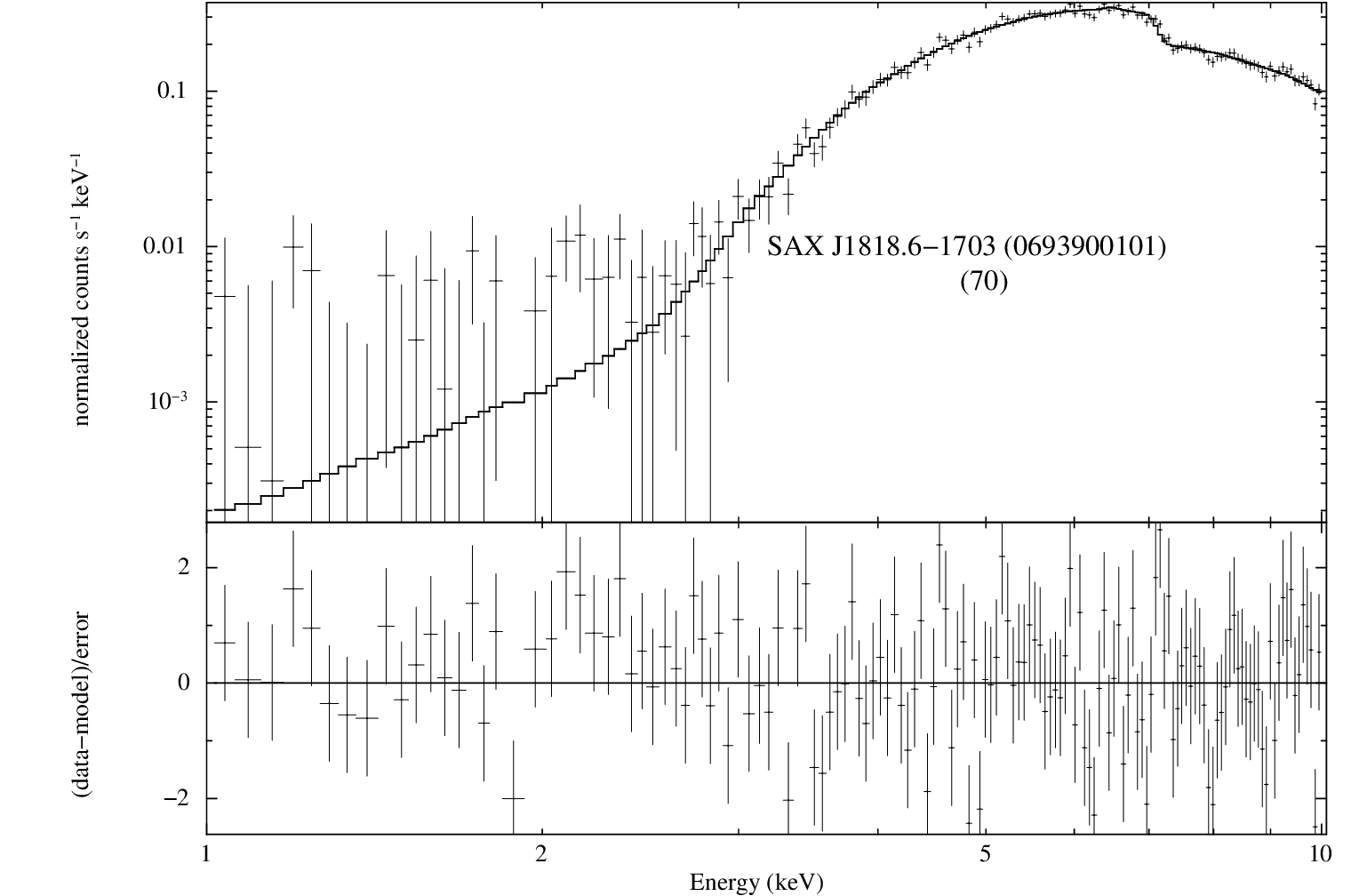}
\includegraphics[height=4.2cm,width=4.3cm,angle=-90]{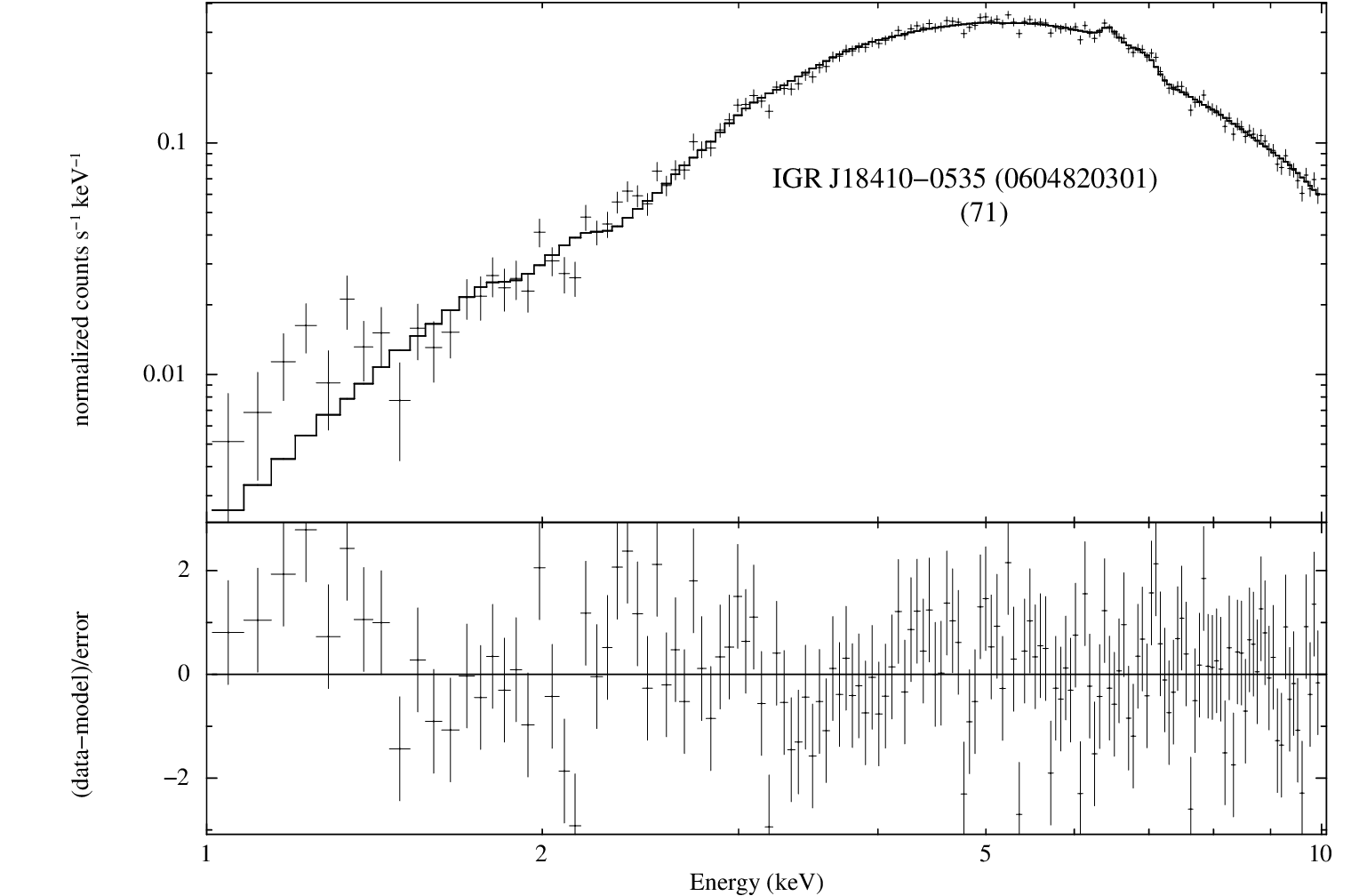}
\includegraphics[height=4.2cm,width=4.3cm,angle=-90]{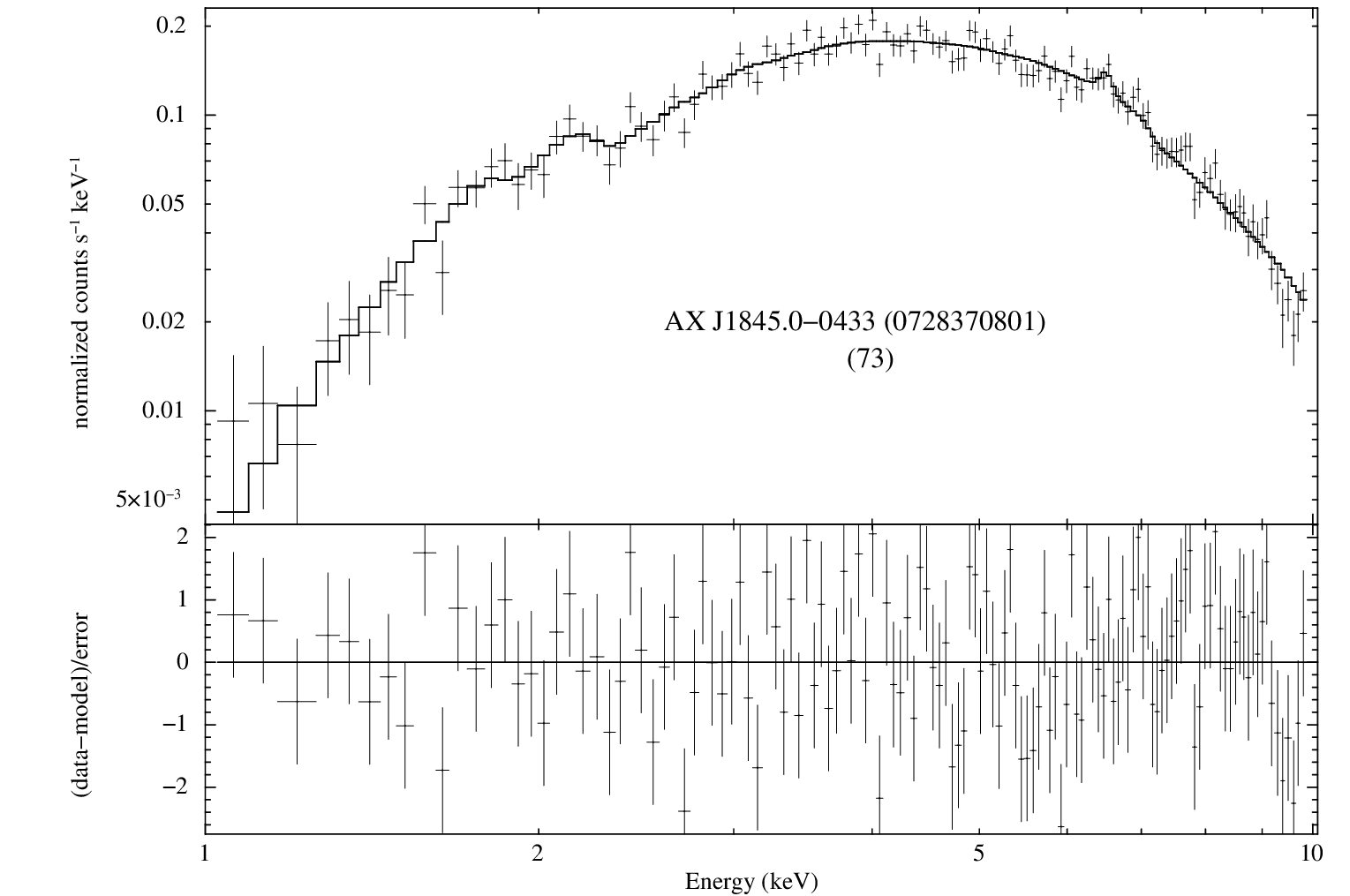}
\includegraphics[height=4.2cm,width=4.3cm,angle=-90]{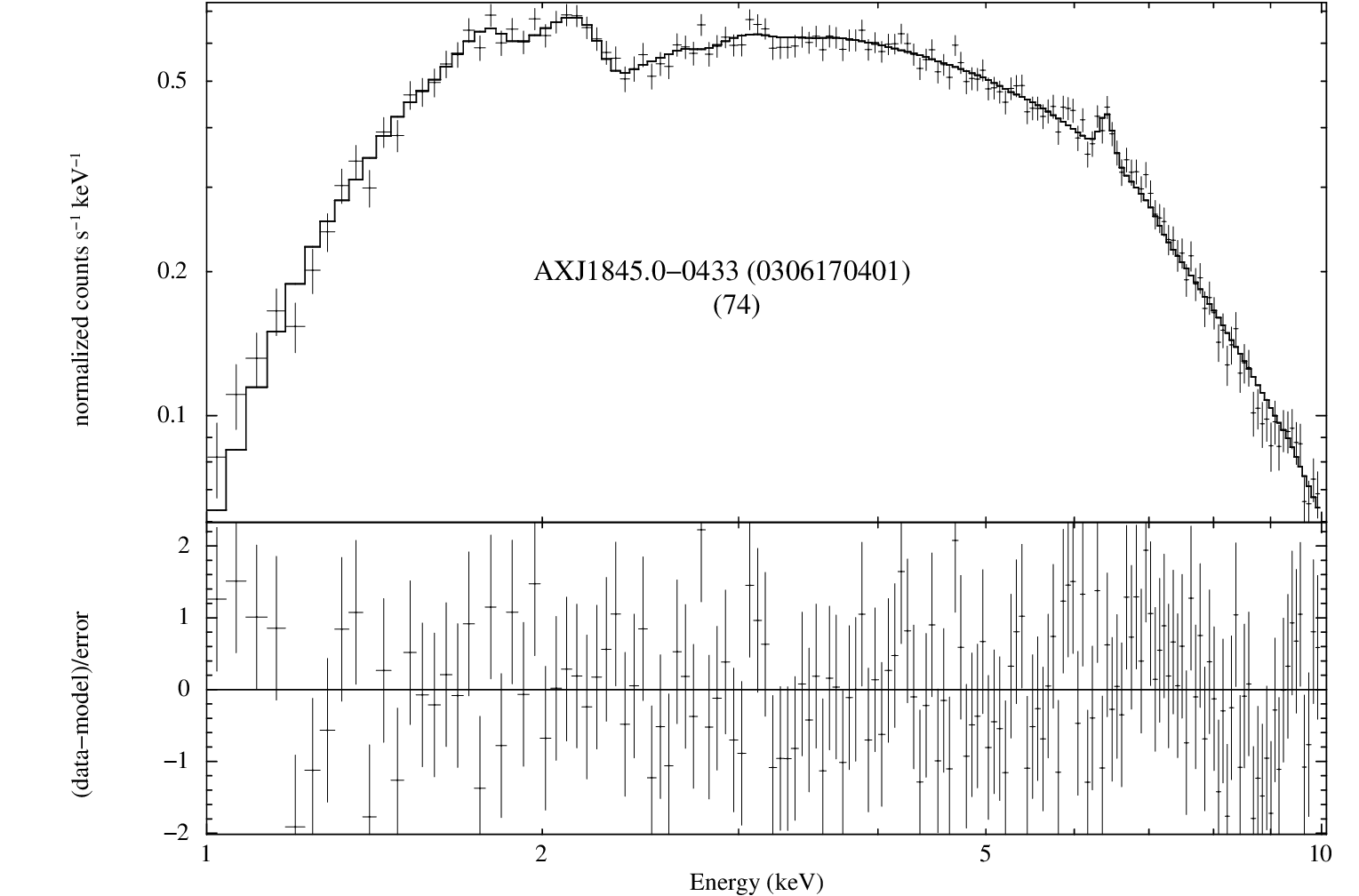}
\includegraphics[height=4.2cm,width=4.3cm,angle=-90]{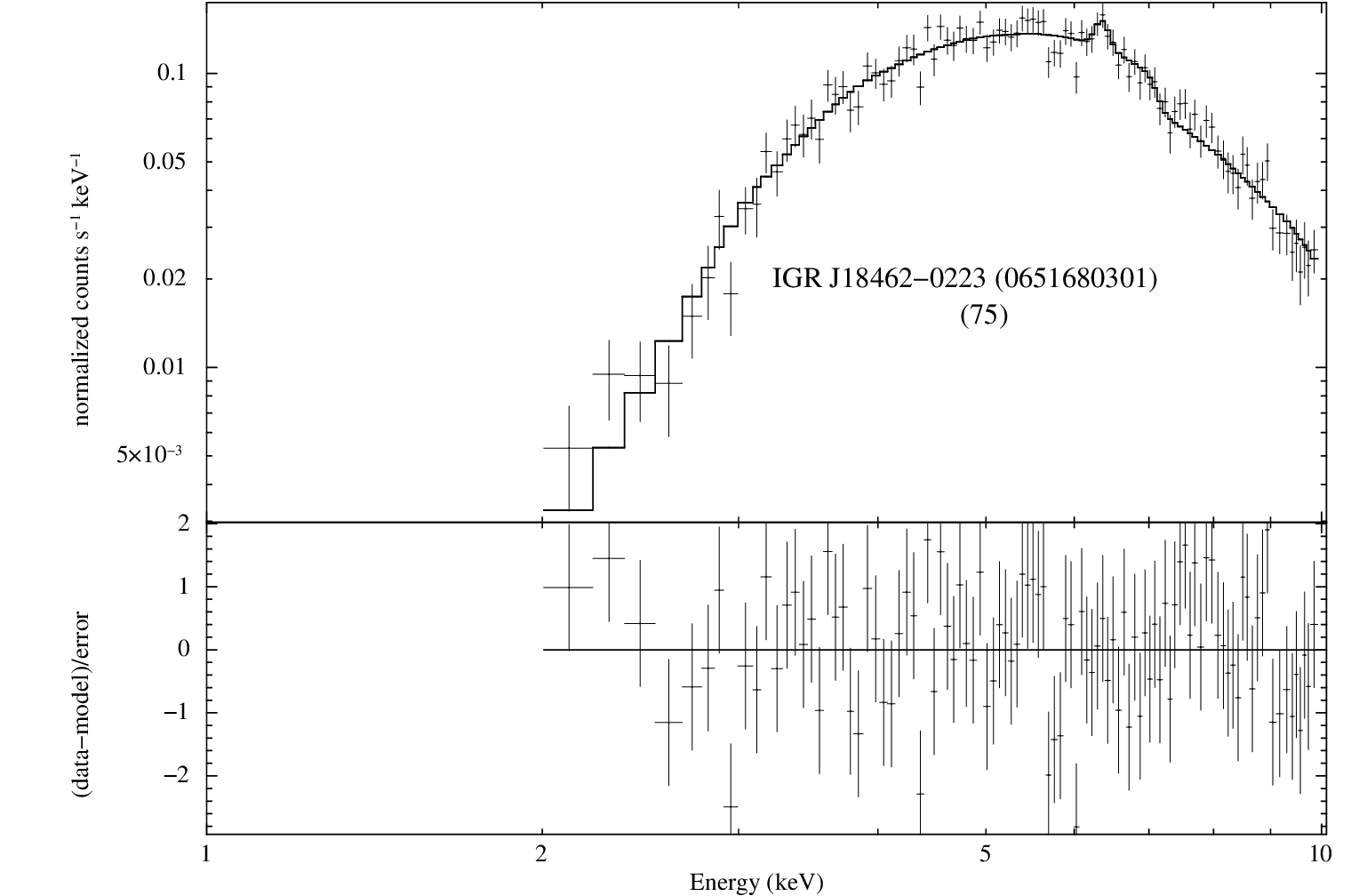}
\includegraphics[height=4.2cm,width=4.3cm,angle=-90]{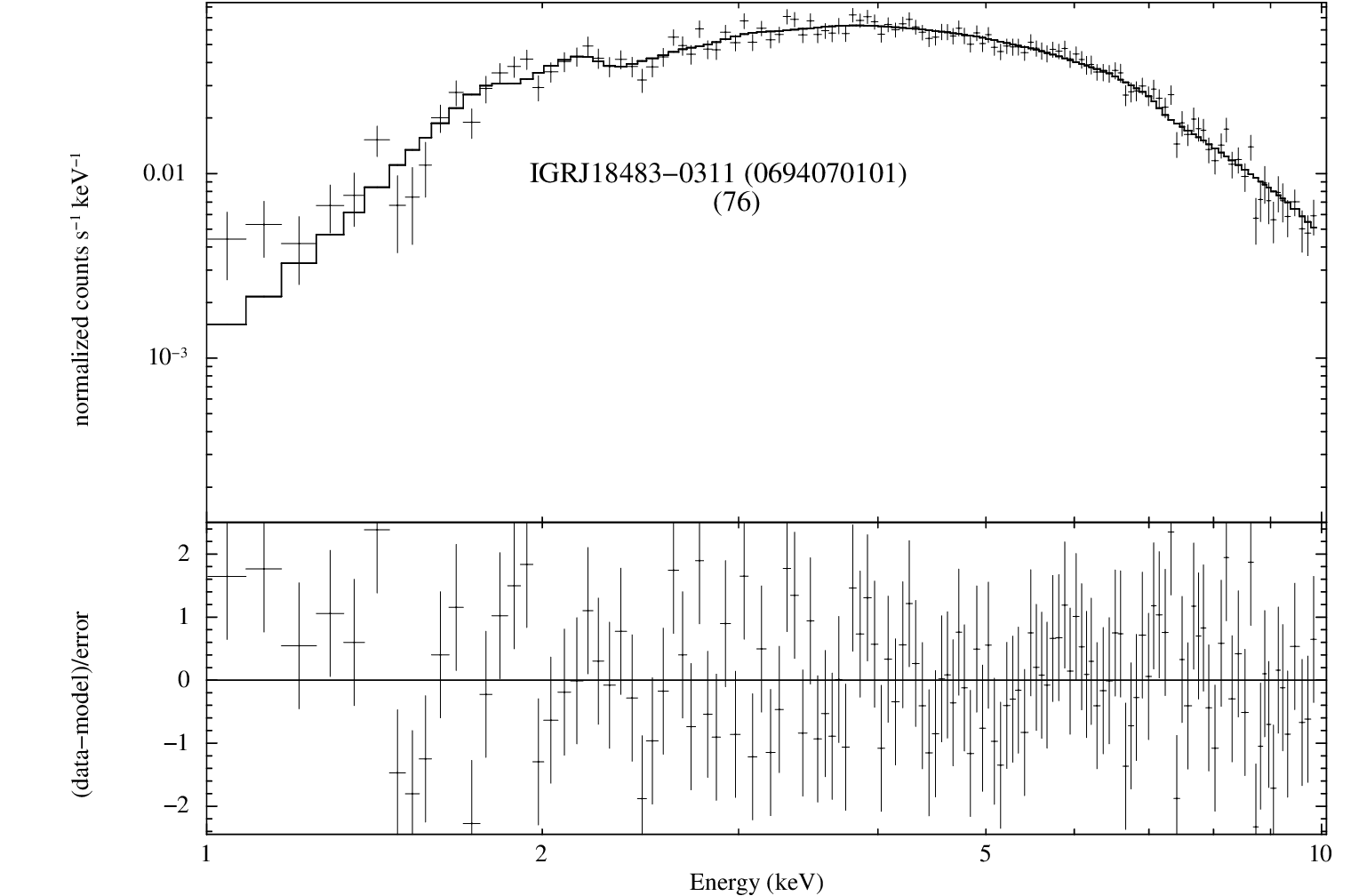}
 \label{spectra5}
\end{figure*}

In this section, we report the details of the spectral fits performed on all observations of classical SgXBs and SFXTs used in this paper. 
We provide all the absorption and power-law parameters used for fits in Table~ A.1 and A.2, while the individual spectra are shown in Fig.~\ref{spectra1}, 
\ref{spectra2}, \ref{spectra3}, \ref{spectra4} and \ref{spectra5} (together with the best-fit models and the residuals from the fit). 
All uncertainties are provided at 90\% confidence limit.
\begin{table*}
\scriptsize
\begin{center}  
\caption{Absorption and power-law parameters used for spectral fits of classical SgXBs (S. No 1-55) and SFXTs (S. No 56-76) used in this paper.}
\resizebox{\textwidth}{!}{ %
\begin{tabular}{@{}ccccccccccccccc@{}} 
\hline
\hline
No & Source & OBSID (Mission)& ${N_{\rm H1}}$ & ${N_{\rm H2}}$ & ${C_{V}}$ & $\Gamma$  & Emission Line  & EW &  $\chi^{2}_{red}$/dof & Flux (1-10 keV) & 
Distance\\
& & & $10^{22}$ atoms cm$^{-2}$ & $10^{22}$ atoms cm$^{-2}$ & & &(\rm{keV})  & (\rm{keV}) & & 10 $^{-11}$ ergs cm$^{-2}$ s$^{-1}$ &(kpc) \\
\noalign{\smallskip}  
\hline
\multicolumn{12}{c}{Classical SgXBs} \\
\hline
\noalign{\smallskip}  
(1) & IGR J00370+6122 & 0501450101 (XMM) & 0.81 $\pm$ 0.03 & 3.20 $\pm$ 0.28 & 0.62 $\pm$ 0.02 & 1.60 $\pm$ 0.03 & 6.42 $\pm$ 0.10 & 
0.014 $\pm$ 0.007 & 1.48/169 & 7.48 $\pm$ 0.05 & 3.3 \citep{dist_igrj00370}\\
 \noalign{\smallskip}  
 
(2) & 4U 0114+65 & 406017010 (Suzaku) & 2.99 $\pm$ 0.15 & 8.00 $\pm$ 0.80 & 0.64 $\pm$ 0.03 & 1.05 $\pm$ 0.03 & 6.437 $\pm$ 0.014 & 
0.0319 $\pm$ 0.004 & 1.01/211 & 14.44 $\pm$ 0.06 & 7.0 $\pm$ 3.6 \citep{dist_4u0114}\\
& & & & & & & 7.08 $\pm$ 0.04 & 0.020 $\pm$ 0.005  \\
 \noalign{\smallskip}  
 
(3) & Vela X-1$^{*}$ & 0406430201 (XMM) & 3.29 $\pm$ 0.25 & 22.43 $\pm$ 2.55 & 0.31 $\pm$ 0.02  & 1.00 $\pm$ 0.01 & 6.419 $\pm$ 0.005 & 
0.067 $\pm$ 0.001 & 1.29/101 & 214.99 $\pm$ 0.91 & 1.9 $\pm$ 0.2 \citep{dist_velax1}\\
& & & & & & &  6.71 $\pm$ 0.02 & 0.007 $\pm$ 0.002 \\
& & & & & & & 7.05 $\pm$ 0.01 & 0.019 $\pm$ 0.001  \\  

(4) & & 0111030101 (XMM) & 8.96 $\pm$ 1.00 & 13.37 $\pm$ 1.30 & 0.95 & 1.04 $\pm$ 0.02 & 6.361 $\pm$ 0.004 & 
0.059 $\pm$ 0.003 & 1.45/90 & 75.23 $\pm$ 0.15 & 1.9 $\pm$ 0.2 \citep{dist_velax1} \\

(5) & & 403045010 (Suzaku) & 2.63 $\pm$ 0.11 & - & - & 1.22 $\pm$ 0.01 & 6.394 $\pm$ 0.003 & 
0.056 $\pm$ 0.002 & 1.42/159 & 126.29 $\pm$ 0.39 & 1.9 $\pm$ 0.2 \citep{dist_velax1} \\
& & & & & & & 7.07 $\pm$ 0.01 & 0.019 $\pm$ 0.001  \\


(6)& GX 301-2$^{*}$ & 0555200301$^{1}$ (XMM) & 30.39 $\pm$ 0.09 & 70.93 $\pm$ 0.40  & 0.84 $\pm$ 0.01 & 0.65 $\pm$ 0.01 & 6.40 $\pm$ 0.01
& 0.604 $\pm$ 0.006  & 1.53/103 &  79.2 $\pm$ 1.33 & 3.04 \citep{dist_gx301-21}\\
& & & & & & & 7.06 $\pm$ 0.002 & 0.129 $\pm$ 0.002 \\
 \noalign{\smallskip}  
 (7) &  & 0555200401$^{2}$ (XMM) & 39.11 $\pm$ 0.11 & 72.14 $\pm$ 0.95 & 0.82 $\pm$ 0.02 &0.74 $\pm$ 0.01 & 6.42 $\pm$ 0.02 
& 0.746 $\pm$ 0.009 & 1.51/1366 &  102.30 $\pm$ 0.50 & 3.04 \citep{dist_gx301-21}\\
& & & & & & & 7.09 $\pm$ 0.01 & 0.181 $\pm$ 0.003 \\
 \noalign{\smallskip}  
 (8) &  & 403044020$^{3}$ (Suzaku) & 22.49 $\pm$ 2.22 & 18.45 $\pm$ -3.00  & 0.61 $\pm$ 0.17 &0.99 $\pm$ 0.02  & 6.39 $\pm$ & 
0.127 $\pm$ & 1.35/155 & 87.044 $\pm$ 0.25 & 3.04 \citep{dist_gx301-21} \\
& & & & & & & 7.09 $\pm$ 0.02 & 0.017 $\pm$ 0.003 \\
 \noalign{\smallskip}  
(9) & & 403044010  (Suzaku) & 14.95 $\pm$ 1.40 & 52.76 $\pm$ 3.48 & 0.95 & 1.03 $\pm$ 0.11 & 6.38 $\pm$ 0.01 & 
0.219 $\pm$ 0.014 & 0.92/160 & 16.31 $\pm$ 0.71 & 3.04 \citep{dist_gx301-21} \\
 \noalign{\smallskip}  

(10) & 4U 1538-522$^{e}$ & 0152780201 (XMM) & 0.64 $\pm$ 0.06 & 20.05 $\pm$ 2.05 & 0.74 $\pm$ 0.04 & 0.64 $\pm$ 0.09 & 6.446 $\pm$ 0.021 & 
0.6067 $\pm$ 0.060 & 1.39/148 & 1.06 $\pm$ 0.01 & 6.4 $\pm$ 1 \citep{dist_4u1538}\\
& & & & & & & 6.92 $\pm$ 0.01 & 0.069 $\pm$ 0.021  \\
 \noalign{\smallskip}  
(11) & & 407068010 (Suzaku) & 1.48 $\pm$ 0.02 & - & - & 1.19 $\pm$ 0.01 & 6.423 $\pm$ 0.009 & 
0.069 $\pm$ 0.005 & 0.98/215 & 43.79 $\pm$ 0.16 & 6.4 $\pm$ 1 \citep{dist_4u1538}\\
 \noalign{\smallskip} 

 (12) & IGR J16207-5129 & 0402920201 (XMM) & 3.03 $\pm$ 1.14 & 6.38 $\pm$ 1.10 & 0.83 $\pm$ 0.15 & 1.11 $\pm$ 0.09 & 6.43 $\pm$ 0.04 &
0.033 $\pm$ 0.012 & 1.28/166 & 2.09 $\pm$ 0.02 & $6.10_{-3.50}^{+8.90}$ \citep{dist_many2}\\
 \noalign{\smallskip}  
(13) & & 402065020 (Suzaku) & 11.30 $\pm$ 0.89 & - & - & 1.23 $\pm$ 0.11 & 6.427 $\pm$ 0.035 & 
0.058 $\pm$ 0.019 & 1.17/116 & 1.75 $\pm$ 0.03 & $6.10_{-3.50}^{+8.90}$ \citep{dist_many2}\\
 \noalign{\smallskip} 

 (14) & IGR J16318-4848 & 0154750401 (XMM) & 153.28 $\pm$ 18.50 & - & - & 0.34 $\pm$ 0.44 & 6.389 $\pm$ 0.003 & 
 2.33 $\pm$ 0.92 & 1.33/59 & 0.725 $\pm$ 0.001 & 2.60 \citep{dist_igrj16318-4848}\\
& & & & & & & 7.24 $\pm$ 0.06 & 0.623 $\pm$ 0.360 \\

(15) & & 0201000201 (XMM) & 120.63 $\pm$ 11.00 & - & - & 0.45 $\pm$ 0.27 & 6.403 $\pm$ 0.004 & 
1.17 $\pm$ 0.18 & 1.36/66 & 0.732 $\pm$ 0.012 & 2.60 \citep{dist_igrj16318-4848} \\
& & & & & & & 7.05 $\pm$ 0.02 & 0.258 $\pm$ 0.052 \\

(16) & & 0201000301 (XMM) & 133.30 $\pm$ 27.50 & - & - & 0.35 $\pm$ 0.64 & 6.41 $\pm$ 0.01 & 
2.55 $\pm$ 1.94 & 0.97/101 & 0.235 $\pm$ 0.002 & 2.60 \citep{dist_igrj16318-4848}\\
& & & & & & & 7.05 $\pm$ 0.01 & 0.258 $\pm$ 0.156 \\

(17) &  & 401094010$^{4}$ (Suzaku) & 124.16 $\pm$ 9.05 & - & - &  1.07 $\pm$ 0.20 & 6.395 $\pm$ 0.005 & 
0.758 $\pm$ 0.050 & 1.64/124 & 0.922 $\pm$ 0.008 & 2.60 \citep{dist_igrj16318-4848} \\
& & & & & & & 7.09 $\pm$ 0.04 & 0.108 $\pm$ 0.008 \\

(18) &   & 0201000401$^{5}$ (XMM) & 114.56 $\pm$ 8.50 & - & - & 0.71 $\pm$ 0.33 & 6.404 $\pm$ 0.001 & 
1.115 $\pm$ 0.213 & 1.12/60 & 0.507 $\pm$ 0.007 & 2.60 \citep{dist_igrj16318-4848}\\
& & & & & & & 7.05 $\pm$ 0.03 & 0.298 $\pm$ 0.055 \\

(19) &  & 0742270201$^{6}$ (XMM) & 146.16 $\pm$ 4.50 & - & - & 0.66 $\pm$ 0.02 & 6.439 $\pm$ 0.003 & 
1.76 $\pm$ 0.04 & 1.26/101 & 0.570 $\pm$ 0.047 & 2.60 \citep{dist_igrj16318-4848}\\
& & & & & & & 7.11 $\pm$ 0.02 & 0.432 $\pm$ 0.077 \\
 \noalign{\smallskip}  
 
(20) & IGR J16320-4751 & 0128531101 (XMM) & 20.36 $\pm$ 2.45 & - & - & 1.09 $\pm$ 0.20 & 6.43 $\pm$ 0.15 & 
0.109 $\pm$ 0.070 & 0.89/105 & 1.392 $\pm$ 0.031 & 3.5 \citep{dist_igrj18483}\\
(21) & & 0556140101 (XMM) & 26.19 $\pm$ 1.15 & - & - & 0.41 $\pm$ 0.08  & 6.420 $\pm$ 0.002 & 
0.243 $\pm$ 0.017 & 1.30/99 & 6.68 $\pm$ 0.08 & 3.5 \citep{dist_igrj18483}\\
& & & & & & & 7.04 $\pm$ 0.03 & 0.049 $\pm$ 0.011 \\
(22) & & 0556140201 (XMM) & 18.88 $\pm$ 0.68 & - & - & 0.309 $\pm$ 0.056 & 6.420 $\pm$ 0.004 & 
0.210 $\pm$ 0.012 & 1.21/123 & 13.75 $\pm$ 0.20 & 3.5 \citep{dist_igrj18483}\\
& & & & & & & 7.01 $\pm$ 0.06 & 0.061 $\pm$ 0.022 \\
(23) & & 0556140301 (XMM) & 18.68 $\pm$ 0.65 & -& - & 0.34 $\pm$ 0.06 & 6.42 $\pm$ 0.01 & 
0.176 $\pm$ 0.014 & 1.18/119 & 0.132 $\pm$ 0.001 & 3.5 \citep{dist_igrj18483}\\
& & & & & & & 7.01 $\pm$ 0.06 & 0.071 \\
 \noalign{\smallskip}  
(24) & & 0556140401 (XMM) & 22.30 $\pm$ 0.95 & - & - & 0.56 & 6.41 $\pm$ 0.01& 
0.157 $\pm$ 0.001 & 1.37/117 & 8.29 $\pm$ 0.07 & 3.5 \citep{dist_igrj18483}\\
& & & & & & & 7.23 $\pm$ 0.30 & 0.108 $\pm$ 0.112 \\
(25) & & 0556140501 (XMM) & 21.92 $\pm$ 1.80 & - & - & 0.56 $\pm$ 0.14 & 6.41 $\pm$ 0.02 & 
0.125 $\pm$ 0.025 & 1.07/89 & 6.71 $\pm$ 0.13 & 3.5 \citep{dist_igrj18483}\\
(26) & & 0556140601 (XMM) & 19.13 $\pm$ 0.28 & - & - & 0.18 $\pm$ 0.03 & 6.420 $\pm$ 0.004 & 
0.216 $\pm$ 0.010 & 1.52/122 & 0.172 $\pm$ 0.001 & 3.5 \citep{dist_igrj18483}\\
& & & & & & & 6.99 $\pm$ 0.03 & 0.117 $\pm$ 0.012 \\
(27) & & 0556140701 (XMM) & 41.95 $\pm$ 2.90 & - & - & 0.022 $\pm$ 0.131 & 6.420 $\pm$ 0.005 & 
0.469 $\pm$ 0.023 & 1.37/96 & 6.23 $\pm$ 0.05 & 3.5 \citep{dist_igrj18483}\\
& & & & & & & 7.03 $\pm$ 0.05 & 0.167 $\pm$ 0.042 \\
(28) & & 0556140801 (XMM) & 19.86 $\pm$ 1.02 & - & - & 0.54 $\pm$ 0.09 & 6.409 $\pm$ 0.004 & 
0.170 $\pm$ 0.010 & 1.16/113 & 8.17 $\pm$ 0.07 & 3.5 \citep{dist_igrj18483}\\
& & & & & & & 7.07 $\pm$ 0.24 & 0.298 $\pm$ 0.202 \\
(29) & & 0556141001 (XMM) & 18.27 $\pm$ 0.94 & - & - & 0.45 $\pm$ 0.08 & 6.408 $\pm$ 0.010 & 
0.151 $\pm$ 0.010 & 0.95/113 & 9.91 $\pm$ 0.20 & 3.5 \citep{dist_igrj18483}\\
& & & & & & & 7.07 $\pm$ 0.17 & 0.286 $\pm$ 0.118 \\
(30) & & 0201700301 (XMM) & 14.41 $\pm$ 0.48 & - & - & 0.49 $\pm$ 0.05 & 6.397 $\pm$ 0.001 &
0.107 $\pm$ 0.007 & 1.50/125 & 6.83 $\pm$ 0.02 & 3.5 \citep{dist_igrj18483}\\
& & & & & & & 7.01 $\pm$ 0.20 & 0.365 $\pm$ 0.308 \\
 \noalign{\smallskip}  
 
 
(31) & IGR J16393-4643 & 0206380201 (XMM) & 25.51 $\pm$ 1.28 & - & - & 0.61 $\pm$ 0.09 & 6.393 $\pm$ 0.027 & 
0.084 $\pm$ 0.017 & 1.05/107 & 4.14 $\pm$ 0.04 & 10.6 \citep{dist_igrj16479-4514} \\
(32) &  & 0604520201 (XMM) & 25.28 $\pm$ 2.95 & - & - & 0.65 $\pm$ 0.21 & 6.48 $\pm$ 0.06 & 
0.069 $\pm$ 0.035 & 0.71/94 & 1.27 $\pm$ 0.02 & 10.6 \citep{dist_igrj16479-4514}\\
(33) & & 404056010 (Suzaku) & 29.36 $\pm$ 1.75 & - & - & 1.12 $\pm$ 0.13 & 6.390 $\pm$ 0.048 & 
0.057 $\pm$ 0.020 & 1.16/129 & 2.33 $\pm$ 0.03 & 10.6 \citep{dist_igrj16479-4514}\\
 \noalign{\smallskip}  
 
(34)& IGR J16418-4532 & 0405180501 (XMM) & 3.51 $\pm$ 0.89 & 5.76 $\pm$ 0.99 & 0.64 $\pm$ 0.17 & 1.47 $\pm$ 0.11 & 6.4 & 
0.008 & 0.95/133 & 0.97 $\pm$ 0.02 & 13 \citep{dist_igrj18483}\\
(35)& & 0679810201 (XMM)& 1.42 $\pm$ 0.14 & 3.93 $\pm$ 0.31 & 0.95 & 0.99 $\pm$ 0.04 & 6.39 $\pm$ 0.20 
& 0.013 $\pm$ 0.009 & 1.19/154 & 6.41 $\pm$ 0.08 & 13 \citep{dist_igrj18483}\\
 \noalign{\smallskip}  
 
(36) & IGR J16493-4348 & 401054010 (Suzaku) & 8.26 $\pm$ 1.85 & - & - & 1.41 $\pm$ 0.21 & 6.452 $\pm$ 0.093 & 
0.046 $\pm$ 0.001 & 1.61/67 & 1.42 $\pm$ 0.03 & 6 \citep{unified2017} \\
 \noalign{\smallskip}  

(37) & OAO 1657-415 & 406011010 (Suzaku) & 17.28 $\pm$ 1.35 & 37.14 $\pm$ 6.57 & 0.57 $\pm$ 0.07 & 0.87 $\pm$ 0.04 & 6.459 $\pm$ 0.002 & 
0.254 $\pm$ 0.008 & 1.31/162 & 14.72 $\pm$ 0.15 & 1.5 \citep{dist_oao} \\
& & & & & & & 7.14 $\pm$ 0.04 & 0.098 $\pm$ 0.007  \\
 \noalign{\smallskip}  
 
(38) & 4U 1700-37 & 0083280101 (XMM) & 3.88 $\pm$ 0.19 & 5.89 $\pm$ 0.44 & 0.67 $\pm$ 0.04 & 1.09 $\pm$ 0.03 & 6.43 $\pm$ 0.01 & 
0.037 $\pm$ 0.004 & 1.59/154 & 81.97 $\pm$ 0.54 & 2.12 $\pm$ 0.34 \citep{dist_4u17001}\\
& & & & & & & 7.10 & 0.021  \\
(39) &  & 0083280201$^{7}$ (XMM) & 1.36 $\pm$ 0.08 & 5.55 $\pm$ 1.20 & 0.27 $\pm$ 0.04 & 0.75 $\pm$ 0.03 & 6.43 $\pm$ 
&  0.083 $\pm$ 0.004 & 1.42/137 & 187.09 $\pm$ 1.39 & 2.12 $\pm$ 0.34 \citep{dist_4u17001}\\
& & & & & & & 7.11 $\pm$ 0.03 & 0.014 $\pm$  0.003 \\
(40) &  & 0083280301 (XMM) & 6.10 $\pm$ 0.25  & 10.74 $\pm$ 0.46 & 0.95 & 1.11 $\pm$ 0.03  & 6.43 $\pm$ 0.01 & 
0.064 $\pm$ 0.007 & 1.38/157 & 84.41 $\pm$ 0.92 & 2.12 $\pm$ 0.34 \citep{dist_4u17001} \\
& & & & & & & 7.16 $\pm$ 0.06 & 0.013 $\pm$ 0.005 & & \\
(41) &  & 401058010$^{8}$ (Suzaku) & 1.91 $\pm$ 0.10 & 3.92 $\pm$ 0.25 & 0.61 $\pm$ 0.03 & 0.23 $\pm$ 0.09 & 6.41 $\pm$ 0.01 & 
0.079 $\pm$ 0.001 & 1.42/214 & 197.03 $\pm$ 0.32 & 2.12 $\pm$ 0.34 \citep{dist_4u17001}\\
& & & & & & & 7.07 $\pm$ 0.02 & 0.018 $\pm$ 0.002 & & \\
 \noalign{\smallskip} 
 
(42) & EXO 1722-363 & 0405640301 (XMM) & 11.77 $\pm$ 0.45 & - & - & 0.82 $\pm$ 0.08  & 6.41 $\pm$ 0.01
& 0.216 $\pm$ 0.025 & 1.07/115 & 5.19 $\pm$ 0.12 & 6-10.5 \citep{unified2017}  \\
& & & & & & & 7.11 $\pm$ 0.05 & 0.049 $\pm$ 0.019 & & \\
(43) &  & 0405640401 (XMM) & 26.96 $\pm$ 2.15 & - & - & 0.64 $\pm$ 0.15 & 6.41 $\pm$ 0.01 
& 0.221 $\pm$ 0.033 & 1.04/84 & 2.31 $\pm$ 0.19 & 6-10.5 \citep{unified2017}\\
& & & & & & & 7.14 $\pm$ 0.12 & 0.057 $\pm$ 0.001 \\
(44) &  & 0405640701 (XMM) & 28.06 $\pm$ 1.20 & - & - & 0.71 $\pm$ 0.09 & 6.45 $\pm$ 0.02 & 
0.100 $\pm$ 0.010 & 1.01/132 & 2.46 $\pm$ 0.06 & 6-10.5 \citep{unified2017}\\
& & & & & & & 7.13 $\pm$ 0.02 & 0.052 $\pm$ 0.017 \\
(45) & & 0405640801 (XMM) & 13.83 $\pm$ 4.28 & 19.85 $\pm$ 4.41 & 0.83 $\pm$ 0.17 & 1.01 $\pm$ 0.12 & 6.42 $\pm$ 0.01 & 
0.153 $\pm$ 0.019 & 0.95/116 & 3.81 $\pm$ 0.22 & 6-10.5 \citep{unified2017} \\
& & & & & & & 7.09 $\pm$ 0.02 & 0.069 $\pm$ 0.034 \\
(46) &  & 0405640901 (XMM) & 19.95 $\pm$ 0.98 & - & - & 0.93 $\pm$ 0.09 & 6.42 $\pm$ 0.05 &
0.046 $\pm$ 0.016 & 0.99/112 & 2.68 $\pm$ 0.10 & 6-10.5 \citep{unified2017} \\
& & & & & & & 7.09 $\pm$ 0.35 & 0.033  \\
(47) &  & 0206380401 (XMM) & 15.63 $\pm$ 3.50 & - & - & 0.42 $\pm$ 0.30 & 6.40 $\pm$ 0.05& 
0.144 $\pm$ 0.072 & 0.97/54 & 5.10 $\pm$ 0.95 & 6-10.5 \citep{unified2017} \\
& & & & & & & 7.07 & 0.058  \\
 
 \noalign{\smallskip}  
(48) & SAX J1802.7-2017 & 0206380601 (XMM) & 7.41 $\pm$ 0.44 &  & & 0.78 $\pm$ 0.08 & 6.50 $\pm$ 0.14 & 
0.116 $\pm$ 0.060 & 1.21/125 & 4.87 $\pm$ 0.05 & 12.4 \citep{unified2017}\\
(49) & & 0745060401 (XMM) & 4.65 $\pm$ 1.11 & 14.64 $\pm$ 4.50 & 0.71 $\pm$ 0.09 & 0.78 $\pm$ 0.11 & 6.38 & 
0.038 $\pm$ 0.012 & 1.38/146 & 1.05 $\pm$ 0.01 & 12.4 \citep{unified2017}\\
(50) & & 0745060501 (XMM) & 2.42 $\pm$ 0.27 & 7.44 $\pm$ 2.11 & 0.56 $\pm$ 0.07 & 1.07 $\pm$ 0.09 & 6.39 & 
0.096 $\pm$ 0.016 & 1.03/146 & 3.38 $\pm$ 0.03 & 12.4 \citep{unified2017}\\
 (51) & & 0745060601 (XMM) & 2.21 $\pm$ 0.85 & 6.49 $\pm$ 0.60 & 0.89 $\pm$ 0.08 & 0.67 $\pm$ 0.06 & 6.39 & 
0.105 $\pm$ 0.017 & 1.07/149 & 4.64 $\pm$ 0.03 & 12.4 \citep{unified2017} \\
(52) & & 0745060801 (XMM) & 1.75 $\pm$ 0.05 & - & - & 0.98 $\pm$ 0.02 & 6.41 $\pm$ 0.04 & 
0.057 $\pm$ 0.015 & 1.02/152 & 4.83 $\pm$ 0.04 & 12.4 \citep{unified2017}\\
 \noalign{\smallskip}  
 
(53) & XTE J1855-026 & 409022010 (Suzaku) & 4.54 $\pm$ 1.75 & 14.75 $\pm$ 3.50 & 0.63 $\pm$ 0.18 & 1.36 $\pm$ 0.08 & 6.408 $\pm$  0.011 & 
0.100 $\pm$ 0.009 & 1.27/215 & 6.74 $\pm$ 0.11 & 8.6 $\pm$ 0.8 \citep{distance_1855} \\
& & & & & & & 7.15 $\pm$ 0.08 & 0.023 $\pm$ 0.009  \\

 \noalign{\smallskip}  
(54) & 4U 1909+07 & 405073010 (Suzaku) & 5.54 $\pm$ 0.60 & 6.87 $\pm$ 2.15 & 0.49 $\pm$ 0.15 & 1.33 $\pm$ 0.05 & 6.424 $\pm$ 0.012 & 
0.075 $\pm$ 0.007 & 1.37/109 & 15.64 $\pm$ 0.26 & 7.00 $\pm$ 3.0 \citep{dist_4u1909} \\
& & & & & & & 7.10 & 0.018 $\pm$ 0.005  \\

\noalign{\smallskip}  
(55) & IGR J19140+0951 & 0761690301 (XMM) & 7.67 $\pm$ 0.46 & - & - & 1.79 $\pm$ 0.11 & 6.4 & 
0.032 & 0.95/116 & 0.29 $\pm$ 0.004 & 2-5 \citep{unified2017}  \\
\hline
\hline

\end{tabular}
}
\tablefoot{\scriptsize 
$^{1}$ Additional Emission lines centroid energy (uncertainty), EW (uncertainty) = 3.72 (0.01), 0.029 (0.001) , 5.41 (0.02), 0.013 (0.002), 6.29 (0.02), 0.033 (0.001), 
7.47 (0.01), 0.065 (0.002), 8.11 (0.04), 0.11 (0.01)   \\   
$^{2}$ Additional Emission lines centroid energy (uncertainty), EW (uncertainty) = 3.73 (0.01), 0.091 (0.004), 5.43 (0.01), 0.015 (0.002), 6.20 (0.01), 0.015 (0.001), 
7.50 (0.01), 0.015 (0.001), 8.23 (0.02), 0.23 (0.01) \\
$^{3}$ Additional Emission lines centroid energy (uncertainty), EW (uncertainty) = 7.46 (0.02), 0.016 (0.002) \\
$^{4}$ Additional Emission line centroid energy (uncertainty), EW (uncertainty) = 7.39 (0.11), 0.138 (0.065) \\
$^{5}$ Additional Emission line centroid energy (uncertainty), EW (uncertainty) = 7.47 (0.06), 0.145 (0.083) \\
$^{6}$ Additional Emission line centroid energy (uncertainty), EW (uncertainty) = 7.52 (0.02), 0.287 (0.047) \\
$^{7}$ bbKT (uncertainty), bbnorm (uncertainty) = 0.10 (0.02), 0.0053 (0.0021) \\
$^{8}$ Ecut (uncertainty) = 7.57 (1.00) \\
$^{*}$ The X-ray spectra of Vela X-1 and GX 301-2 are dominated below 3 keV by orbital-dependent emission lines and soft excesses which origin is still
largely debated (see \citealt[]{martinez2014,suchy2011,islam2014} and references therein). As the interpretation of these features
is beyond the scope of the current paper, we excluded the energy range 1-3 keV from the fit of the spectra of these two sources. We verified that this has
no quantitative impact on the reported results and on their average absorption column density measured. \\
$^{e}$ Eclipse times are excluded. }
\end{center} 
\label{obslog2}
\end{table*}

\begin{table*}
\scriptsize
\begin{center}  
\caption{Continuation of Table~A.1.}
\resizebox{\textwidth}{!}{ %
\begin{tabular}{@{}ccccccccccccccc@{}} 
\hline
\hline
No & Source & OBSID (Mission)& ${N_{\rm H1}}$ & ${N_{\rm H2}}$ & ${C_{V}}$ & $\Gamma$  & Emission Line  & EW &  $\chi^{2}_{red}$/dof & Flux (1-10 keV) & 
Distance\\
& & & $10^{22}$ atoms cm$^{-2}$ & $10^{22}$ atoms cm$^{-2}$ & & &(\rm{keV})  & (\rm{keV}) & & 10 $^{-11}$ ergs cm$^{-2}$ s$^{-1}$ &(kpc) \\
\hline
\multicolumn{12}{c}{SFXTs} \\
\hline
(56)& IGR J11215-5952 & 0405181901 (XMM) & 0.66 $\pm$ 0.04 & 2.27 $\pm$ 0.49 & 0.46 $\pm$ 0.05 & 1.19 $\pm$ 0.05 & 6.46 $\pm$ 0.06 & 
0.042 $\pm$ 0.012 & 1.16/163 & 3.77 $\pm$ 0.04 & 6.2 \citep{unified2017}\\
 \noalign{\smallskip}  
 
(57) & IGR J16195-4945 & 401056010 (Suzaku) & 8.63 $\pm$ 0.83 & - & - & 1.19 $\pm$ 0.11 & 6.38 $\pm$ 0.06 & 
0.052 $\pm$ 0.019 & 1.10/171 & 1.74 $\pm$ 0.08 & 5 \citep{dist_igrj16195-4945}\\
 \noalign{\smallskip} 
 
(58)& IGR J16328-4726 & 0728560201 (XMM) & 4.23 $\pm$ 0.55 & 12.06 $\pm$ 0.88 & 0.97 & 1.36 $\pm$ 0.06 & 6.41 $\pm$ 0.05 & 0.029 $\pm$ 0.011 & 1.14/133 & 
1.65 $\pm$ 0.02 &  3-10 \citep{unified2017}\\
(59)&  & 0728560301 (XMM) & 12.86 $\pm$ 0.63 & - & - & 1.40 $\pm$ 0.09 & 6.42 $\pm$ 0.07 & 
0.075 $\pm$ 0.020 & 0.98/118 & 1.47 $\pm$ 0.03 & 3-10 \citep{unified2017}\\
(60)&  & 0654190201 (XMM) & 16.88 $\pm$ 1.10 & - & - & 1.45 $\pm$ 0.12 & 6.4 & 0.015 $\pm$ 0.010 & 1.09/106 & 
0.79 $\pm$ 0.04 & 3-10 \citep{unified2017} \\
 \noalign{\smallskip}  
 
(61) & IGR J16479-4514$^{e}$ & 406078010 (Suzaku) & 2.33 $\pm$ 1.11 & 6.06 $\pm$ 2.16 & 0.90 $\pm$ 0.09 & 1.44 $\pm$ 0.14 & 6.34 & 
0.081 $\pm$ 0.026 & 0.73/47 & 1.30 $\pm$ 0.03 & \citep{dist_igrj16479-4514} \\
 \noalign{\smallskip}  
 
(62)& IGR J17354-3255 & 0701230101 (XMM) & 2.17 $\pm$ 0.53 & 5.73 $\pm$ 1.12 & 0.96 & 1.33 $\pm$ 0.09 & 6.41 $\pm$ 0.04 
& 0.036 $\pm$ 0.019 & 0.94/145 & 9.80 $\pm$ 0.53 & 8 \citep{dist_igrj17354} \\
(63)& & 0701230701 (XMM) & 5.22 $\pm$ 0.21 & - &-& 1.20 $\pm$ 0.05 & 6.407 $\pm$ 0.065 
& 0.094 $\pm$ 0.045 & 1.10/138 & 1.86 $\pm$ 0.43 & 8 \citep{dist_igrj17354} \\
 \noalign{\smallskip}  
 
(64)& XTE 1739-302  & 0554720101 (XMM) & 2.08 $\pm$ 0.44 & 3.87 $\pm$ 1.30 & 0.67 $\pm$ 0.16 & 1.96 $\pm$ 0.15 & 6.4 
& 0.004 & 0.89/122 & 0.25 $\pm$ 0.01 & 2.7 \citep{dist_igrj18483} \\
(65) & & 0561580101 (XMM) & 3.28 $\pm$ 0.55 & 7.03 $\pm$ 3.35 & 0.52 $\pm$ 0.15 & 1.95 $\pm$ 0.18 & 6.42 $\pm$ 0.11 
& 0.038 $\pm$ 0.027 & 0.89/117 & 0.37 $\pm$ 0.04 & 2.7 \citep{dist_igrj18483} \\
 \noalign{\smallskip}
 
(66)& IGR J17544-2619 & 0744600101 (XMM) & 0.36 $\pm$ 0.08 & 1.23 $\pm$ 0.13 & 0.95 & 1.19 $\pm$ 0.04 & 6.4 
& 0.013 & 1.21/168 &  0.71 $\pm$ 0.01 & 3.20 $\pm$ 1.00 \citep{dist_igrj17544-2619}\\
(67)&  & 0679810401 (XMM) & 1.59 $\pm$ 0.57 & 3.70 $\pm$ 1.71 & 0.76 $\pm$ 0.15 & 2.60 $\pm$ 0.31 & 6.40 
& 0.034 $\pm$ 0.029 & 0.79/70 & 0.19 $\pm$ 0.03 & 3.20 $\pm$ 1.00 \citep{dist_igrj17544-2619}\\
(68) &  & 402061010 (Suzaku) & 1.24 $\pm$ 0.21 & 1.92 $\pm$ 0.30 & 0.71 $\pm$ 0.11 & 1.26 $\pm$ 0.03 & 6.32 $\pm$ 0.08 
& 0.019 $\pm$ 0.011 & 0.99/150 & 5.55 $\pm$ 0.52 & 3.20 $\pm$ 1.00 \citep{dist_igrj17544-2619}\\
(69)&  & 0154750601 (XMM) & 1.31 $\pm$ 0.45 & 3.76 $\pm$ 1.21 & 0.78 $\pm$ 0.10  & 2.31 $\pm$ 0.27 & 6.40 
& 0.06 $\pm$ 0.05 &0.92/126 & 2.17 $\pm$ 0.43 & 3.20 $\pm$ 1.00 \citep{dist_igrj17544-2619}\\
 \noalign{\smallskip}  
 
(70) & SAX J1818.6-1703 & 0693900101 (XMM) & 27.93 $\pm$ 1.15 & - & - & 0.49 $\pm$ 0.07 & 6.4 & 
0.0069 $\pm$ & 1.16/139 & 2.99 $\pm$ 0.06 & 2.1 $\pm$ 0.1 \citep{dist_saxj1818}\\
 \noalign{\smallskip}  
 
(71)& IGR J18410-0535 & 0604820301(XMM) &2.87 $\pm$ 0.18 & 10.81 $\pm$ 0.55 & 0.95 & 1.13 $\pm$ 0.05 
&6.44 $\pm$ 0.05 & 0.026 $\pm$ 0.010&1.34/154 & 2.36 $\pm$ 0.02 & $3.2_{-1.5}^{+2.0}$ \citep{dist_many2} \\
(72)&  & 505090010(Suzaku) & 1.91 $\pm$ 0.28  & 5.44 $\pm$ 1.21 & 0.69 $\pm$ 0.07 & 1.70 $\pm$ 0.07 & 6.40 $\pm$ 0.02 & 0.051 $\pm$ 0.008 
 & 0.96/169 & 4.75 $\pm$ 0.12 & $3.2_{-1.5}^{+2.0}$ \citep{dist_many2} \\
 \noalign{\smallskip}  

 (73)& IGR J18450-0435 & 0728370801 (Suzaku) & 2.78 $\pm$ 0.76 & 6.41 $\pm$ 1.15 & 0.84 $\pm$ 0.10 & 1.39 $\pm$ 0.11 & 6.49 $\pm$ 0.09 
& 0.036 $\pm$ 0.003 & 1.19/132 & 1.13 $\pm$ 0.04 & 4 \citep{dist_igrj18450}\\
(74)&  & 0306170401 (XMM)  & 1.79 $\pm$ 0.18 & 4.39 $\pm$ 1.20 & 0.55 $\pm$ 0.07 & 1.33 $\pm$ 0.07 & 6.41 $\pm$ 0.03 
& 0.048 $\pm$ 0.003 & 0.85/148 & 3.54 $\pm$ 0.14 & 4 \citep{dist_igrj18450} \\
 \noalign{\smallskip}  

 (75) & IGR J18462-0223 & 0651680301 (XMM) & 16.91 $\pm$ 1.20 & - & - & 1.15 $\pm$ 0.14 & 6.34 $\pm$ 0.06 & 
0.052 $\pm$ 0.035 & 0.98/97 & 1.09 $\pm$ 0.02 & 11 \citep{igrjJ18462} \\
 \noalign{\smallskip}  

 (76)& IGR J18483-0311 & 0694070101 (XMM) & 4.45 $\pm$ 0.75 & 9.09 $\pm$ 3.65 & 0.62 $\pm$ 0.12 & 2.04 $\pm$ 0.17 & 6.40 
& 0.045 & 1.05/128 & 0.32 $\pm$ 0.03 & 3-4 \citep{dist_igrj18483} \\
\hline
\hline
\end{tabular}
}
\tablefoot{\scriptsize \\
$^{e}$ Eclipse times are excluded.}
\end{center} 
\label{obslog3}
\end{table*} 

\end{document}